\documentclass[fleqn,usenatbib,useAMS]{mnras}

\usepackage{graphicx}	
\usepackage{amsmath}	
\usepackage{amssymb}	
\usepackage{multicol}        
\usepackage{bm}		
\usepackage{pdflscape}	
\usepackage{units}   
\usepackage{xcolor} 

\newcommand{\0}{\phantom{0}}
\newcommand{\ntgt}[0]{twelve }

\usepackage[T1]{fontenc}
\usepackage{ae,aecompl}

\usepackage{newtxtext,newtxmath}

\title[Aperture polarimetry of debris disc host stars]{Multi-wavelength aperture polarimetry of debris disc host stars}

\author[J. P. Marshall \textit{et al.}]{Jonathan P. Marshall$^{1,2}$\thanks{Contact e-mail: \href{mailto:jmarshall@asiaa.sinica.edu.tw}{jmarshall@asiaa.sinica.edu.tw}}, Daniel V. Cotton$^{3,4}$, Kimberly Bott$^{5}$, Jeremy Bailey$^{4,6}$, \newauthor Lucyna Kedziora-Chudczer$^{2}$ and Emma L. Brown$^{2}$
\\
$^{1}$Academia Sinica Institute of Astronomy and Astrophysics, 11F of AS/NTU Astronomy-Mathematics Building, No. 1, Sect. 4, Roosevelt Rd,\\ Taipei 10617, Taiwan\\
$^{2}$Centre for Astrophysics, University of Southern Queensland, Toowoomba, QLD 4350, Australia\\
$^{3}$Monterey Institute for Research in Astronomy, 200 Eighth Street, Marina, CA, 93933, USA.\\
$^{4}$Western Sydney University, Locked Bag 1797, Penrith-South DC, NSW 1797, Australia.\\
$^{5}$Department of Earth and Planetary Science, University of California, Riverside, CA, 92521, USA.\\
$^{6}$School of Physics, University of New South Wales, Sydney, NSW 2052, Australia}

\date{Last updated \today; in original form \today}

\pubyear{2023}

\begin{document}
\label{firstpage}
\pagerange{\pageref{firstpage}--\pageref{lastpage}}
\maketitle

\begin{abstract}
Debris discs around main sequence stars have been extensively characterised from infrared to millimetre wavelengths through imaging, spectroscopic, and total intensity (scattered light and/or thermal emission) measurements. Polarimetric observations have only been used sparingly to interpret the composition, structure, and size of dust grains in these discs. Here we present new multi-wavelength aperture polarisation observations with parts-per-million sensitivity of a sample of \ntgt bright debris discs, spanning a broad range of host star spectral types, and disc properties. These measurements were mostly taken with the HIgh Precision Polarimetric Instrument on the Anglo-Australian Telescope. We combine these polarisation observations with the known disc architectures and geometries of the discs to interpret the measurements. We detect significant polarisation attributable to circumstellar dust from HD~377 and HD~39060, and find tentative evidence for HD~188228 and HD~202628.
\end{abstract}





\section{Introduction}
\label{sec:intro}

Almost all stars form with attendant gas- and dust-rich protoplanetary discs which are processed into planetary systems \citep{2015Wyatt}. Within those primordial discs, sub-micron sized dust grains are grown into kilometre-sized planetesimals within a few Myr \citep{2021Lovell}, probably at ice lines due to the enhancement in solid material density favouring their growth \citep{2018Matra,2021Marshall}. These planetesimals are the raw material from which planetary companions aggregate, and abundant small dust grains are generated in mutual collisions. The small dust grains are quickly removed by radiative and collisional processes on timescales much shorter than the lifetime of the host stars  \citep{2010Krivov}, meaning that they must be continually generated by attrition of planetesimals. As such, these evolved circumstellar discs are called ``debris discs'' \citep[for recent reviews see e.g.][]{2014Matthews,2018Hughes}. They are typically observed at infrared to millimetre wavelengths, where the continuum emission from the dust grains dominates over that of the host star \citep[e.g.][]{2013Eiroa,2014Thureau,2017Holland,2018Sibthorpe}. These systems are therefore larger, more massive analogues to the Solar system's Asteroid and Edgeworth-Kuiper belts \citep{2020Horner}. 

Alternatively, or additionally, debris discs can be detected at optical and near-infrared wavelengths by light from the star scattered by the dust grains. Whilst several hundred debris discs have been identified through continuum emission \citep[see e.g.][]{2016Cotten}, the number of debris discs observed in scattered light (either in total intensity or polarisation) is far fewer due to the twin requirements of high contrast and high angular resolution to resolve a disc from its host star. Furthermore, scattered light detection of debris discs is confounded by the orientation of the disc and the scattering properties of the constituent dust grains -- a disc that is bright in thermal emission may not yield a similarly clear detection in scattered light \citep{2014Schneider}. Multiple ground-based facilities, such as VLT/SPHERE and Gemini/GPI (mainly geared toward exoplanet imaging), along with space-based observations by \textit{HST} have continued to expand the menagerie of scattered light discs over the past decade \citep[e.g.][]{2015Perrin,2014Soummer,2016Choquet,2020Esposito}.

Polarimetry provides a mechanism to constrain the properties of scattering particles within debris discs \citep[e.g.][]{2017Milli,2019Milli}. Unpolarised light from stars is scattered by surrounding small dust grains, which induces a polarisation, leading to a measurable imprint of the presence of dust around another star. The magnitude of the scattering-induced polarisation is related to not only a disc's optical depth but also the size, shape and albedo of the dust grains responsible for the scattering \citep[e.g.][]{2000Krivova}. Polarisation is a consequence of scattering and therefore the geometric considerations for scattered light detection of debris discs equally affect polarimetric detection.  Measurement of the polarisation signal from a debris disc host star in multiple wavebands is therefore useful in determining the optical properties of small grains in the debris disc \citep[e.g.][]{2000Krivova,2007Graham}. 

However, the interstellar medium (ISM) also polarises light passing through it at a low level. At the parts-per-million sensitivities of modern instruments the interstellar medium within 50 pc of the Sun is influenced by the local interstellar magnetic field and the line of sight to it \citep{2015Frisch, 2017Cotton}, structures like the Loop I Superbubble \citep{2014Frisch, 2017Cotton, 2020Piirola}, as well as dust clouds \citep{1982Tinbergen}, and filamentary structures \citep{2015Frisch, 2020Piirola}. On average the ISM within 50~pc polarises at 0.37$\pm$0.14 $\times10^{-6}~{\rm pc}^{-1}$  \citep{2020Piirola}. However, the inhomogeneity of the ISM means this value is highly direction dependant \citep{2016Cotton, 2017Cotton, 2020Piirola}. \citet{2010Bailey} and \citet{2020Piirola} both note distinct regions with negligible or very low polarisation to the north. In contrast, stars in the southern sky are generally more polarised with \citet{2017Cotton} finding around 1.6 $\times10^{-6}~{\rm pc}^{-1}$ for stars below a Galactic latitude of 30 degrees ($b < +30$), albeit with significant scatter, and with stars closer than 14.5 pc being less polarised than this. The wavelength dependence of the ISM also varies, with closer regions showing a bluer peak polarisation than is typical \citep{2016Marshall, 2019Cotton}. The signal from any circumstellar dust must be disentangled from this contaminating foreground in aperture polarimetry measurements.

Previous efforts to survey stars for polarised light have thus been strongly limited by instrument sensitivity, typically at the $10^{-4}$ to $10^{-5}$ level \citep[see e.g.][]{1982Tinbergen,1993Leroy}, rendering all but the brightest and closest discs undetectable \citep[e.g.][]{2000Bhatt,2006Chavero,2015GarciaGomez}. Direct measurement of the dust polarisation has been obtained for an increasing number of bright discs through imaging polarisation which sidesteps the need to characterize the intervening ISM at the cost of requiring high contrast imaging with small inner working angles to disentangle the star and disc \citep[see e.g.][]{1991Gledhill,2006Tamura,2007Graham,2020Esposito}. Aperture polarimetry has also been moderately successful as a method to characterize circumstellar dust, but the measurements obtained are plagued by low significance of the detections, particularly in broad band photometric filters \citep[see e.g.][]{2000Bhatt,2006Chavero,2006Hales,2010bWiktorowicz,2010Bailey,2015GarciaGomez}. Despite its drawbacks, aperture polarimetry has potential as a valuable tool in the characterisation of debris discs which are too faint, or too close to their host star to be directly imaged \citep{2019Vanderportal}, or as a complementary technique to monochromatic imaging polarimetry to provide colour information \citep{2020Marshall}.  

In this work we present a summary and analysis of aperture polarimetry observations to date, focusing on recent measurements taken by the HIgh Precision Polarimetric Instrument \citep[HIPPI/-2;][]{2015Bailey,2020Bailey} with precision at the few~$\times10^{-6}$ level, that is sufficient to detect debris-induced polarisation. The remainder of the paper is laid out as follows: in Section \ref{sec:observations} we present the sample of stars combined in this analysis and the polarimetric observations used to characterise the discs' properties. The methods used to model the dust scattered light and account for a polarisation component from the interstellar medium are summarised along with the results in Section \ref{sec:results}, comparing the inferred dust grain properties from our polarisation measurements to analyses based on modelling the dust continuum emission. In Section \ref{sec:discussion} we discuss the results and their utility in the context of identifying and interpreting emission from debris disc host stars. Finally, in Section \ref{sec:conclusions}, we present our conclusions.

\section{Observations}
\label{sec:observations}

\subsection{Sample}
\label{sec:sample}

We observed \ntgt debris disc host stars for the polarimetric debris disc survey with HIPPI or HIPPI-2. The targets spanned a broad range of stellar spectral types (A0~V to M1~V), and disc fractional luminosities $L_{\rm dust}/L_{\star}$ (as a proxy for expected polarisation level, 0.5 to 500~$\times10^{-5}$), and architectures. All but one of the targets had been spatially resolved in continuum emission and/or scattered light, providing constraint on the discs' orientations and extents to assist in modelling the unresolved polarimetric observations obtained here. Our targets for this study were selected in 2014, before many of the developments described in the introduction. A brief summary of relevant target properties is provided below and the relevant target properties are given in table \ref{tab:sample}.

\subsubsection{HD~377}

HD~377 is a young, Sun-like star, spectral type G2~V, lying at a distance of 38.523~$\pm$~0.086~pc \citep{2018Gaia}. The star is somewhat active with $|B_\ell|_{\rm max}$ of 5.1~$\pm$~1.7~G \citep{2014Petit, 2014Marsden}, but its early spectral type is likely to blunt activity contamination of the linear polarisation. The debris disc was spatially resolved in scattered light by  \textit{HST}/NICMOS \citep{2016Choquet}, and at millimetre wavelengths with the Sub-Millimeter Array \citep{2016Steele}, revealing its edge-on orientation. There exists a slight difference in the inferred radii for the disc at the two wavelengths (60 vs 80 au), which can be explained by low signal-to-noise of mm-wavelength observation. Regardless, the disc lies fully within the HIPPI aperture (diameter 6\farcs6). 

\subsubsection{HD~39060}

HD~39060 ($\beta$ Pictoris) is an A6~V type star at a distance of 19.44~$\pm$~0.05~pc \citep{2007vanLeeuwen}. It is one of the archetypal debris disc systems and eponymous member of the $\beta$ Pic moving group giving it a well defined age of 20~$\pm$~11~Myr \citep{2014MamajekCameron}. \citet{2009Schroeder} found 0.231 for the S-index of the star \citep{1978Vaughan,1991Duncan}. This is a fairly low value for activity, but no conversion to an expected $B_\ell$ is available for A-type stars. A non-detection for the magnetic field with a value of $B_{\rm z} =$ -79~$\pm$~53~G was recorded by \cite{2006Hubrig}. In any case, we expect the star to have a low activity induced broadband polarisation on account of there being few spectral lines. The debris disc is oriented edge-on to us and hosts two known Jovian mass planets \citep{2010Lagrange,2019Lagrange,2021Brandt} along with a substantial debris disc which has been extensively imaged at near-infrared to millimetre wavelengths \citep[e.g.][]{1984SmithTerrile,2010Vandenbussche,2014Dent,2019Matra}. The disc is dynamically active, exhibiting evidence of exocometary activity in both spectroscopic \citep{2014Kiefer} and photometric observations \citep{2019Zieba,2022Pavlenko}, an asymmetric clump of cold CO emission at mm-wavelengths \citep{2014Dent,2017Matra}, and evidence of a dynamically excited planetesimal population \citep{2019Matra}. Previous polarimetric observations of the disc have been made, revealing an asymmetric contribution from the two sides of the disc and a polarisation fraction of 15 to 20 per cent in scattered light \citep{1991Gledhill,2000Krivova}, and 0.51~$\pm$~0.19 per cent at millimetre wavelengths \citep{2022Hull}.

\subsubsection{HD~92945}

HD~92945 is a K1~V star at a distance of 21.54~$\pm$~0.02~pc \citep{2018Gaia}. The star is very active BY~Dra type variable \citep{2006Kazarovets} -- and its K spectral type suggests a potentially large polarisation \citep{2017Cotton}. A spectropolarimetric determination of the global magnetic field is the best method for gauging likely linear polarisation contamination, unfortunately none exists. \citet{2018BoroSaikia} measure an S-index of 0.641. Converting the S-index to $\log(R^\prime_{HK})$ using the relations in \citet{2015SarezMascare} (assuming $B$-$V$ from \citealp{2000Hog}) gives $-4.284$ (similar to \citet{2018BoroSaikia} who give values of between $-4.32$ and $-4.40$), this then allows a \textit{mean} $|B_\ell|$ value of $\sim$32--100~G to be estimated using the relations in \citet{2022Brown}. We might expect values of linear polarisation in ppm around four times the value of $|B_{\ell}|$ in Gauss,  based on \citet{2019bCotton}. A comparable star, HD~189733, was observed by \citet{2021Bailey} to sometimes produce polarisations of 80-100~ppm in the 500SP band ($\lambda_{\rm eff}=$ 450~nm), which is somewhat less but still likely to obfuscate other polarisation signals. The star hosts a substantial debris disc viewed at a moderate inclination, which has been spatially resolved in both scattered light \citep{2011Golimowski}, and thermal emission \citep{2019Marino}. The disc architecture is broad, with evidence of a gap within the belt from mm-wavelength observations, suggestive of interaction with an unseen planetary companion \citep{2019Marino}.

\subsubsection{HD~105211}

HD~105211 ($\eta$ Crucis) is an F2~V star located at a distance of 19.75~$\pm$~0.04 \citep{2007vanLeeuwen}, near to the galactic plane. The debris disc was marginally resolved in \textit{Herschel}/PACS observations \citep{2016DodRob,2017Hengst}. ALMA mm-wavelength observations show the outer belt architecture is a narrow ring (Matra et al. in prep.), similar to those of many F-type stars \citep{2021Pawellek}. The HIPPI field of view (6\farcs6), combined with the extent and inclination of the disc, omits almost all of the outer disc (diameter 15\farcs6), such that in effect we probe the inner regions of the system for polarised light from an asteroid belt analogue.

\subsubsection{HD~109573A}

HD~109573A (HR~4796A) is a bright, young A0~V spectral type star at a distance of 71.909~$\pm$~0.691~pc \citep{2018Gaia}. The system is the most distant target in the sample but has a high fractional luminosity ($L_{\rm dust}/L_{\star} \simeq 10^{-3}$) and compact angular extent ($\phi_{\rm disc} \simeq 1\arcsec$), therefore lying fully within the HIPPI aperture. Due to its brightness, high inclination and narrow width, the disc is an excellent target for imaging in scattered light \citep{1999Schneider,2015Perrin,2017Milli,2019Milli}. The system has also been spatially resolved at millimetre wavelengths \citep{2018Kennedy}. Previous studies of the system have inferred the presence of a planetary companion to the system on the basis of the eccentricity of the debris disc \citep{1999Wyatt} which induces a wavelength dependent asymmetric brightening of the disc \citep{2011Moerchen}, but a detection has not yet been made.

\subsubsection{HD~115892}

HD~115892 ($\iota$ Centauri) is an A3~V type star, 17.840~$\pm$~0.348~pc from the Sun \citep{2018Gaia}. The stellar magnetic field has been measured by \cite{2006Hubrig}, obtaining a value of $B_{\rm z} =$ 77~$\pm$~30~G. Their method of measuring the field uses the hydrogen Balmer lines, which is better suited to an A-type star, and produces values comparable with $|B_\ell|$ \citep{2010Hubrig}. Although this is technically a non-detection, the value is quite high, and at 2.6-$\sigma$, it is well worth noting. The star hosts a faint, asteroid belt analogue debris disc ($L_{\rm dust}/L_{\star} = 7\times10^{-6}$, $T_{\rm dust} \sim 195~$K) \citep{2016Cotten}. Unlike the rest of the sample examined here, this target has not been spatially resolved either in continuum or scattered light. It was included in the original sample to test the capability of HIPPI to detect the polarimetric signature of debris dust in very faint disc systems where scattered light imaging would be impractical. 

\subsubsection{HD~161868}

HD~161868 ($\gamma$ Ophiuchi), a bright A-type star at 31.52~$\pm$~0.21~pc \citep{2007vanLeeuwen}, hosts a large, faint debris disc at moderate inclination ($i \simeq 60\degr$). The system was spatially resolved in thermal emission by \textit{Herschel} \citep{2014Thureau}, but a scattered light detection remains elusive, likely due to the disc's low fractional luminosity and large extent. Given the disc extent and the star's proximity much of the disc is expected to lie beyond the HIPPI aperture. The system was previously observed by the PlanetPol instrument at longer wavelengths than the HIPPI observations presented here \citep{2010Bailey}. A small polarimetric signal was detected in that measurement, and we include that observation in our modelling.

\subsubsection{HD~181327}

HD~181327 is another $\beta$ Pic moving group member, an F6~V star at a distance of 48.213~$\pm$~0.133~pc \citep{2018Gaia}. The star is not active, and \citet{2007Scholz} found this star to be chromospherically inactive according to its $H\alpha$ equivalent width of 5.09~$\pm$~0.05. The disc has been spatially resolved in scattered light and thermal emission, revealing a moderately inclined $\sim~30\degr$, bright ring structure \citep{2006Schneider,2008Chen,2012Lebreton}. An extended halo of small grains seen in scattered light along with sub-structure within the disc has been interpreted as being the result of recent collisional events \citep{2014Stark}. At millimetre wavelengths, the disc is more compact, consistent with the expected segregation of grain sizes acted on by radiation forces, and a molecular gas disc is present \citep{2019Marino}. The disc around HD~181327 is also viewed at a moderate inclination angle which could impact its detectability due to increased self-cancellation of the polarimetric signal from the disc, despite its apparent brightness. 

\subsubsection{HD~188228}

HD~188228 ($\epsilon$ Pavonis), an A0~V star lying at a distance of 32.22~$\pm$~0.18~pc \citep{2007vanLeeuwen}, is a member of the 30~Myr old Argus association \citep{2008Torres}. The debris disc is faint ($L_{\rm dust}/L_{\star} = 5\times10^{-6}$), and was marginally resolved at far-infrared wavelengths by \textit{Herschel}/PACS \citep{2013Booth}. Whilst relatively extended, the disc should lie fully within the HIPPI aperture, making this system a good test of the assumption that disc brightness in thermal emission and scattered light/polarisation are uncorrelated.

\subsubsection{HD~197481}

HD~197481 (AU Mic), an M1~V star at a distance of 9.725~$\pm$~0.005~pc \citep{2018Gaia}, is also a member of the $\beta$ Pic moving group. The star is very active, with $|B_\ell|_{\rm max}$ values determined by \citet{2014Petit} and \citet{2020Martioli} being similar and both large, with the potential to produce linear polarisation contamination of hundreds of ppm \citep{2017Cotton, 2019Cotton}, though stars with M spectral types have been little studied by precision instruments. The strong flaring activity has impacted the interpretation of radio wavelength observations \citep{2020Macgregor}. The debris disc has an edge-on orientation imaged in scattered light \citep[both polarimetric and total intensity,][]{2007Graham} and continuum emission \citep{2013Macgregor,2015Matthews}. Scattered light imaging reveals radially migrating, transient features in the disc, perhaps radially migrating dust clumps \citep{2015Boccaletti}. In addition to the disc, two low mass planetary companions have been identified in the system based on radial velocities and TESS time series observations \citep{2020Plavchan,2021Addison,2021Martioli}.   

\subsubsection{HD~202628}

HD~202628 is another young, Sun-like star with a G5~V spectral type, located at a distance of 23.831~$\pm$~0.026~pc \citep{2018Gaia}. The star exhibits some activity; \citet{2018BoroSaikia} measure an S-index of 0.217, converting this to $log(R^{\prime}_{HK})$ using the relations in \citet{2015SarezMascare} (assuming $B-V$ from \citealp{2000Hog}), then allows a \textit{mean} $|B_\ell|$ value of $\sim$3~G to be estimated using the relations in \citet{2022Brown}. The system's debris disc is radially narrow and eccentric and has been imaged in both scattered light with \textit{HST}/NICMOS \citep{2010Krist} and continuum emission at far-infrared wavelengths by \textit{Herschel} \citep{2021Marshall} and millimetre wavelengths by ALMA \citep{2019Faramaz}. The asymmetric architecture of the debris disc has been used to infer the presence of a perturbing planetary companion \citep{2016Thilliez}. The imaged disc has a large angular extent ($\sim 6$ to $8\arcsec$ in radius) and moderate inclination, such that most of it will lie outside the aperture of HIPPI.

\subsubsection{HD~216956}

HD~216956 (Fomalhaut) is another of the archetypical debris discs, an A4~V type star at a distance of 7.70~$\pm$~0.03~pc \citep{2007vanLeeuwen}. The cool, outer disc has been extensively imaged from optical to millimetre wavelengths \citep{2005Kalas,2012Acke,2017Macgregor,2017Holland}, revealing a relatively faint, narrow and eccentric at moderate inclination. Evidence for a spatially unresolved warm component close to the star exists from the spectral energy distribution; the nature of that warm excess is as yet unknown. A proposed planetary companion thought to be responsible for the eccentric disc was identified in \textit{HST} observations \citep{2008Kalas}, but subsequent observations determined the candidate object's projected orbit was inconsistent with the disc architecture \citep{2014Beust}, and the companion itself appeared to expand and dissipate over time leading it to be reclassified as an expanding dust cloud from the aftermath of a planetesimal collision interior to the debris belt \citep{2020Gaspar,2020Markus}. Similar to HD~105211 and HD~202628, the HIPPI field of view omits the outer belt from its regard, such that the observations presented here probe the inner regions of the system that are believed to host an asteroid belt analogue.

\begin{table*}
    \centering
    \caption{Stellar sample and their properties. $\phi_{\rm disc}$ is the disc angular radius, which should be compared to the HIPPI aperture radius of 3.3\arcsec. The `Images' column refers to the existence of spatially resolved thermal emission (`T') or scattered light (`S') imaging data for each disc. \label{tab:sample}}
    \begin{tabular}{llcccccccccccc}
        \hline\hline
               && \multicolumn{5}{c}{Star} & \multicolumn{6}{c}{Debris disc} & \\ \cline{3-7} \cline{9-14}
        \multicolumn{2}{c}{Target} & Spectral & Distance &  $V$   &  $L_{\star}$  & $|B_\ell|_{\rm max}$ & &  $L_{\rm dust}/L_{\star}$ & $R_{\rm disc}$ & $\phi_{\rm disc}$ & $i$ & $\theta$ &  Images \\
        HD      & Other Name     & Type     & (pc)     & (mag)  & ($L_{\odot}$) & (G) & &  ($\times10^{-5}$)        &     (au) & (\arcsec)   & (\degr)& (\degr) & \\
        \hline
        377     &                & G2~V & 38.523~$\pm$~0.086 & 7.59 & 1.00 & \05.1 $\pm$ \01.7 & & 20 & 86 & 2.23 & 85~$\pm$~5 & 47~$\pm$~4 & T,S \\
        39060   & $\beta$ Pic    & A6~V & 19.44\0~$\pm$~0.05\0 & 3.86   & 10.48 & ND (a) & & 243 & 85 & 4.37 & 89~$\pm$~1 & 30~$\pm$~1 & T,S \\
        92945   & V419 Hya       & K1~V & 21.540~$\pm$~0.020 & 7.72 & 0.38 & 32--100 (a) & & 76 & 65 & 3.02 & 27~$\pm$~1 & 100~$\pm$~1 & T,S \\
        105211  & $\eta$ Cru     & F2~V & 19.75\0~$\pm$~0.04\0 & 4.14   & 7.30 & --- & & 5.4 & 154 & 7.80 & 56~$\pm$~5 & 32~$\pm$~2 & T \\
        109573 & HR 4796A       & A0~V & 71.909~$\pm$~0.691 & 5.77 & 22.80 & ---& & 470 & 70 & 0.97 & 77~$\pm$~1 & 26~$\pm$~1 & T,S \\
        115892  & $\iota$ Cen    & A3~V & 17.840~$\pm$~0.348 & 2.73 & 21.84 & 77 $\pm$ 30 (a) &  & 0.7 & 16 & 0.73 & --- & --- & --- \\
        161868  & $\gamma$ Oph   & A1~V & 31.52\0~$\pm$~0.21\0 & 3.75   & 28.56 & ND & & 10 & 143 & 4.54 & 60~$\pm$~10 & 118~$\pm$~3 & T \\
        181327  &                & F6~V & 48.213~$\pm$~0.133 & 7.04 & 3.33 & ND (a) & & 293 & 86 & 1.78 & 32~$\pm$~2 & 107~$\pm$~2 & T,S \\
        188228  & $\epsilon$ Pav & A0~V & 32.22\0~$\pm$~0.18\0 & 3.94  & 25.60 & --- & & 0.5 & 106 & 3.29 & 49~$\pm$~6 & 11~$\pm$~15 & T \\
        197481  & AU Mic         & M1~V & \09.725~$\pm$~0.005 & 8.63  & 0.06 & 72.2 $\pm$ 10.3 & & 39 & 38 & 3.90 & 89~$\pm$~1 & 129~$\pm$~1 & T,S \\
        202628  &                & G5~V & 23.831~$\pm$~0.026 & 6.74 & 1.15 & 3 (a) & & 14 & 162 & 6.81 & 64~$\pm$~2 & 134~$\pm$~2 & T,S \\
        216956  & Fomalhaut      & A4~V & \07.70\0~$\pm$~0.03\0 & 1.16    & 15.50 & ND & & 8 & 123 & 15.9 & 66~$\pm$~1 & 156~$\pm$~1 & T,S \\
        \hline 
    \end{tabular}
    \begin{flushleft}
    \textbf{References:} Spectral types from SIMBAD \citep{1982Houk,2006Torres,2006Gray}. Distances are from \textit{Gaia} DR2 \citep{2016Gaia,2018Gaia} or \textit{Hipparcos} \citep{2007vanLeeuwen}. V magnitudes are from the Tycho-2 catalogue \citep{2000Hog}. Stellar activity: all from \citet{2014Petit}, with additional sources for HD 377 \citet{2014Marsden}, and HD 197481 \citet{2020Martioli}; ND indicates a non-detection and thus a low activity level; (a) the value has been derived from $\log R^{\prime}_{\rm HK}$ according to \cite{2022Brown}, or through other activity indicators (see text). Debris disc properties: HD~377 \citet{2016Steele,2016Choquet}. HD~10700 \citet{2014Lawler}. HD~39060 \citet{2010Vandenbussche,2014Dent}. HD~92945 \citet{2011Golimowski,2019Marino}. HD~102647 \citet{2011Churcher}. HD~105211 \citet{2016DodRob,2017Hengst}. HD~109573 \citet{2011Thalmann}. HD~115892 \citet{2016Morales}. HD~161868 \citet{2014Thureau}. HD~181327 \citet{2012Lebreton}. HD188228 \citet{2013Booth}. HD~197481 \citet{2007Graham,2015Schuppler}, HD~202628 \citet{2012Krist,2019Faramaz}. HD~216956 \citet{2005Kalas,2012Acke,2017Macgregor}.
    \end{flushleft}
\end{table*}

\subsubsection{Interstellar controls}
\label{sec:interstellar_controls}

In addition to any intrinsic polarisation a disc system might possess, each of them also has an associated interstellar polarisation -- imparted by aligned dust grains between the target and observer. A common way to gauge the level of interstellar polarisation is to look at nearby intrinsically unpolarised stars. There are a number of fairly recent studies of the nearby stars, with sufficient precision, that can be a source of data for this purpose; these include small surveys by \citet{2010Bailey, 2016Cotton, 2017Cotton, 2019Cotton}, a larger one conducted more recently by \citet{2020Piirola}, and also other work that has collected such data incidentally \citep{2015Bailey, 2016Marshall, 2017bCotton, 2020Bailey, 2020Marshall, 2021Bailey, 2022Lewis}\footnote{Some of these works also derive the interstellar component for intrinsically polarised stars -- an additional study is \citet{2023Howarth} -- this data can also be used for mapping interstellar polarisation.}. Despite the numerous works just listed, the vastness of space means that the coverage they provide is sub-optimal. Consequently we report in section \ref{sec:hippi_obs} a number of additional controls. Some of these were observed specifically for this purpose, others were observed as controls for other projects, which have not yet been brought to fruition. In each case these stars are not known to be close binaries, active, host substantial discs, be variable, or rapid rotators; they have spectral types between A0 and K2. Consequently we avoid the stars with significant intrinsic polarisation \citep{2016bCotton,2016Cotton}. 

\subsection{HIPPI/-2 observations}
\label{sec:hippi_obs}

Between May of 2014 and August of 2018 we made multiband high precision polarimetric observations of twelve debris disc systems, along with single band (SDSS $g^\prime$) observations of thirty interstellar control stars for them. Altogether, the observations are spread over thirteen observing runs -- when one also includes the required calibration observations. Every run, but one, was carried out with the Equatorial 3.9-m Anglo-Australian Telescope (AAT) located at Siding Spring Observatory in Australia. The 2018JUN run used the 8.1-m Alt-Azimuth Gemini North Telescope (GMT), located at Mauna Kea, Hawaii. For the first nine runs, up until 2017AUG we used the HIgh Precision Polarimetric Instrument (HIPPI; \citealp{2015Bailey}), for the subsequent observations its successor, HIPPI-2 \citep{2020Bailey} was used. 

HIPPI was always used at the AAT's $f/8$ Cassegrain focus where its 1\,mm aperture subtends 6.6$\arcsec$ on the sky. HIPPI-2 was designed for Gemini's $f/16$ focus, but may be used at $f/8$ at the AAT with the aid of a $2\times$ negative achromatic (Barlow) lens to effectively achieve $f/16$. HIPPI-2 has the ability to select different sized apertures through an aperture wheel; for observations of the debris disc systems our selection was designed to match the earlier HIPPI observations, for later control star observations a larger aperture was used. HIPPI-class polarimeters achieve their very high precision (4.3\,ppm and $<$3\,ppm in SDSS $g^\prime$ respectively for HIPPI and HIPPI-2) through the combination of modern photomultiplier tube (PMT) detectors and ferroelectric liquid crystal (FLC) modulators -- which operate at 500\,Hz to beat seeing noise. Three different FLCs were used: one each from Micron Technologies (MT), Boulder Nonlinear Systems (BNS), and Meadowlark (ML). The BNS modulator's performance drifted over time, and so has been characterised into different performance eras as described by \citet{2020Bailey}. A summary of the instrument and telescope set-up for each observing run is given in table \ref{tab:runs}.

\begin{table}
\caption{Summary of observing run set-ups}
\label{tab:runs}
\tabcolsep 1.5 pt
\begin{tabular}{ll|ccrcccc|ccc|rrr}
\hline\hline
\multicolumn{1}{c}{Run} & \multicolumn{1}{c|}{Date Range} & Instr. & Tel. & \multicolumn{1}{c}{$f/$} & Ap. & Mod. \\
 &  \multicolumn{1}{c|}{(UT)} &  &   & & ($\arcsec$) &  \\
\hline
2005APR & 2005-04-25 to 05-08  &   PlanetPol   &   WHT &   11\phantom{*} &   5.2 &   PEM     \\ 
2014MAYC & 2014-05-11 to 05-12 &   HIPPI   &   AAT &   8\phantom{*}   &   6.6 &   MT      \\
2014AUG & 2014-08-28 to 09-02  &   HIPPI   &   AAT &   8\phantom{*}   &   6.6 &   BNS-E1  \\   
2015MAY & 2015-05-22 to 05-26  &   HIPPI   &   AAT &   8\phantom{*}   &   6.6 &   BNS-E1  \\
2015JUN & 2015-06-26 to 06-29  &   HIPPI   &   AAT &   8\phantom{*}   &   6.6 &   BNS-E1  \\
2015OCT & 2015-10-14 to 10-20  &   HIPPI   &   AAT &   8\phantom{*}   &   6.6 &   BNS-E1  \\
2015NOV & 2015-10-29 to 11-02  &   HIPPI   &   AAT &   8\phantom{*}   &   6.6 &   BNS-E1  \\
2016DEC & 2016-11-30 to 12-07  &   HIPPI   &   AAT &   8\phantom{*}   &   6.6 &   BNS-E2  \\
2017JUN & 2017-06-22 to 07-05  &   HIPPI   &   AAT &   8\phantom{*}   &   6.6 &   BNS-E2  \\
2017AUG & 2017-08-07 to 08-20  &   HIPPI   &   AAT &   8\phantom{*}   &   6.6 &   BNS-E2  \\
2018MAR & 2018-03-28 to 04-06  &   HIPPI-2 &   AAT &   8*  &   5.3 &   BNS-E3  \\
2018JUN & 2018-07-05 to 08-07  &   HIPPI-2 &   GN  &   16\phantom{*}  &   6.4 &   ML-E1   \\
2018JUL & 2018-07-10 to 07-25  &   HIPPI-2 &   AAT &   8*  &  11.9 &   BNS-E4  \\
2018AUG & 2018-08-16 to 09-02  &   HIPPI-2 &   AAT &   8*  &  11.9 &   BNS-E5/6/7 \\
\hline
\end{tabular}
\begin{flushleft}
\textbf{Notes:} * Uses a $2\times$ negative achromatic lens to give an effective $f/16$.
\end{flushleft}
\end{table}

For the most part the standard operating procedures for HIPPI and HIPPI-2, as described by \citet{2015Bailey} and \citet{2020Bailey} were followed, with the reduction of all of the data following the updated procedures in \citet{2020Bailey}. For this work we deviated from standard HIPPI-2 centering procedure: HIPPI-2 has its own built-in instrument rotator, so that it does not have to rely on the telescope's Cassegrain unit as HIPPI does. The resulting improvement in centering precising means that, usually, the target only has to be acquired once per observation, and not at each different position angle (0, 45, 90, and 135$^\circ$). However, for the small aperture observations described here we did reacquire at each different position angle. This precaution was taken since debris discs are extended objects, and we wanted to ensure that each measurement was acquired with the target in exactly the same position, especially where not all of the disc lies within the aperture. The centering procedure is carried out manually, with fixed offset buttons in RA and Dec, using real-time visual feedback from the instrument PMTs, and is accurate to $\sim0.5\arcsec$. For the debris disc observations, which demand a high level of consistency, we always used the same operator (JB) for this task. Despite these precautions, given the extended nature of debris  discs, the centering precision is a source of error that depends on both the seeing and the nature of the object that we cannot easily estimate -- consequently we expect the uncertainty in measurements of polarisation presented here to be larger than the nominal errors.

Each debris disc system was observed with between three and five filter bands. The filters used varied, and included Clear (no filter), 425 and 500\,nm short pass filters (425SP and 500SP), SDSS $g^\prime$ and $r^\prime$ filters and a 650\,nm long pass filter (650LP). Two different versions of the SDSS filters were used, with the later HIPPI-2 observations using Astrodon versions with squarer profiles (see \citealp{2015Bailey} and \citealp{2020Bailey} for their the filter transmission profiles). Different PMTs were used depending on the filter band. By default blue-sensitive (B) Hamamatsu H10720-210 modules which have ultrabialkali photocathodes were used, with red sensitive (R) Hamamatsu H10720-20 modules with extended red multialkali photocathodes used for all 650LP and some $r^\prime$ observations. Because of manufacturing tolerances across the face of the FLCs, there is a wavelength dependant positioning error that is larger for bluer wavelengths. Values for this error determined by passband are reported in \citet{2020Bailey}, and are added in quadrature with the photon count statistics based errors for every observation reported here.

The raw observations are rotated from the instrumental frame to equatorial co-ordinates by reference to measurements of high polarisation standard stars. The standards, described in \citet{2020Bailey}, have uncertainties of about a degree. These high polarisation standard observations are made in either $g^\prime$ or Clear. A small wavelength dependant telescope polarisation (TP), must be determined by the measurement of unpolarised standard stars, and then subtracted. The TP is determined as the straight mean of all standards observed in a given band using the same aperture and procedure. TP errors are incorporated into the errors in the science observations by adding them in quadrature. Usually the TP is determined for each run individually, but it is possible to combine runs that are close together if the telescope is not realuminised, and there is minimal dust accumulation in the meanwhile. A summary of standard observations relevant to the observations made here and some additional details are presented in Appendix \ref{sec:std_obs}.

\begin{table*}
\caption{Polarisation observations of debris disc systems}
\label{tab:observations1}
\fontsize{8.25pt}{9pt}\selectfont
\tabcolsep 2.5 pt
\begin{tabular}{lllrrccrrrrrr}
\hline
\hline
\multicolumn{1}{c}{Target} & \multicolumn{1}{c}{Run}    & \multicolumn{1}{c}{UT}                  & Dwell & Exp. & Filt. & Det. & $\lambda_{\rm eff}$ & Eff. & \multicolumn{1}{c}{$q$} & \multicolumn{1}{c}{$u$} & \multicolumn{1}{c}{$p$} & \multicolumn{1}{c}{$\theta$}\\
        &               &        & \multicolumn{1}{c}{(s)} & \multicolumn{1}{c}{(s)} & & & \multicolumn{1}{c}{(nm)} & \multicolumn{1}{c}{(\%)} & \multicolumn{1}{c}{(ppm)} & \multicolumn{1}{c}{(ppm)} & \multicolumn{1}{c}{(ppm)} & \multicolumn{1}{c}{($^\circ$)}\\
\hline
HD    377 & 2015JUN & 2015-06-28 19:15:58 & 2583 & 1280 & 425SP & B & 402.6 & 58.4 &   15.1 $\pm$  61.5 &  108.0 $\pm$  60.4 &  109.1 $\pm$  60.9 &   41.0 $\pm$  19.8 \\
HD    377 & 2017AUG & 2017-08-10 17:30:49 & 2137 & 1280 & 425SP & B & 402.6 & 51.8 &   14.4 $\pm$  78.8 &  $-$14.5 $\pm$  81.3 &   20.4 $\pm$  80.0 &  157.4 $\pm$  46.7 \\
HD    377 & 2017JUN & 2017-07-05 17:31:30 & 2859 & 1920 & 425SP & B & 402.8 & 52.0 &  151.2 $\pm$  66.9 &  $-$35.7 $\pm$  66.1 &  155.4 $\pm$  66.5 &  173.4 $\pm$  14.2 \\
HD    377 & 2015JUN & 2015-06-28 16:57:01 & 1963 & 1280 & 425SP & B & 403.5 & 58.9 &  119.0 $\pm$  65.8 &  168.0 $\pm$  66.5 &  205.9 $\pm$  66.2 &   27.3 $\pm$   \phantom{0}9.7 \\
HD    377 & 2016DEC & 2016-12-03 12:21:35 & 2054 & 1280 & 425SP & B & 403.7 & 52.4 &  187.4$\pm$152.0 & $-$145.6$\pm$136.0 &  237.3$\pm$144.0 &  161.1$\pm$  21.6 \\
HD    377 & 2017JUN & 2017-07-02 18:02:16 & 3032 & 1920 & 500SP & B & 444.7 & 79.3 &   $-$4.0 $\pm$  25.9 &   61.0 $\pm$  27.9 &   61.1 $\pm$  26.9 &   46.9 $\pm$  14.8 \\
HD    377 & 2015JUN & 2015-06-27 19:31:20 & 1261 & 640 & $g^\prime$ & B & 473.2 & 91.3 &   14.0 $\pm$  28.1 &   83.7 $\pm$  28.0 &   84.9 $\pm$  28.0 &   40.3 $\pm$  10.0 \\
HD    377 & 2016DEC & 2016-12-03 11:39:43 & 1973 & 1280 & $g^\prime$ & B & 473.8 & 89.5 &    9.3 $\pm$  37.7 &   88.6 $\pm$  32.6 &   89.1 $\pm$  35.1 &   42.0 $\pm$  12.7 \\
HD    377 & 2015JUN & 2015-06-28 16:28:36 & 1272 & 640 & $g^\prime$ & B & 474.8 & 91.6 &  148.9 $\pm$  30.3 &    9.4 $\pm$  30.3 &  149.2 $\pm$  30.3 &    1.8 $\pm$   \phantom{0}5.9 \\
HD    377 & 2017AUG & 2017-08-08 16:57:32 & 2215 & 1280 & $r^\prime$ & R & 624.3 & 81.8 &    7.0 $\pm$  25.7 & $-$122.7 $\pm$  27.0 &  122.9 $\pm$  26.4 &  136.6 $\pm$   \phantom{0}6.3 \\
HD    377 & 2015JUN & 2015-06-28 17:46:27 & 1992 & 1280 & $r^\prime$ & R & 624.4 & 78.6 &  $-$20.1 $\pm$  26.6 &  $-$60.0 $\pm$  26.3 &   63.3 $\pm$  26.5 &  125.7 $\pm$  13.8 \\
HD    377 & 2015JUN & 2015-06-28 18:23:56 & 1950 & 1280 & $r^\prime$ & R & 624.4 & 78.7 &   $-$1.9 $\pm$  26.6 &   17.5 $\pm$  26.7 &   17.6 $\pm$  26.7 &   48.1 $\pm$  38.5 \\
HD    377 & 2017AUG & 2017-08-08 14:59:09 & 3576 & 2560 & 650LP & R & 722.4 & 65.2 &   $-$0.5 $\pm$  32.0 & $-$154.6 $\pm$  31.7 &  154.6 $\pm$  31.9 &  134.9 $\pm$   \phantom{0}6.0 \\
HD    377 & 2017AUG & 2017-08-08 16:01:43 & 3943 & 2560 & 650LP & R & 722.4 & 65.2 &    5.3 $\pm$  30.6 & $-$144.5 $\pm$  30.6 &  144.6 $\pm$  30.6 &  136.1 $\pm$   \phantom{0}6.2 \\
\hline
HD  39060 & 2017AUG & 2017-08-10 18:14:31 & 1476 & 640 & 425SP & B & 401.1 & 52.1 &  $-$99.2 $\pm$  20.2 & $-$218.5 $\pm$  21.3 &  240.0 $\pm$  20.7 &  122.8 $\pm$   \phantom{0}2.4 \\
HD  39060 & 2015JUN & 2015-06-28 19:57:22 & 1170 & 640 & 425SP & B & 401.8 & 59.3 &  $-$79.8 $\pm$  19.8 &  $-$71.6 $\pm$  19.8 &  107.2 $\pm$  19.8 &  110.9 $\pm$   \phantom{0}5.3 \\
HD  39060 & 2017AUG & 2017-08-10 18:38:58 & 1350 & 640 & 500SP & B & 438.3 & 76.1 &  $-$73.8 $\pm$  10.3 & $-$109.0 $\pm$  10.0 &  131.6 $\pm$  10.1 &  117.9 $\pm$   \phantom{0}2.2 \\
HD  39060 & 2014AUG & 2014-08-29 19:19:28 & 1167 & 640 & $g^\prime$ & B & 466.1 & 89.7 &  $-$85.0 $\pm$   \phantom{0}6.0 &  $-$87.8 $\pm$   \phantom{0}5.9 &  122.2 $\pm$  \phantom{0}5.9 &  113.0 $\pm$   \phantom{0}1.4 \\
HD  39060 & 2014AUG & 2014-08-30 19:17:50 & 1029 & 640 & $r^\prime$ & B & 598.9 & 83.5 &  $-$74.0 $\pm$  11.4 &  $-$67.5 $\pm$  11.1 &  100.2 $\pm$  11.3 &  111.2 $\pm$   \phantom{0}3.2 \\
HD  39060 & 2017AUG & 2017-08-07 18:16:29 & 2533 & 1000 & $r^\prime$ & R & 621.1 & 82.4 & $-$108.4 $\pm$   \phantom{0}9.3 & $-$126.5 $\pm$  11.4 &  166.6 $\pm$  10.4 &  114.7 $\pm$   \phantom{0}1.8 \\
HD  39060 & 2017AUG & 2017-08-08 17:46:22 & 1896 & 640 & $r^\prime$ & R & 621.1 & 82.3 &  $-$78.4 $\pm$   \phantom{0}9.3 &  $-$96.4 $\pm$   \phantom{0}9.1 &  124.3 $\pm$   \phantom{0}9.2 &  115.4 $\pm$   \phantom{0}2.1 \\
HD  39060 & 2018MAR & 2018-04-01 11:33:25 & 1215 & 640 & $r^\prime$ & R & 622.6 & 81.3 &  $-$57.4 $\pm$   \phantom{0}8.3 &  $-$95.2 $\pm$   \phantom{0}8.3 &  111.2 $\pm$   \phantom{0}8.3 &  119.5 $\pm$   \phantom{0}2.1 \\
HD  39060 & 2017AUG & 2017-08-07 19:04:36 & 2455 & 1440 & 650LP & R & 719.1 & 65.6 &  $-$78.3 $\pm$  12.7 &  $-$89.3 $\pm$  12.7 &  118.8 $\pm$  12.7 &  114.4 $\pm$   \phantom{0}3.1 \\
HD  39060 & 2017AUG & 2017-08-08 17:46:40 & 2572 & 640 & 650LP & R & 719.1 & 65.6 &  $-$72.8 $\pm$  13.4 &  $-$92.6 $\pm$  13.4 &  117.8 $\pm$  13.4 &  115.9 $\pm$   \phantom{0}3.3 \\
\hline
HD  92945 & 2015JUN & 2015-06-29 09:23:39 & 7397 & 1280 & 425SP & B & 406.4 & 59.7 &  657.3 $\pm$  81.8 &   10.0 $\pm$  80.0 &  657.4 $\pm$  80.9 &    0.4 $\pm$   \phantom{0}3.5 \\
HD  92945 & 2017JUN & 2017-07-02 09:07:49 & 3434 & 2560 & 425SP & B & 406.4 & 53.3 &   $-$6.6 $\pm$  65.8 & $-$216.0 $\pm$  65.0 &  216.1 $\pm$  65.4 &  134.1 $\pm$   \phantom{0}9.0 \\
HD  92945 & 2015JUN & 2015-06-27 11:26:31 & 2044 & 1280 & $g^\prime$ & B & 478.3 & 92.4 &  168.7 $\pm$  24.5 &  $-$63.6 $\pm$  24.6 &  180.3 $\pm$  24.5 &  169.7 $\pm$   \phantom{0}3.9 \\
HD  92945 & 2015JUN & 2015-06-28 08:43:53 & 1184 & 640 & $r^\prime$ & R & 625.8 & 78.4 &   34.2 $\pm$  47.5 &  $-$30.3 $\pm$  48.0 &   45.7 $\pm$  47.8 &  159.2 $\pm$  32.7 \\
HD  92945 & 2015JUN & 2015-06-28 10:20:06 & 1167 & 640 & $r^\prime$ & R & 626.1 & 78.3 &  116.4 $\pm$  48.6 &  $-$53.2 $\pm$  48.1 &  128.0 $\pm$  48.4 &  167.7 $\pm$  12.0 \\
\hline
HD 105211 & 2015JUN & 2015-06-26 09:02:42 & 2086 & 1280 & 425SP & B & 401.0 & 58.3 &  $-$20.0 $\pm$  18.0 &   $-$6.4 $\pm$  18.0 &   21.0 $\pm$  18.0 &   98.9 $\pm$  28.9 \\
HD 105211 & 2015JUN & 2015-06-26 08:32:19 & 1350 & 640 & $g^\prime$ & B & 468.8 & 90.3 &  $-$15.7 $\pm$   \phantom{0}6.3 &    6.6 $\pm$   \phantom{0}6.4 &   17.0 $\pm$   \phantom{0}6.3 &   78.6 $\pm$  11.7 \\
\hline
HD 109573A & 2015MAY & 2015-05-23 12:41:45 & 2114 & 1280 & 425SP & B & 400.4 & 58.6 & $-$142.6 $\pm$  24.6 &  $-$37.8 $\pm$  24.9 &  147.5 $\pm$  24.7 &   97.4 $\pm$   \phantom{0}4.9 \\
HD 109573A & 2015MAY & 2015-05-23 11:31:37 & 1322 & 640 & $g^\prime$ & B & 463.4 & 89.1 & $-$186.4 $\pm$  11.0 &  $-$74.9 $\pm$  11.1 &  200.9 $\pm$  11.1 &  100.9 $\pm$   \phantom{0}1.6 \\
HD 109573A & 2015MAY & 2015-05-23 12:02:52 & 2173 & 1280 & $r^\prime$ & B & 597.9 & 83.7 & $-$155.7 $\pm$  21.8 &  $-$86.6 $\pm$  22.0 &  178.2 $\pm$  21.9 &  104.5 $\pm$   \phantom{0}3.5 \\
\hline
HD 115892 & 2015JUN & 2015-06-29 08:17:47 & 1944 & 1280 & 425SP & B & 400.9 & 58.0 &  $-$45.9 $\pm$  15.4 &    9.5 $\pm$  15.4 &   46.9 $\pm$  15.4 &   84.2 $\pm$  10.0 \\
HD 115892 & 2014AUG & 2014-09-02 09:37:54 & 1229 & 640 & $g^\prime$ & B & 471.9 & 91.0 &  $-$21.4 $\pm$   \phantom{0}3.8 &   27.6 $\pm$   \phantom{0}3.8 &   34.9 $\pm$   \phantom{0}3.8 &   63.9 $\pm$   \phantom{0}3.1 \\
HD 115892 & 2015JUN & 2015-06-28 08:09:14 & 1715 & 640 & $r^\prime$ & R & 622.8 & 79.0 &   $-$0.8 $\pm$   \phantom{0}7.4 &   23.5 $\pm$   \phantom{0}7.5 &   23.5 $\pm$   \phantom{0}7.4 &   46.0 $\pm$   \phantom{0}9.5 \\
\hline
HD 161868 & 2015JUN & 2015-06-26 13:31:30 & 1366 & 640 & 425SP & B & 400.5 & 58.7 &   19.0 $\pm$  17.7 &   17.0 $\pm$  17.6 &   25.5 $\pm$  17.7 &   20.9 $\pm$  24.5 \\
HD 161868 & 2015JUN & 2015-06-28 13:12:25 & 2010 & 1280 & 425SP & B & 400.5 & 58.7 &    3.2 $\pm$  16.3 &   34.4 $\pm$  16.3 &   34.5 $\pm$  16.3 &   42.3 $\pm$  16.2 \\
HD 161868 & 2015JUN & 2015-06-29 10:22:03 & 5392 & 640 & 425SP & B & 401.1 & 59.1 &  $-$11.5 $\pm$  18.4 &   69.0 $\pm$  18.5 &   70.0 $\pm$  18.4 &   49.7 $\pm$   \phantom{0}7.7 \\
HD 161868 & 2015JUN & 2015-06-26 13:06:51 & 1266 & 640 & $g^\prime$ & B & 464.2 & 89.3 &   27.0 $\pm$   \phantom{0}5.1 &   47.1 $\pm$   \phantom{0}5.1 &   54.3 $\pm$   \phantom{0}5.1 &   30.1 $\pm$   \phantom{0}2.7 \\
HD 161868 & 2018JUN & 2018-07-06 07:48:29 & 593 & 192 & $g^\prime$ & B & 466.9 & 93.1 &   60.0 $\pm$  25.1 &   44.6 $\pm$  25.2 &   74.8 $\pm$  25.1 &   18.3 $\pm$  10.3 \\
HD 161868 & 2018JUN & 2018-07-05 11:31:11 & 532 & 192 & $g^\prime$ & B & 467.1 & 93.1 &    2.8 $\pm$  25.0 &   29.7 $\pm$  25.2 &   29.8 $\pm$  25.1 &   42.3 $\pm$  28.6 \\
HD 161868 & 2018JUN & 2018-07-06 11:38:51 & 474 & 192 & $g^\prime$ & B & 467.2 & 93.1 &  $-$13.1 $\pm$  25.0 &   54.6 $\pm$  25.2 &   56.1 $\pm$  25.1 &   51.7 $\pm$  15.0 \\
HD 161868 & 2018JUN & 2018-07-05 11:40:29 & 480 & 192 & $r^\prime$ & B & 602.5 & 61.6 &   24.6 $\pm$  13.8 &   34.8 $\pm$  13.8 &   42.6 $\pm$  13.8 &   27.4 $\pm$   \phantom{0}9.8 \\
HD 161868 & 2018JUN & 2018-07-06 07:57:33 & 463 & 192 & $r^\prime$ & B & 602.5 & 61.7 &   53.5 $\pm$  13.2 &   36.3 $\pm$  13.2 &   64.7 $\pm$  13.2 &   17.1 $\pm$   \phantom{0}6.0 \\
HD 161868 & 2018JUN & 2018-07-06 11:47:37 & 499 & 192 & $r^\prime$ & B & 602.5 & 61.6 &   18.1 $\pm$  13.8 &   54.0 $\pm$  13.2 &   57.0 $\pm$  13.5 &   35.7 $\pm$   \phantom{0}6.9 \\
HD 161868 & 2017JUN & 2017-07-04 12:13:36 & 1232 & 800 & $r^\prime$ & R & 619.6 & 82.6 &   44.4 $\pm$   \phantom{0}6.4 &   22.7 $\pm$   \phantom{0}6.2 &   49.9 $\pm$   \phantom{0}6.3 &   13.5 $\pm$   \phantom{0}3.6 \\
\hline
HD 181327 & 2017JUN & 2017-07-05 13:20:48 & 3362 & 2560 & 425SP & B & 401.2 & 51.4 &   28.6 $\pm$  36.9 &  $-$61.1 $\pm$  37.1 &   67.5 $\pm$  37.0 &  147.5 $\pm$  19.4 \\
HD 181327 & 2017JUN & 2017-07-02 13:56:41 & 2032 & 1280 & 500SP & B & 441.2 & 77.4 &   23.0 $\pm$  19.2 & $-$133.3 $\pm$  19.0 &  135.3 $\pm$  19.1 &  139.9 $\pm$   \phantom{0}4.1 \\
HD 181327 & 2015JUN & 2015-06-27 12:07:35 & 2017 & 1280 & $g^\prime$ & B & 470.7 & 90.8 &   35.9 $\pm$  14.8 &  $-$70.2 $\pm$  14.8 &   78.8 $\pm$  14.8 &  148.5 $\pm$   \phantom{0}5.4 \\
HD 181327 & 2015JUN & 2015-06-28 09:43:43 & 2007 & 1280 & $r^\prime$ & R & 623.5 & 78.8 &   88.1 $\pm$  29.1 & $-$130.5 $\pm$  29.1 &  157.5 $\pm$  29.1 &  152.0 $\pm$   \phantom{0}5.3 \\
HD 181327 & 2017AUG & 2017-08-09 09:27:53 & 3397 & 2560 & 650LP & R & 721.3 & 65.3 &   56.4 $\pm$  24.3 &  $-$98.2 $\pm$  23.8 &  113.2 $\pm$  24.0 &  149.9 $\pm$   \phantom{0}6.2 \\
\hline
HD 188228 & 2015JUN & 2015-06-26 14:57:40 & 2015 & 1280 & 425SP & B & 400.6 & 58.8 &   26.6 $\pm$  16.5 &  $-$18.7 $\pm$  16.5 &   32.5 $\pm$  16.5 &  162.4 $\pm$  17.7 \\
HD 188228 & 2017AUG & 2017-08-10 15:48:04 & 1455 & 640 & 425SP & B & 400.8 & 52.2 &   55.1 $\pm$  19.5 &   $-$0.6 $\pm$  19.1 &   55.1 $\pm$  19.3 &  179.7 $\pm$  10.8 \\
HD 188228 & 2015MAY & 2015-05-23 19:33:30 & 1322 & 640 & $g^\prime$ & B & 464.2 & 89.3 &   61.7 $\pm$   \phantom{0}5.3 &  $-$14.3 $\pm$   \phantom{0}5.4 &   63.3 $\pm$   5.4 &  173.5 $\pm$   \phantom{0}2.5 \\
HD 188228 & 2015JUN & 2015-06-28 10:51:08 & 1915 & 1280 & $r^\prime$ & R & 619.7 & 79.5 &   81.2 $\pm$   \phantom{0}8.6 &  $-$14.5 $\pm$   \phantom{0}8.6 &   82.5 $\pm$   8.6 &  174.9 $\pm$   \phantom{0}3.0 \\
HD 188228 & 2017AUG & 2017-08-07 13:22:32 & 2937 & 1920 & 650LP & R & 717.7 & 65.9 &   48.2 $\pm$  10.8 &  $-$40.6 $\pm$  11.3 &   63.0 $\pm$  11.1 &  159.9 $\pm$   \phantom{0}5.1 \\
HD 188228 & 2017AUG & 2017-08-08 13:45:14 & 2067 & 1280 & 650LP & R & 717.7 & 65.9 &   43.2 $\pm$  10.2 &  $-$39.1 $\pm$  10.3 &   58.3 $\pm$  10.3 &  158.9 $\pm$   \phantom{0}5.1 \\
\hline
HD 197481 & 2015JUN & 2015-06-28 13:53:59 & 2002 & 1280 & 500SP & B & 451.2 & 86.6 & $-$115.4 $\pm$  50.7 & $-$281.3 $\pm$  50.8 &  304.1 $\pm$  50.7 &  123.8 $\pm$   \phantom{0}4.9 \\
HD 197481 & 2014AUG & 2014-08-29 13:32:21 & 3644 & 2560 & Clear & B & 515.7 & 87.1 &    7.9 $\pm$  22.0 & $-$281.1 $\pm$  22.6 &  281.2 $\pm$  22.3 &  135.8 $\pm$   \phantom{0}2.2 \\
HD 197481 & 2014AUG & 2014-08-30 14:06:33 & 3523 & 2560 & $r^\prime$ & B & 606.0 & 82.1 &  $-$55.9 $\pm$  38.4 &  $-$80.4 $\pm$  39.3 &   97.9 $\pm$  38.9 &  117.6 $\pm$  12.9 \\
\hline
\end{tabular}
\end{table*}

\begin{table*}
\contcaption{}
\fontsize{8.25pt}{9pt}\selectfont
\tabcolsep 2.5 pt
\begin{tabular}{lllrrccrrrrrr}
\hline
\hline
\multicolumn{1}{c}{Target} & \multicolumn{1}{c}{Run}    & \multicolumn{1}{c}{UT}                  & Dwell & Exp. & Filt. & Det. & $\lambda_{\rm eff}$ & Eff. & \multicolumn{1}{c}{$q$} & \multicolumn{1}{c}{$u$} & \multicolumn{1}{c}{$p$} & \multicolumn{1}{c}{$\theta$}\\
        &               &        & \multicolumn{1}{c}{(s)} & \multicolumn{1}{c}{(s)} & & & \multicolumn{1}{c}{(nm)} & \multicolumn{1}{c}{(\%)} & \multicolumn{1}{c}{(ppm)} & \multicolumn{1}{c}{(ppm)} & \multicolumn{1}{c}{(ppm)} & \multicolumn{1}{c}{($^\circ$)}\\
\hline
HD 202628 & 2015JUN & 2015-06-27 15:52:18 & 1209 & 640 & 425SP & B & 403.1 & 58.6 &  $-$83.4 $\pm$  54.3 & $-$172.2 $\pm$  54.7 &  191.3 $\pm$  54.5 &  122.1 $\pm$   \phantom{0}8.4 \\
HD 202628 & 2015JUN & 2015-06-27 16:14:49 & 1246 & 640 & 425SP & B & 403.1 & 58.6 & $-$111.0 $\pm$  56.8 & $-$224.1 $\pm$  56.7 &  250.1 $\pm$  56.7 &  121.8 $\pm$   \phantom{0}6.6 \\
HD 202628 & 2017JUN & 2017-06-29 19:11:22 & 1436 & 800 & 425SP & B & 403.4 & 52.1 &  $-$93.3 $\pm$  74.2 & $-$272.2 $\pm$  74.4 &  287.7 $\pm$  74.3 &  125.5 $\pm$   \phantom{0}7.5 \\
HD 202628 & 2017AUG & 2017-08-10 16:33:43 & 3744 & 2560 & 425SP & B & 403.4 & 52.2 &  $-$85.6 $\pm$  39.1 & $-$227.8 $\pm$  38.6 &  243.4 $\pm$  38.9 &  124.7 $\pm$   \phantom{0}4.6 \\
HD 202628 & 2017AUG & 2017-08-13 18:01:18 & 3664 & 2560 & 500SP & B & 446.8 & 80.5 &  $-$48.6 $\pm$  14.8 & $-$150.9 $\pm$  14.3 &  158.5 $\pm$  14.6 &  126.1 $\pm$   \phantom{0}2.7 \\
HD 202628 & 2015JUN & 2015-06-26 15:37:58 & 1991 & 1280 & $g^\prime$ & B & 473.9 & 91.5 & $-$148.7 $\pm$  13.1 & $-$124.2 $\pm$  13.3 &  193.7 $\pm$  13.2 &  109.9 $\pm$   \phantom{0}1.9 \\
HD 202628 & 2015JUN & 2015-06-28 11:57:10 & 1992 & 1280 & $r^\prime$ & R & 625.1 & 78.5 &  $-$25.7 $\pm$  24.2 & $-$163.3 $\pm$  23.5 &  165.3 $\pm$  23.8 &  130.5 $\pm$   \phantom{0}4.2 \\
HD 202628 & 2017AUG & 2017-08-09 10:27:00 & 3398 & 2560 & 650LP & R & 722.8 & 65.1 & $-$166.9 $\pm$  20.2 & $-$118.2 $\pm$  19.9 &  204.5 $\pm$  20.0 &  107.7 $\pm$   \phantom{0}2.8 \\
\hline
HD 216956 & 2017JUN & 2017-06-22 18:06:42 & 2863 & 640 & 425SP & B & 400.3 & 51.7 &    0.4 $\pm$  16.6 &  $-$19.0 $\pm$  15.4 &   19.0 $\pm$  16.0 &  135.6 $\pm$  28.6 \\
HD 216956 & 2017JUN & 2017-06-26 19:08:43 & 1353 & 640 & 425SP & B & 400.3 & 51.7 &  $-$44.4 $\pm$  15.0 &  $-$31.6 $\pm$  15.2 &   54.5 $\pm$  15.1 &  107.7 $\pm$   \phantom{0}8.1 \\
HD 216956 & 2016DEC & 2016-12-05 11:28:08 & 1475 & 640 & 425SP & B & 400.8 & 52.0 &  $-$17.1 $\pm$  17.2 &   16.9 $\pm$  16.4 &   24.0 $\pm$  16.8 &   67.7 $\pm$  24.7 \\
HD 216956 & 2014AUG & 2014-08-28 14:18:31 & 1429 & 640 & 500SP & B & 436.4 & 79.3 &  $-$22.1 $\pm$   \phantom{0}8.0 &   $-$5.9 $\pm$   \phantom{0}7.9 &   22.9 $\pm$   \phantom{0}8.0 &   97.5 $\pm$  10.8 \\
HD 216956 & 2017JUN & 2017-07-01 20:05:16 & 859 & 320 & 500SP & B & 436.6 & 75.1 &  $-$17.0 $\pm$   \phantom{0}8.0 &  $-$13.2 $\pm$   \phantom{0}8.0 &   21.5 $\pm$   \phantom{0}8.0 &  108.9 $\pm$  11.8 \\
HD 216956 & 2014AUG & 2014-08-28 14:40:54 & 1164 & 640 & $g^\prime$ & B & 464.9 & 89.4 &  $-$15.3 $\pm$   \phantom{0}2.8 &  $-$11.0 $\pm$   2.8 &   18.8 $\pm$   \phantom{0}2.8 &  107.9 $\pm$   \phantom{0}4.2 \\
HD 216956 & 2017AUG & 2017-08-17 11:57:51 & 913 & 320 & $g^\prime$ & B & 465.7 & 87.1 &  $-$21.6 $\pm$   \phantom{0}3.6 &   $-$2.5 $\pm$   3.6 &   21.7 $\pm$   \phantom{0}3.6 &   93.3 $\pm$   \phantom{0}4.7 \\
HD 216956 & 2015NOV & 2015-11-02 12:38:31 & 1221 & 640 & $r^\prime$ & R & 620.4 & 79.4 &  $-$14.6 $\pm$   \phantom{0}3.8 &   $-$3.8 $\pm$   4.0 &   15.1 $\pm$   \phantom{0}3.9 &   97.3 $\pm$   \phantom{0}7.6 \\
\hline
\end{tabular}
\end{table*}

\begin{table*}
\caption{Observations of interstellar control stars.}
\label{tab:controls}
\tabcolsep 3.5 pt
\begin{tabular}{llllrrrrrrrr}
\hline
\hline
\multicolumn{1}{c}{Control}      & SpT   &   Run     & UT                  & Dwell & Exp. & $\lambda_{\rm eff}$ & Eff. & \multicolumn{1}{c}{$q$} & \multicolumn{1}{c}{$u$} & \multicolumn{1}{c}{$p$} & \multicolumn{1}{c}{$\theta$}\\
\multicolumn{1}{c}{} &        &            &           & \multicolumn{1}{c}{(s)} & \multicolumn{1}{c}{(s)} & \multicolumn{1}{c}{(nm)} & \multicolumn{1}{c}{(\%)} & \multicolumn{1}{c}{(ppm)} & \multicolumn{1}{c}{(ppm)} & \multicolumn{1}{c}{(ppm)} & \multicolumn{1}{c}{($^\circ$)}\\
\hline
HD   8350 & F5/6\,V & 2015OCT & 2015-10-18 10:58:53 & 1992 & 1280 & 470.3 & 90.7 &   21.7 $\pm$  18.4 &  $-$27.8 $\pm$  20.0 &   35.3 $\pm$  19.2 &  154.0 $\pm$  19.2 \\
HD  39014 & A7\,V & 2018MAR & 2018-03-23 11:46:13 & 1136 & 640 & 465.9 & 82.0 &    \phantom{00}8.4 $\pm$   \phantom{0}5.9 &  $-$27.8 $\pm$   \phantom{0}5.9 &   \phantom{0}29.0 $\pm$   \phantom{0}5.9 &  143.4 $\pm$   \phantom{0}5.9 \\
HD  64185 & F4\,V & 2018MAR & 2018-03-26 10:59:12 & 1770 & 1280 & 468.3 & 83.0 &  $-$38.1 $\pm$   \07.4 &   38.1 $\pm$   \07.4 &   53.9 $\pm$   \07.4 &   67.5 $\pm$   \03.9 \\
 & & 2018MAR & 2018-03-26 11:28:55 & 1750 & 1280 & 468.3 & 83.0 &  $-$30.7 $\pm$   \07.6 &   43.5 $\pm$   \07.3 &   53.2 $\pm$   \07.4 &   62.6 $\pm$   \04.0 \\
HD  65907 & F9.5\,V & 2018MAR & 2018-03-26 10:28:01 & 1717 & 1280 & 470.8 & 84.0 &   \phantom{0}17.3 $\pm$   \phantom{0}7.3 &   \phantom{0}12.9 $\pm$   \phantom{0}7.5 &   \phantom{0}21.6 $\pm$   \phantom{0}7.4 &   \phantom{0}18.4 $\pm$  10.5 \\
HD  88742 & G0\,V & 2017JUN & 2017-07-01 08:32:02 & 2631 & 1920 & 472.4 & 89.1 &  $-$16.2 $\pm$  10.1 &  $-$16.1 $\pm$   \phantom{0}9.9 &   \phantom{0}22.8 $\pm$  10.0 &  112.4 $\pm$  14.7 \\
HD  90589 & F3\,V & 2018MAR & 2018-03-23 12:20:38 & 1168 & 640 & 468.2 & 83.0 &  $-$17.0 $\pm$   \05.2 &    \phantom{00}1.3 $\pm$   \05.1 &   17.0 $\pm$   \05.1 &   87.8 $\pm$   \09.0 \\
HD 100407 & G7\,IIIb & 2018MAR & 2018-04-02 16:51:05 & 981 & 640 & 474.6 & 85.6 &  $-$29.1 $\pm$   \phantom{0}4.9 &   \phantom{0}55.1 $\pm$   \phantom{0}5.0 &   \phantom{0}62.3 $\pm$   \phantom{0}5.0 &   \phantom{0}58.9 $\pm$   \phantom{0}2.3 \\
HD 113415 & F8.5\,V & 2018AUG & 2018-08-20 09:16:36 & 996 & 640 & 471.6 & 74.6 &  $-$13.4 $\pm$  11.7 &   20.6 $\pm$  11.5 &   24.6 $\pm$  11.6 &   61.5 $\pm$  16.2 \\
HD 131342 & K2\,III & 2018MAR & 2018-03-26 14:22:44 & 1715 & 1280 & 475.8 & 86.1 &   \phantom{0}17.9 $\pm$   \phantom{0}6.8 &   \phantom{0}29.9 $\pm$   \phantom{0}6.9 &   \phantom{0}34.8 $\pm$   \phantom{0}6.9 &   \phantom{0}29.5 $\pm$   \phantom{0}5.7 \\
HD 135235 & A2\,IIIs & 2018MAR & 2018-04-05 17:16:49 & 1788 & 1280 & 462.7 & 80.6 &  $-$28.1 $\pm$   \phantom{0}7.9 &   \phantom{0}93.5 $\pm$   \phantom{0}7.9 &   \phantom{0}97.6 $\pm$   \phantom{0}7.9 &   \phantom{0}53.4 $\pm$   \phantom{0}2.3 \\
HD 136351 & F6\,III-IV & 2018MAR & 2018-04-02 17:15:37 & 1677 & 1280 & 468.9 & 83.2 &  $-$59.8 $\pm$   \phantom{0}5.7 &   \phantom{0}84.5 $\pm$   \phantom{0}5.7 &  103.5 $\pm$   \phantom{0}5.7 &   \phantom{0}62.6 $\pm$   \phantom{0}1.6 \\
HD 157347 & G3\,V & 2017AUG & 2017-08-10 12:11:10 & 3525 & 2560 & 473.7 & 89.5 &  138.9 $\pm$   \phantom{0}8.9 &   \phantom{0}47.1 $\pm$   \phantom{0}8.9 &  146.7 $\pm$   \phantom{0}8.9 &    \phantom{00}9.4 $\pm$   \phantom{0}1.7 \\
HD 162917 & F4\,IV-V & 2015OCT & 2015-10-15 09:24:11 & 1665 & 960 & 471.1 & 90.8 &   \phantom{0}54.7 $\pm$  11.8 &   \phantom{0}32.4 $\pm$  12.0 &   \phantom{0}63.6 $\pm$  11.9 &   \phantom{0}15.3 $\pm$   \phantom{0}5.4 \\
HD 171802 & F5\,III & 2018AUG & 2018-08-19 10:23:03 & 1533 & 1120 & 469.0 & 73.1 &   11.3 $\pm$   \08.4 &   27.6 $\pm$   \07.6 &   29.8 $\pm$   \08.0 &   33.9 $\pm$   \07.8 \\
HD 181391 & G8/K0\,IV & 2018AUG & 2018-08-19 15:22:44 & 978 & 640 & 475.9 & 77.1 &  $-$27.7 $\pm$  10.3 &    \03.1 $\pm$  10.0 &   27.9 $\pm$  10.2 &   86.8 $\pm$  11.5 \\
HD 183414 & G3\,V & 2015OCT & 2015-10-16 09:44:13 & 1859 & 960 & 473.7 & 91.4 &  164.9 $\pm$  28.9 &  $-$80.8 $\pm$  28.0 &  183.6 $\pm$  28.4 &  166.9 $\pm$   \phantom{0}4.4 \\
HD 188114 & K0\,II/III & 2018MAR & 2018-04-06 18:51:29 & 988 & 640 & 474.9 & 85.8 &   \phantom{0}89.5 $\pm$   \phantom{0}6.3 &    \phantom{00}1.6 $\pm$   \phantom{0}6.3 &   \phantom{0}89.5 $\pm$   \phantom{0}6.3 &    \phantom{00}0.5 $\pm$   \phantom{0}2.0 \\
HD 188887 & K2\,IV & 2018MAR & 2018-04-03 17:49:40 & 1667 & 1280 & 476.3 & 86.2 &   \phantom{0}67.5 $\pm$   \phantom{0}7.8 &  $-$53.1 $\pm$   \phantom{0}7.9 &   \phantom{0}85.9 $\pm$   \phantom{0}7.8 &  160.9 $\pm$   \phantom{0}2.6 \\
HD 190248 & G8\,IV & 2018MAR & 2018-04-03 18:15:46 & 975 & 640 & 474.7 & 85.7 &    \phantom{00}5.4 $\pm$   \phantom{0}4.8 &  $-$17.3 $\pm$   \phantom{0}5.0 &   \phantom{0}18.1 $\pm$   \phantom{0}4.9 &  143.7 $\pm$   \phantom{0}7.9 \\
 & & 2018JUL & 2018-07-14 19:50:48 & 1052 & 640 & 475.5 & 85.2 &    \phantom{00}0.8 $\pm$   \phantom{0}4.7 &  $-$10.8 $\pm$   \phantom{0}4.7 &   \phantom{0}10.8 $\pm$   \phantom{0}4.7 &  137.1 $\pm$  14.6 \\
HD 194640 & G8\,V & 2015OCT & 2015-10-17 11:02:42 & 1735 & 960 & 475.1 & 91.8 &   \phantom{0}26.2 $\pm$  17.0 &   \phantom{0}$-$7.4 $\pm$  16.9 &   \phantom{0}27.2 $\pm$  17.0 &  172.1 $\pm$  22.2 \\
 (Clear) & & 2015OCT & 2015-10-20 10:00:19 & 1988 & 1280 & 495.0 & 86.5 &   \phantom{0}$-$0.3 $\pm$  11.6 &   10.2 $\pm$  11.7 &   \phantom{0}10.2 $\pm$  11.6 &   \phantom{0}45.8 $\pm$  34.2 \\
HD 197157 & A9\,IV & 2018MAR & 2018-04-04 18:05:26 & 1064 & 640 & 466.6 & 82.3 &  $-$31.7 $\pm$   \phantom{0}6.2 &  $-$38.2 $\pm$   \phantom{0}6.7 &   \phantom{0}49.6 $\pm$   \phantom{0}6.5 &  115.2 $\pm$   \phantom{0}3.7 \\
HD 199288 & G2\,V & 2017JUN & 2017-06-26 17:08:19 & 2778 & 1920 & 472.6 & 89.2 &  $-$33.3 $\pm$  13.6 &  $-$11.4 $\pm$  12.2 &   \phantom{0}35.2 $\pm$  12.9 &   \phantom{0}99.4 $\pm$  11.5 \\
HD 212132 & F0\,V & 2018AUG & 2018-09-01 16:21:35 & 992 & 640 & 466.4 & 59.4 & $-$169.3 $\pm$  13.7 &  $-$12.7 $\pm$  14.9 &  169.8 $\pm$  14.3 &   92.1 $\pm$   \02.5 \\
HD 212330 & G2\,IV-V & 2018AUG & 2018-09-02 15:05:04 & 1011 & 640 & 471.8 & 62.7 &  $-$54.4 $\pm$  12.7 &  -18.5 $\pm$  13.0 &   57.5 $\pm$  12.9 &   99.4 $\pm$   \06.5 \\
HD 216385 & F6\,V & 2018AUG & 2018-08-18 12:49:58 & 1335 & 960 & 470.3 & 73.9 &    \phantom{00}0.8 $\pm$   \phantom{0}8.0 &  $-$16.0 $\pm$   \phantom{0}8.5 &   \phantom{0}16.0 $\pm$   \phantom{0}8.2 &  136.4 $\pm$  17.9 \\
HD 217364 & K1\,III & 2018AUG & 2018-09-01 16:40:31 & 991 & 640 & 475.4 & 64.9 & $-$109.5 $\pm$   \08.1 &  $-$13.6 $\pm$   \08.3 &  110.3 $\pm$   \08.2 &   93.5 $\pm$   \02.1 \\
HD 218687 & G0\,V & 2018JUL & 2018-07-24 17:48:26 & 1796 & 1280 & 471.7 & 83.5 &  $-$79.8 $\pm$  11.3 &  $-$45.2 $\pm$  11.3 &   \phantom{0}91.7 $\pm$  11.3 &  104.8 $\pm$   \phantom{0}3.5 \\
 & & 2018JUL & 2018-07-24 18:18:15 & 1774 & 1280 & 471.9 & 83.5 &  $-$29.3 $\pm$  11.0 &   \phantom{0}23.1 $\pm$  11.2 &   \phantom{0}37.3 $\pm$  11.1 &   \phantom{0}70.9 $\pm$   \phantom{0}8.9 \\
 & & 2018AUG & 2018-08-19 16:13:15 & 1708 & 1280 & 471.7 & 74.7 &  $-$52.5 $\pm$  13.5 &    \phantom{00}9.0 $\pm$  14.2 &   \phantom{0}53.3 $\pm$  13.9 &   \phantom{0}85.1 $\pm$   \phantom{0}7.6 \\
 & & 2018AUG & 2018-08-19 16:42:56 & 1689 & 1280 & 471.9 & 74.8 &  $-$45.4 $\pm$  12.9 &  $-$22.5 $\pm$  13.4 &   \phantom{0}50.7 $\pm$  13.2 &  103.2 $\pm$   \phantom{0}7.6 \\
HD 222345 & A7\,IV & 2018JUL & 2018-07-14 18:34:24 & 1054 & 640 & 464.7 & 80.5 &  $-$12.8 $\pm$   \phantom{0}7.1 &   \phantom{0}$-$1.7 $\pm$   \phantom{0}6.8 &   \phantom{0}12.9 $\pm$   \phantom{0}7.0 &   \phantom{0}93.8 $\pm$  19.0 \\
HD 224617 & F4\,V & 2015OCT & 2015-10-19 09:34:39 & 1250 & 640 & 470.7 & 90.8 &    \phantom{00}0.7 $\pm$   \phantom{0}6.9 &  $-$26.7 $\pm$   \phantom{0}6.7 &   \phantom{0}26.7 $\pm$   \phantom{0}6.8 &  135.8 $\pm$   \phantom{0}7.5 \\
HD 225003 & F1\,V & 2016DEC & 2016-12-03 10:11:18 & 1472 & 800 & 468.6 & 88.0 &    \phantom{00}7.3 $\pm$  15.6 &  $-$27.4 $\pm$  14.9 &   \phantom{0}28.4 $\pm$  15.3 &  142.5 $\pm$  19.0 \\
\hline
\end{tabular}
\begin{flushleft}
\textbf{Notes:} * All control star observations were made with the SDSS g$^\prime$ filter and the B PMT as the detector, with the exception of HD\,194640, for which the second observation was made with no filter (Clear) * Spectral types are from SIMBAD, as is all position and distance information presented later.\\
\end{flushleft}
\end{table*}

The details of the debris disc observations are given in table \ref{tab:observations1} and the control stars in table \ref{tab:controls}. The dwell time for each observation is longer than the exposure time because 40\,s sky observations are taken for each position angle in sequence. Also included in the table are the modulator's polarisation efficiency, $\rm Eff.$ and the effective wavelength of the observation, $\lambda_{\rm eff}$; both of which are calculated with a bandpass model that takes account of target spectral type, airmass and all of the optical elements (see \citealp{2020Bailey} for details).

\section{Analysis and Results}
\label{sec:results}

\begin{figure*}
    \includegraphics[width=\textwidth]{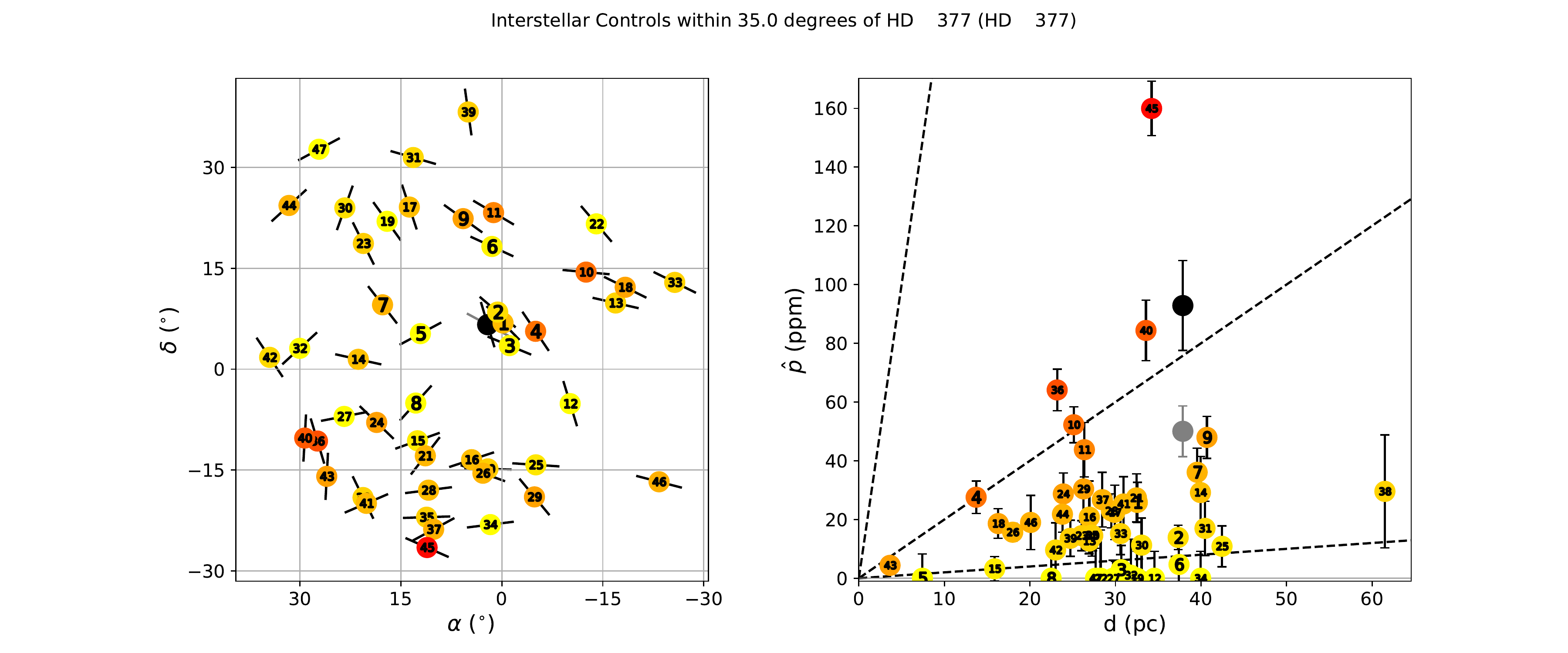}
    \includegraphics[width=\textwidth]{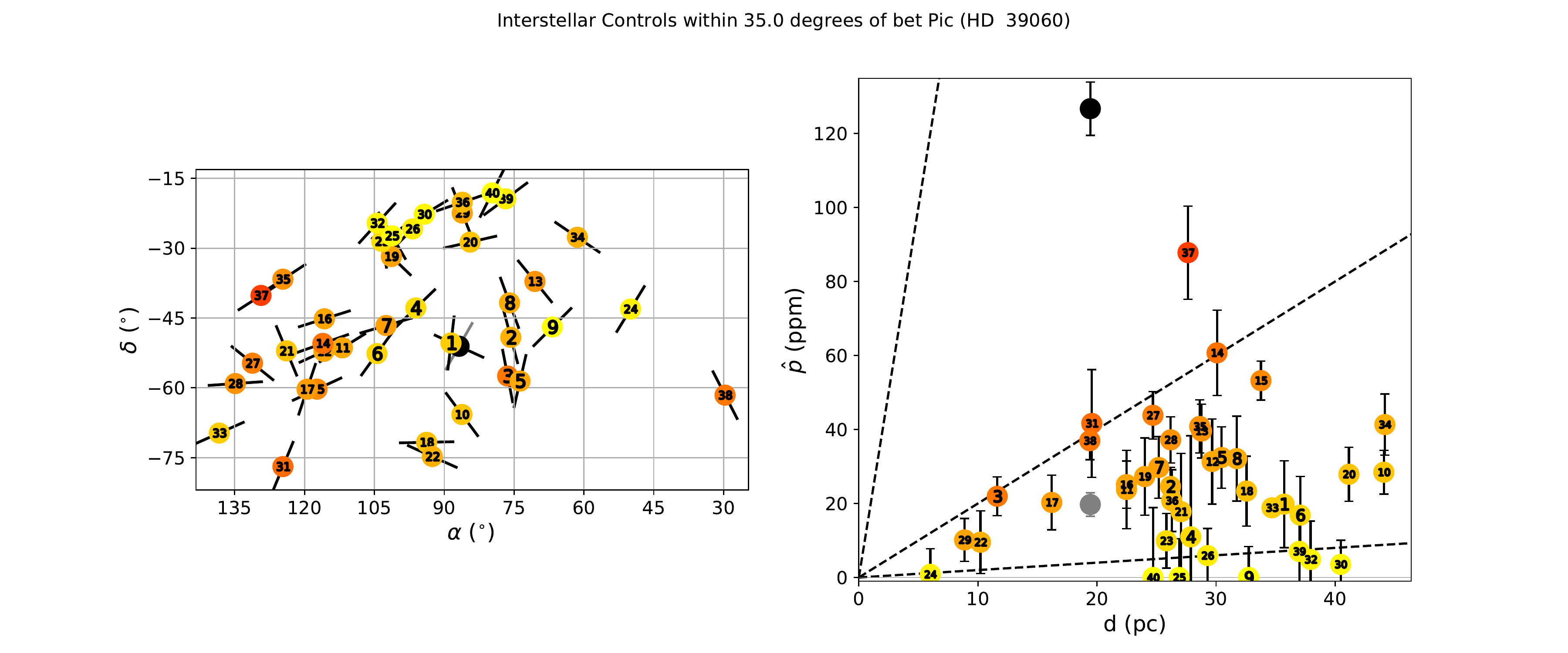}
    \includegraphics[width=\textwidth]{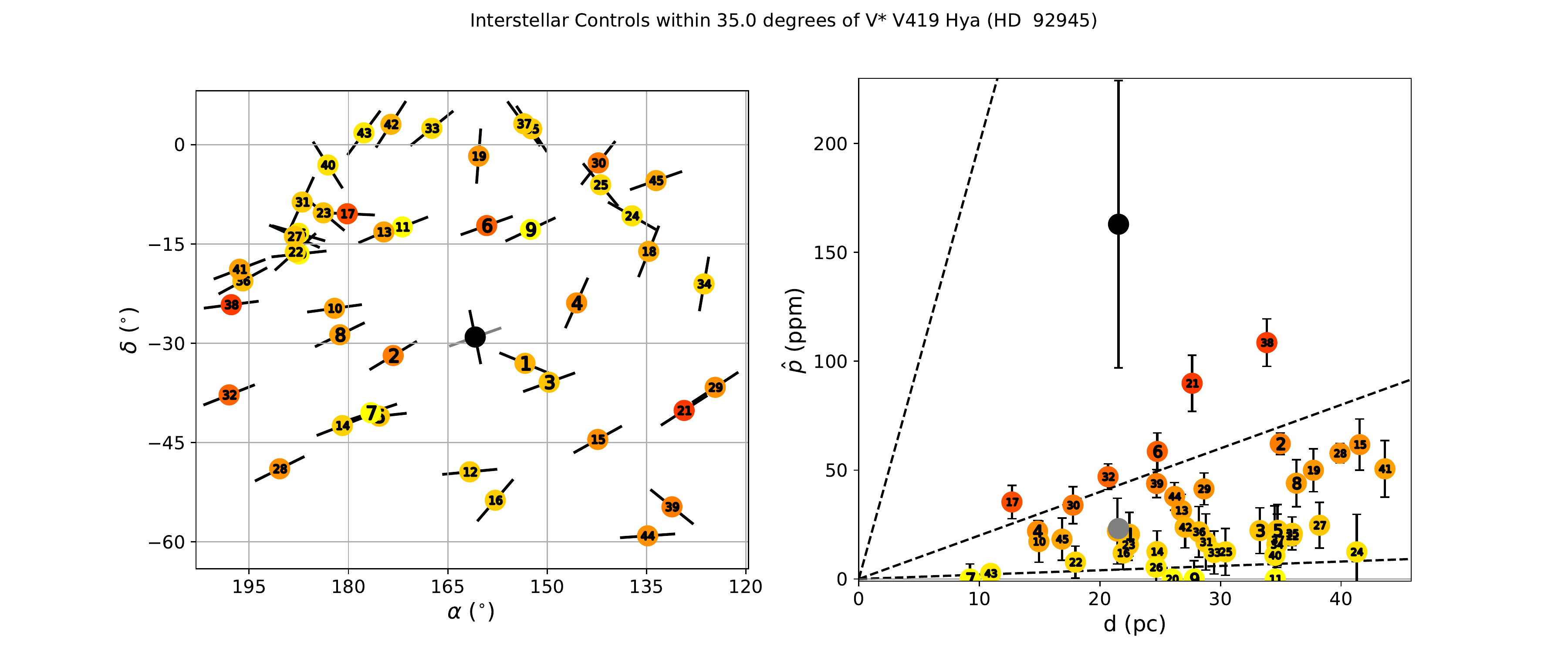}
    \vspace{-12pt}
    \caption{Plots of polarisation in $g^{\prime}$ for the target (black data point), predicted interstellar polarisation (grey data point), and nearby control stars (coloured data points, with colour denoting magnitude of polarisation per distance) as projected onto the sky (left) and as a function of distance (right). Numbering of control stars is consistent between panels. A complete list of control stars shown is given in Appendix \ref{apx:controls}. In this figure we show HD~377 (top), HD~39060 (middle), and HD~92945 (bottom). We note that for stars with $p < \sigma_{p}$ the PA's in the left hand panels are not well
defined ($\sigma_{\rm PA} > 28\degr$). \label{fig:ism1}}
\end{figure*}

\begin{figure*}
    \includegraphics[width=\textwidth]{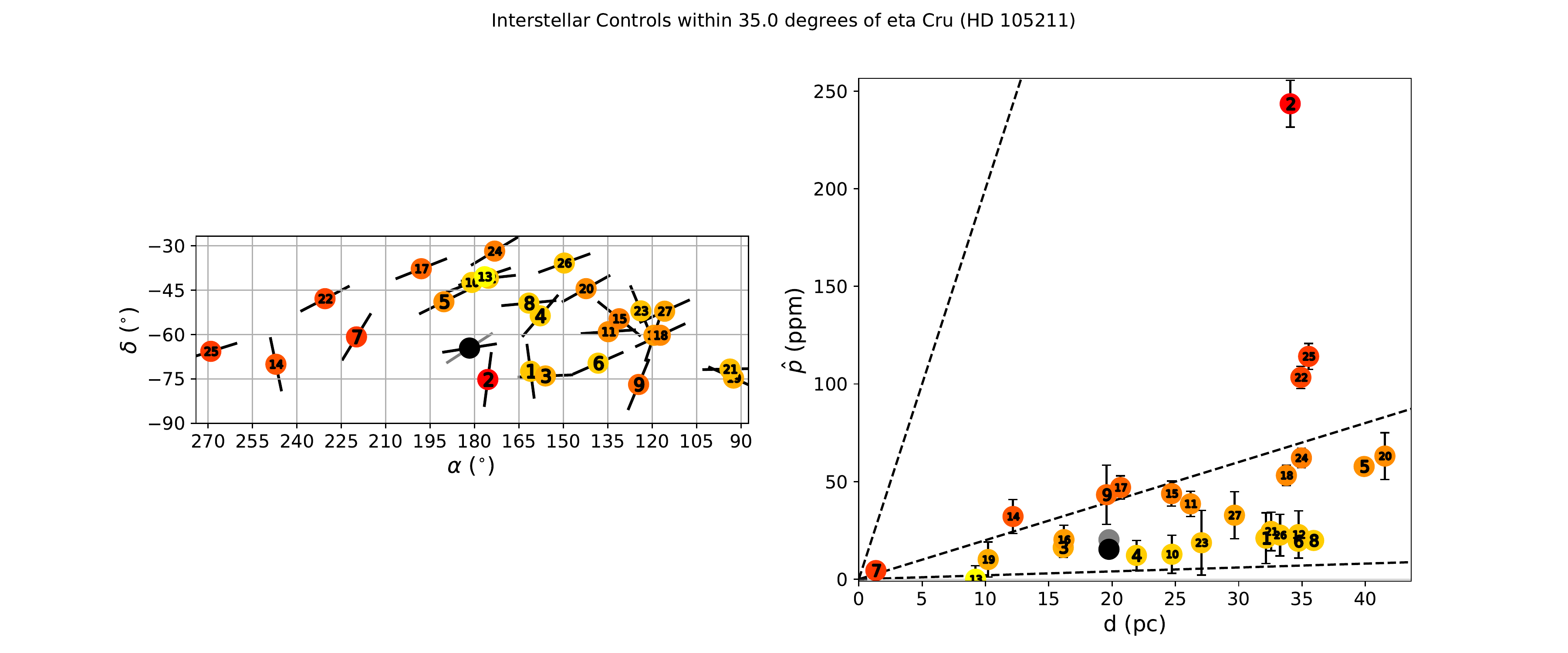}
    \includegraphics[width=\textwidth]{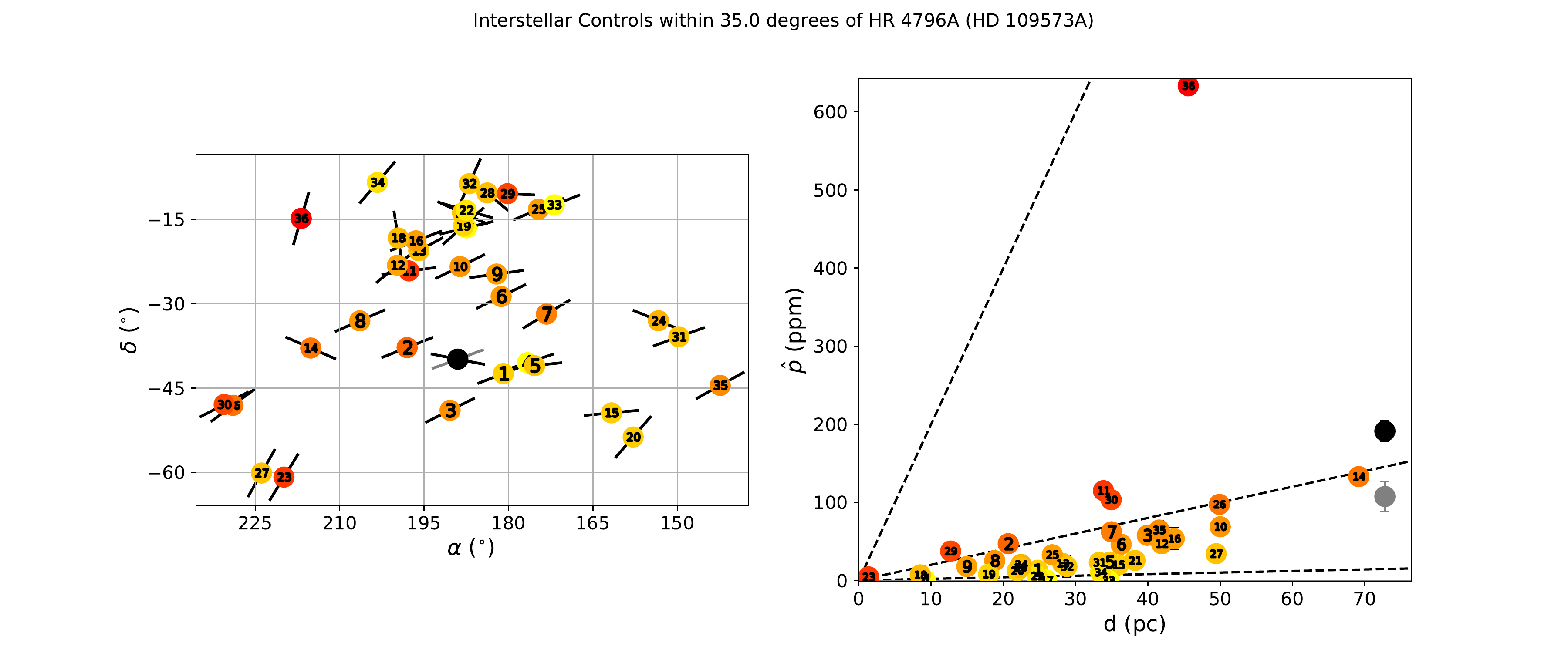}
    \includegraphics[width=\textwidth]{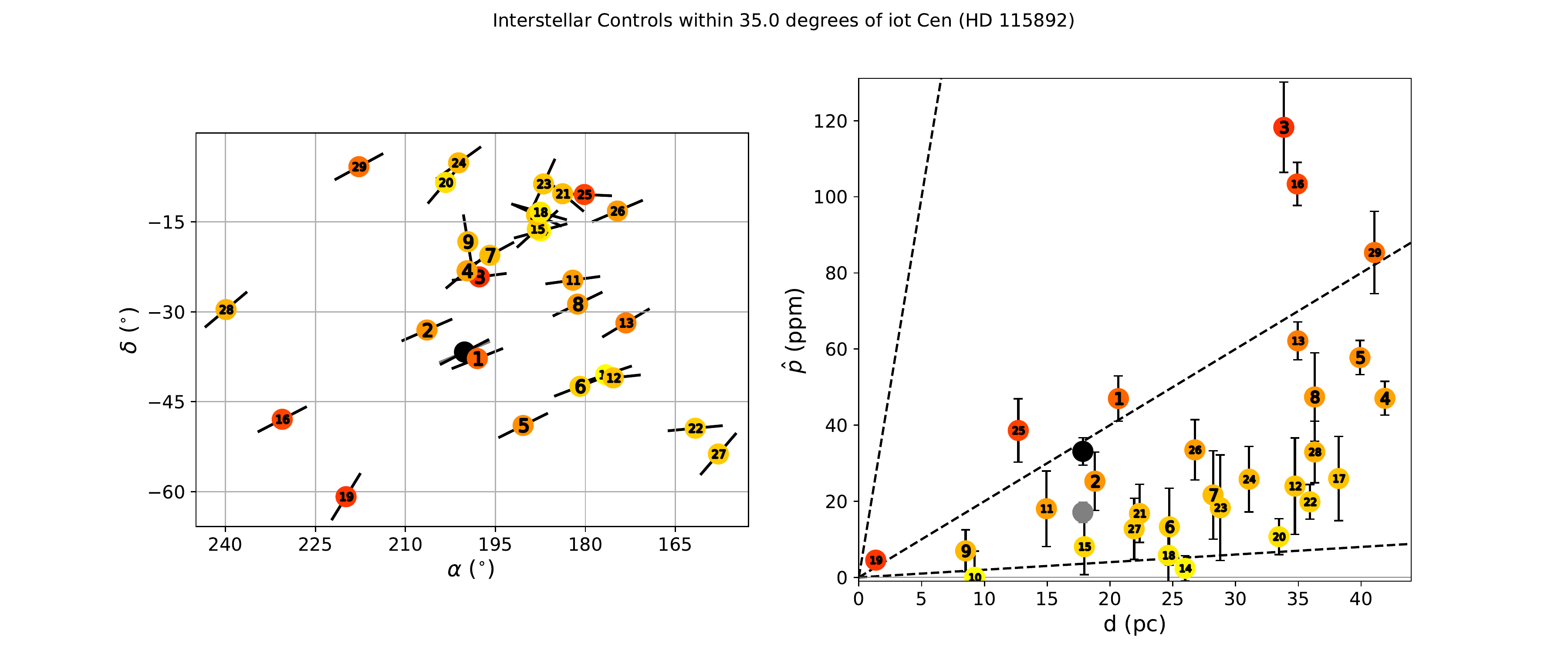}
    \caption{As per figure \ref{fig:ism1}, for targets HD~105211 (top), HD~109573A (middle), and HD~115892 (bottom). \label{fig:ism2}}
\end{figure*}

\begin{figure*}
    \includegraphics[width=\textwidth]{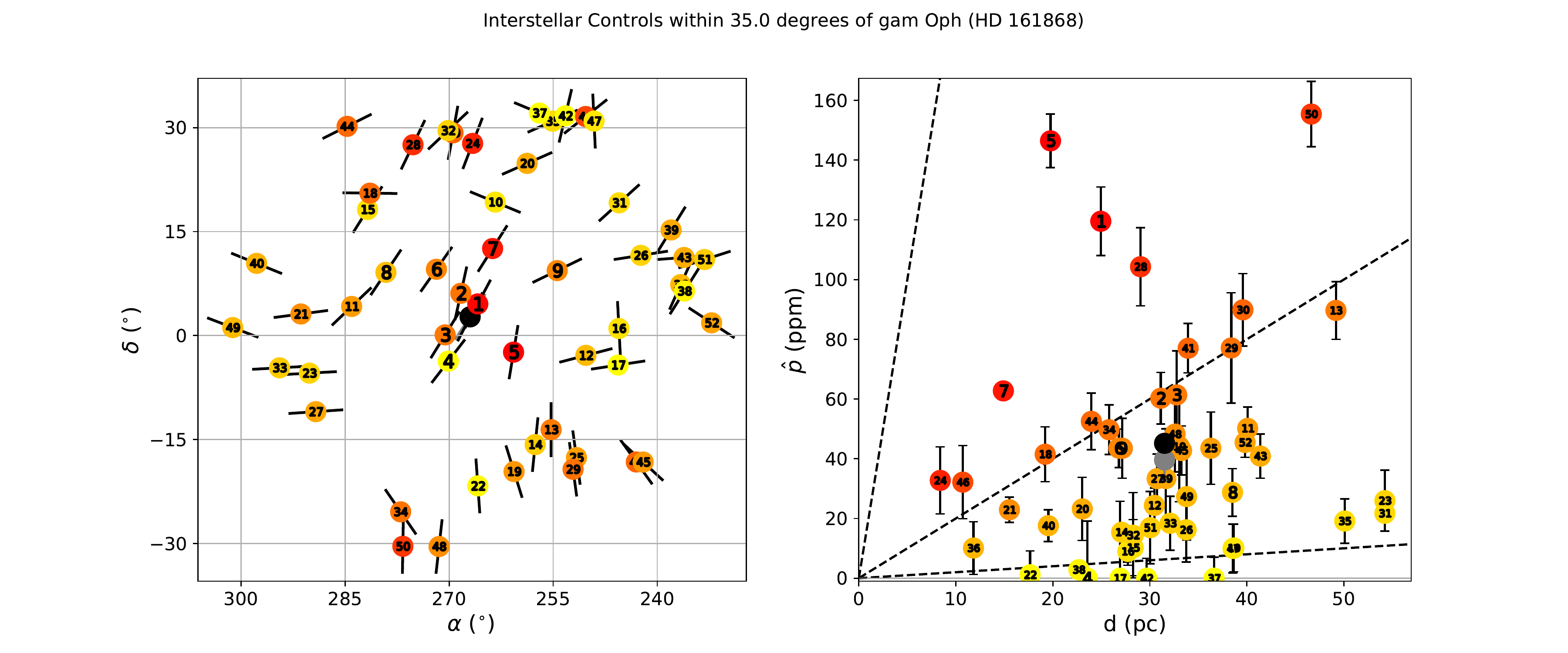}
    \includegraphics[width=\textwidth]{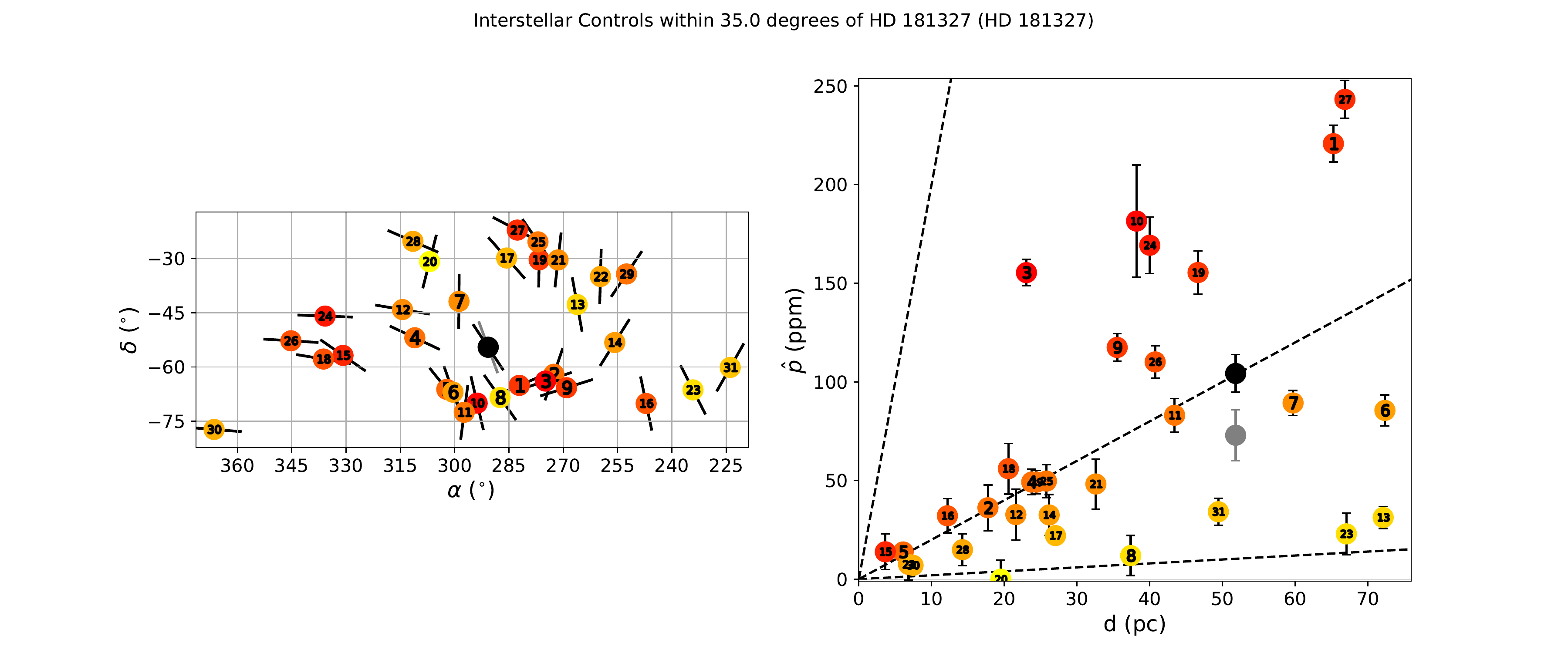}
    \includegraphics[width=\textwidth]{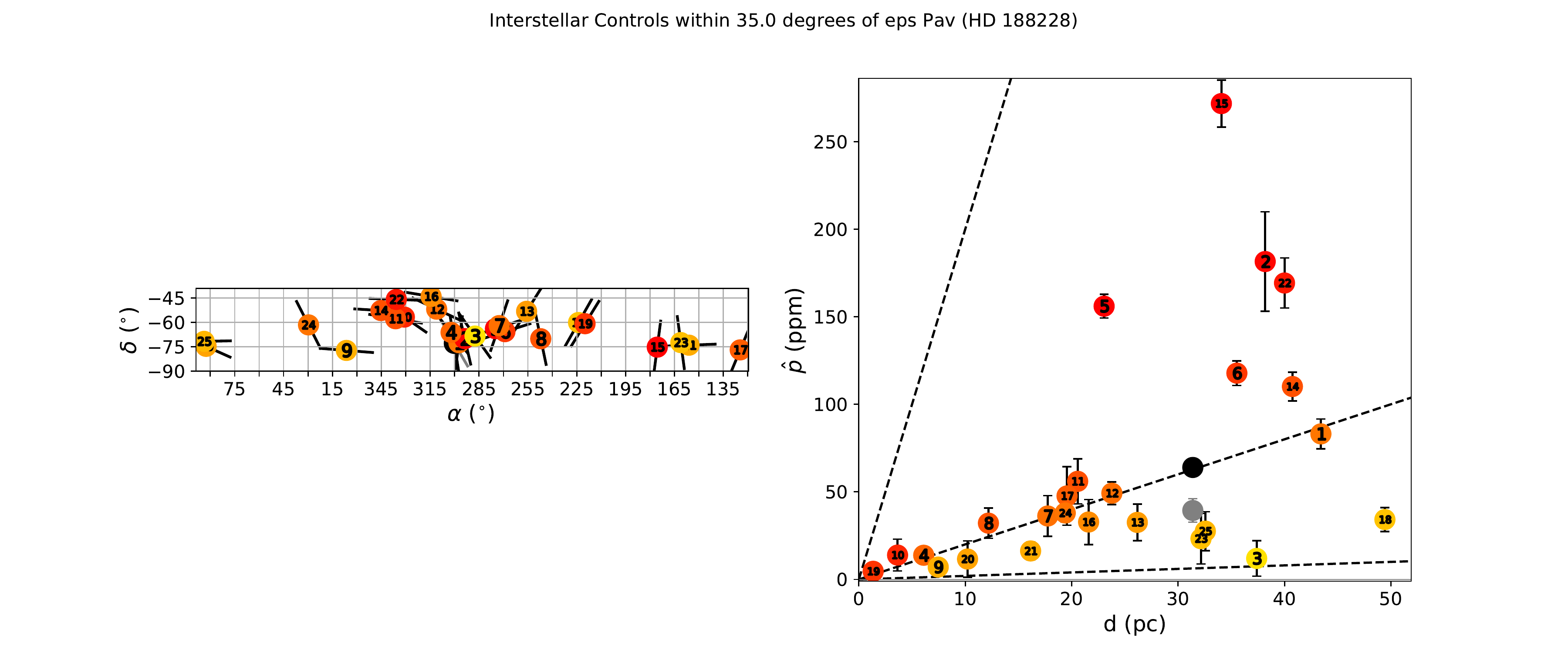}
    \caption{As per figure \ref{fig:ism1}, for targets HD~161868 (top), HD~181327 (middle), and HD~188228 (bottom). \label{fig:ism3}}
\end{figure*}

\begin{figure*}
    \includegraphics[width=\textwidth]{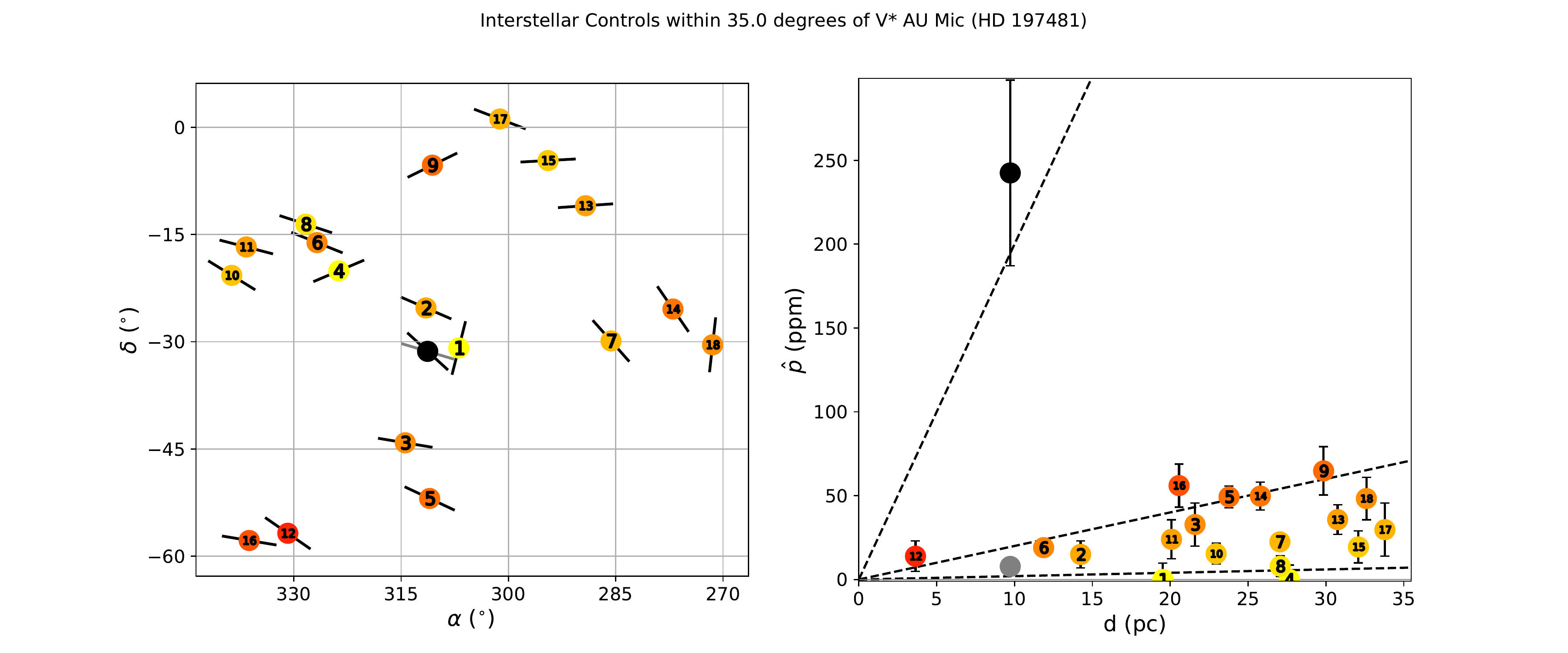}
    \includegraphics[width=\textwidth]{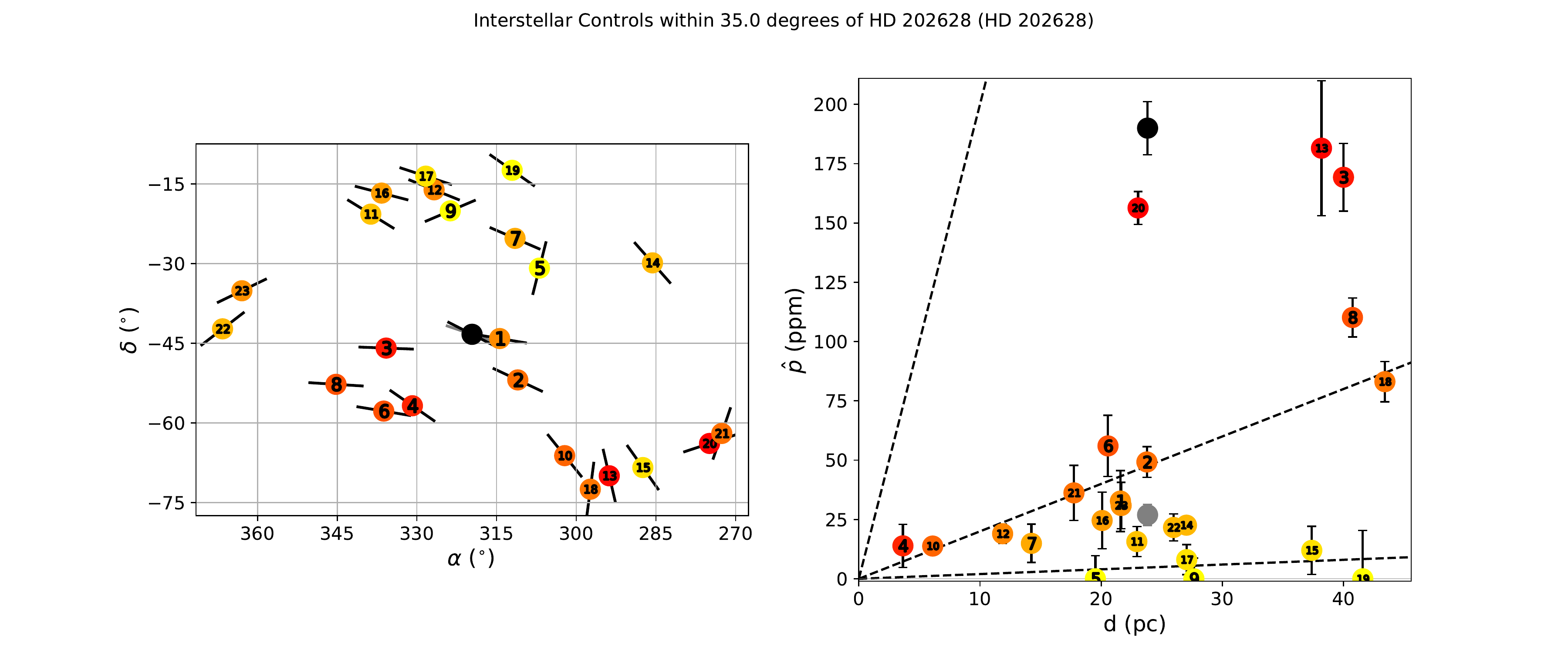}
    \includegraphics[width=\textwidth]{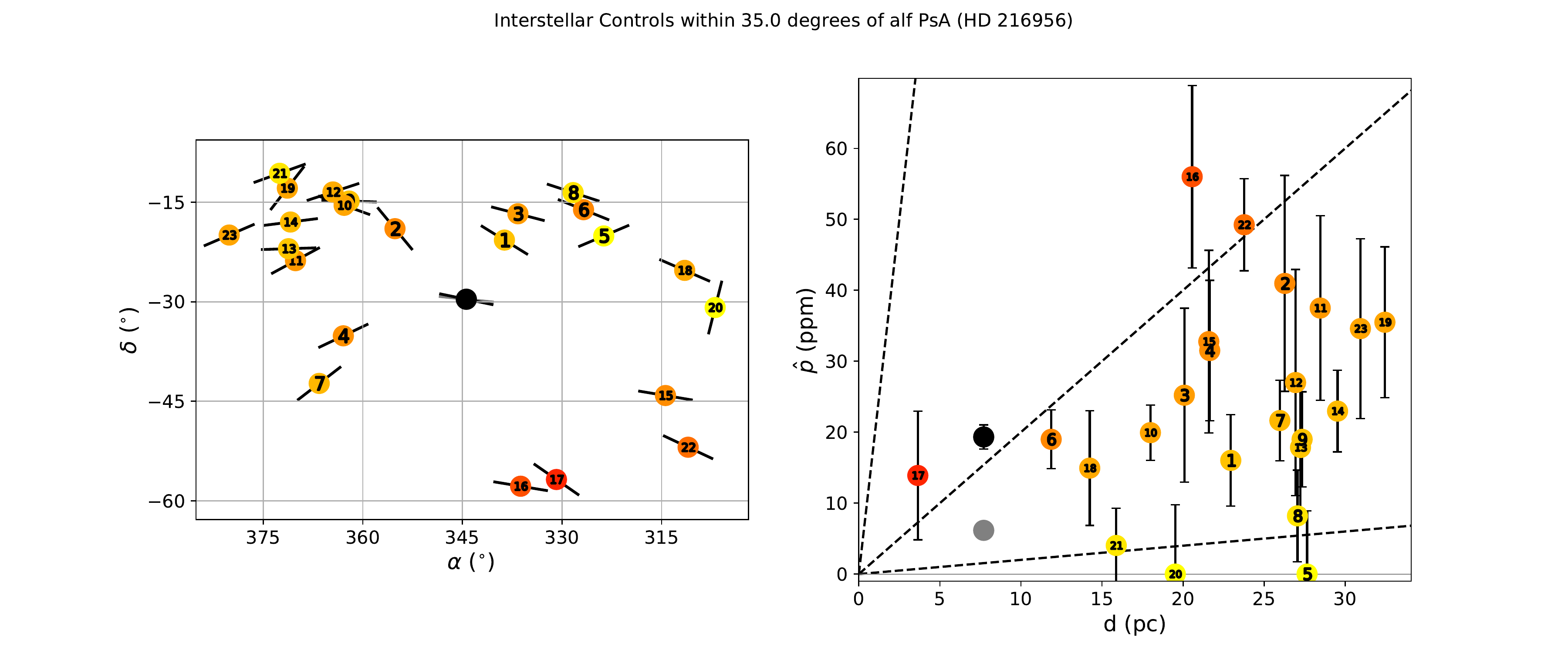}
    \caption{As per figure \ref{fig:ism1}, for targets HD~197481 (top), HD~202628 (middle), and HD~216956 (bottom). \label{fig:ism4}}
\end{figure*}

Our ambition with this work was to examine how polarisation changes with wavelength in debris disc systems; this has turned out to be more difficult than first envisioned. When the program was initiated, our expectation was the polarisation from the disc systems might be as much as half of the reflected light signal (c.f. table \ref{tab:sample}); it is evident from the data presented in table \ref{tab:observations1}, that for most systems, disc polarisation is much less than this. It has also become apparent that intrinsic stellar polarisation might be a significant obfuscating factor (e.g. \citealp{2017Cotton, 2019Cotton}), contrary to expectations \citep{1982Tinbergen, 1993Leroy}. Furthermore, at the time of our first observations, the best data on interstellar polarisation close to the Sun came from the survey conducted in red light with PlanetPol in the Northern hemisphere ($\sim0.2$~ppm/pc, \citealp{2010Bailey}). Since then, it has become clear that interstellar polarisation is 10 times higher per distance in the Southern hemisphere \citep{2016Cotton, 2017Cotton}, and that within the Local Hot Bubble (LHB) it is greater at blue wavelengths \citep{2019Cotton}. The nearby ISM also turns out to be rather heterogeneous, making calibration with control stars difficult. So in many of the systems we have observed, the interstellar polarisation is large compared to the debris disc polarisation, and was not well defined. To combat this difficulty we observed additional control stars (table \ref{tab:controls}), and \citet{2020Piirola} have recently added much more data on the ISM, but the reality is that determining debris disc polarisation with aperture polarimetry is not straight forward. Before we can hope to characterise the polarimetric properties of the debris discs, we must first determine in which systems it is significant and if it can be separated from the other polarisation sources. \citet{2015GarciaGomez} sought to do this by identifying disc systems with larger polarisations than non-disc systems, and then attempting to fit a Serkowski Law to those systems so identified; those poorly fit they regarded as candidates for significant disc-induced polarisation. On the other hand \citet{2019Vanderportal} considered trends in interstellar polarisation with distance. As we describe below, we have refined and built on these approaches.

In figures \ref{fig:ism1} to \ref{fig:ism4} we present maps and plots comparing the polarisation of nearby stars to each of the debris disc systems. The data for these maps comes from the target observations presented in table \ref{tab:observations1}, and interstellar controls presented in table \ref{tab:controls} as well as the literature as described in Section \ref{sec:interstellar_controls} (a full list of the controls for each map is given in Appendix \ref{apx:controls}). In these figures, the target position and polarisation position angle are given by the black data point, the mean ISM polarisation is given by the grey data point, and the individual ISM control orientations are associated with the numbered data points shaded in yellow to red. The position angle of the polarisation in the left hand panels is presented as a rotation of the vector clockwise from north. This is consistent with the presentation of RA increasing to the left in the same panels, as we show the polarisation vector projected onto the sky. The magnitude of the polarisation is shown in the right hand panels, with the same definitions for the data points. The dashed lines in the right hand panel show the magnitude of polarisation for the Northern hemisphere within the LHB (0.2 ppm pc$^{-1}$), Southern hemisphere within the LHB (2.0 ppm pc$^{-1}$), and bulk Milky Way ISM (20 ppm pc$^{-1}$). 

Each map gives an indication of how much polarisation might be attributed to the ISM but alone these are inadequate, owing to the sometimes high degree of scatter and number of outliers. For instance while the interstellar is fairly homogeneous aligned and smoothly increases with distance near to HD~115892 (figure \ref{fig:ism2}, bottom), there is significant scatter in the magnitude of interstellar polarisation near to HD~188228 (figure \ref{fig:ism3}, bottom), and likewise position angle scatter in the wider region around HD~377 (figure \ref{fig:ism1}, top). To more definitively assess the likely contribution interstellar polarisation, we next carry out some statistical tests.

\begin{figure*}
    \includegraphics[width=\textwidth]{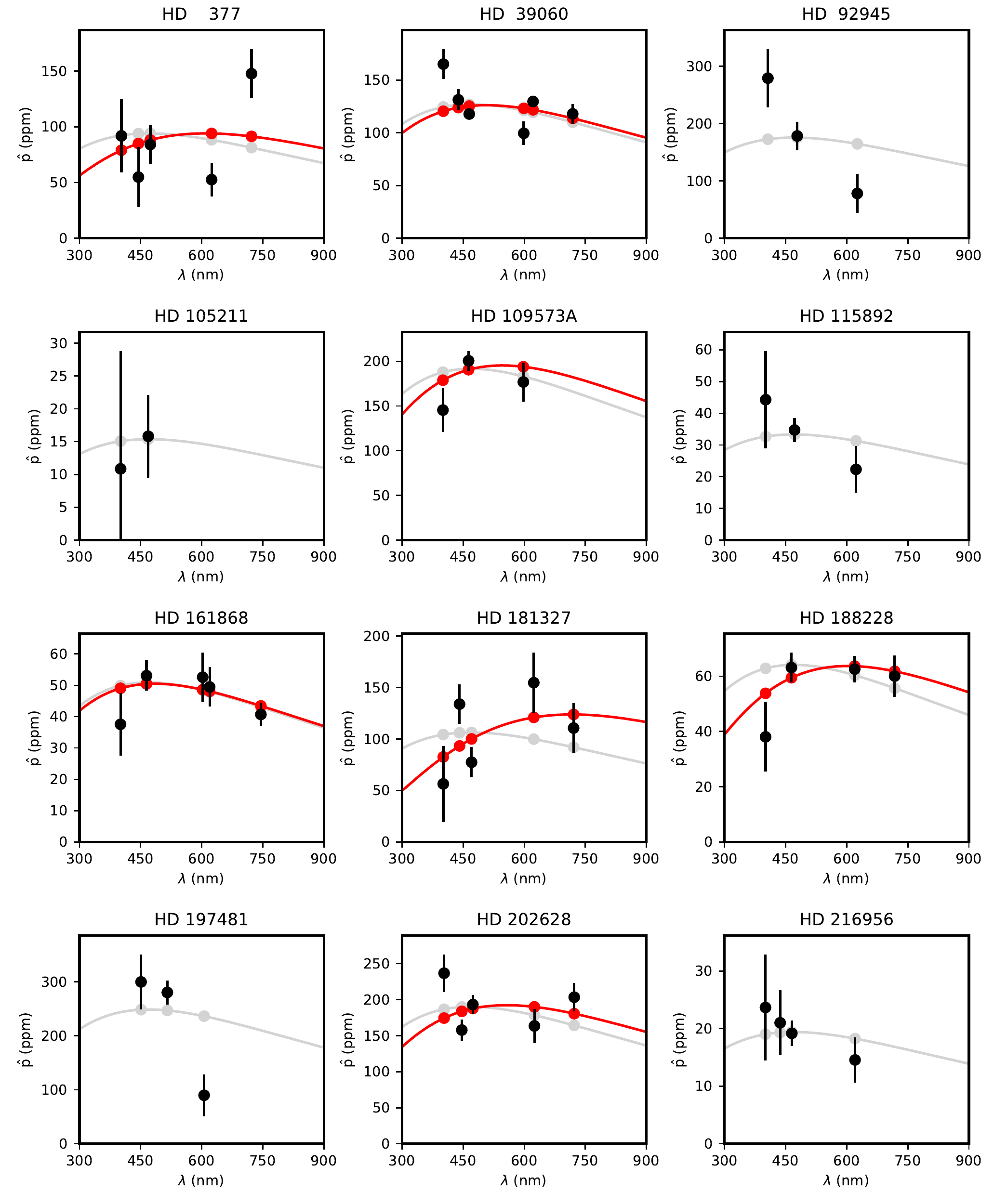}
    \caption{Serkowski fits to the polarisation of each disc system. Band-averaged data (black points with error bars) are plotted alongside 2-parameter (red) and 1-parameter ($\lambda_{\rm max}$ fixed at 470~nm, grey) Serkowski curves (lines) and band predictions (points). Note that the fits are made to the individual data points rather than the band averages. 2-parameter fits are shown only for systems where a fitted $\lambda_{\rm max}$ was found between 200 and 1200~nm -- values consistent with interstellar polarisation. The 1-parameter curves correspond to the expectation for interstellar polarisation close to the Sun.}
    \label{fig:serk}
\end{figure*}

\begin{figure*}
    \includegraphics[width=\textwidth]{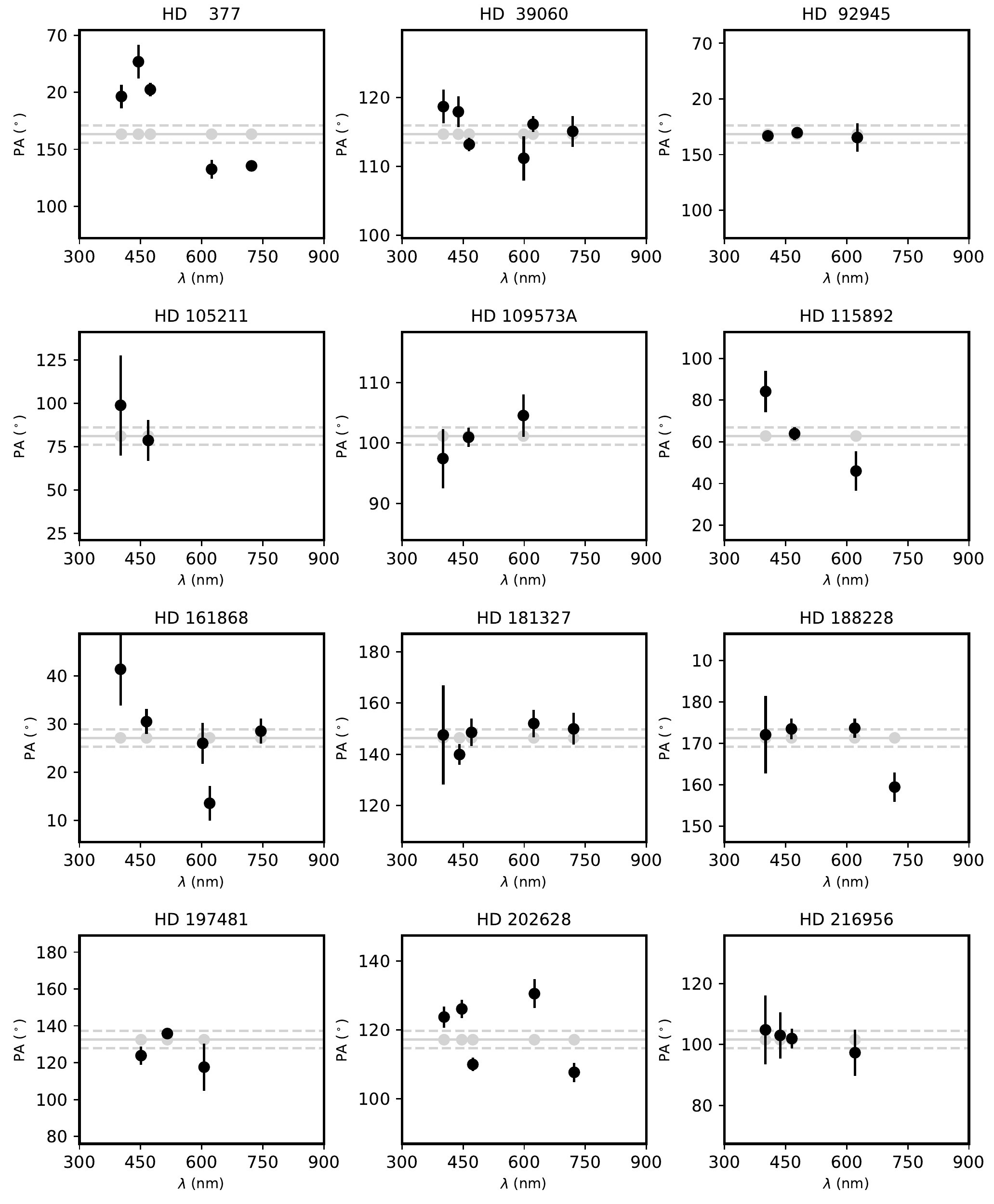}
    \caption{Band-averaged polarisation position angle, PA, data. The data are shown as black data points with nominal error bars. The solid grey lines show the weighted-average position angle, with the solid grey points corresponding to the band effective wavelengths; the dashed grey lines correspond to the uncertainty from fitting the individual observations to those position angles. If the system polarisation is dominated by either disc or interstellar polarisation alone, the PA is expected to be constant with wavelength.}
    \label{fig:PA}
\end{figure*}

\begin{table*}
	\centering
	\caption{Serkowski Fits}
	\label{tab:2p1p_Serk}
	\tabcolsep 4 pt
	\begin{tabular}{rrcccrrlccrlccc} 
		\hline
		        &&   \multicolumn{6}{c}{2-parameter Serkowski Fit}                              &   \multicolumn{4}{c}{1-parameter Serkowski Fit}   &   \multicolumn{1}{c}{ISM Pred.} & $\nicefrac{1}{2} \times \nicefrac{L_{\rm disc}}{L_{\odot}}$ & Act. Pred.\\
		HD\phantom{00}    & $n_o$     &   $p_{\rm max}$     &   $\lambda_{\rm max}$&  $K$ & $\chi^2_r$ & \multicolumn{2}{c}{$H_0$} &   $p_{\rm max}$        & $\chi^2_r$   &   \multicolumn{2}{c}{$H_0$}    & $p_{i}$ & & $4\times|B_\ell|_{\rm max}$\\
		        &                       &   (ppm)             &   (nm)               &          &         & \multicolumn{2}{c}{(S/R)}  &   (ppm)             &             &   \multicolumn{2}{c}{(S/R)}     & (ppm)    & (ppm) &   (ppm)\\
		\hline
		377\phantom{A}    & 14  &  \094.2 $\pm$ 17.8  &   609.7 $\pm$ 255.8  &   1.02   &   2.66  & R   & >99$\%$  &  \094.1 $\pm$ 15.3  & \02.49  & R    &  >99$\%$  &   \050 $\pm$ \09  &   \0100\phantom{.0}   & \0\020\\ 
		39060\phantom{A}  & 10  &   126.1 $\pm$ \07.6 &   507.0 $\pm$ 119.8  &   0.85   &   6.96  & R   & >99$\%$  &   126.9 $\pm$ \07.2 & \06.32  & R    &  >99$\%$  &   \020 $\pm$ \03  &    1215\phantom{.0}   &\\
		92945\phantom{A}  &  5  &                     &                      &          &         &     &          &   175.8 $\pm$ 66.0  &  12.08  & R    &  >99$\%$  &   \023 $\pm$ \04  &   \0380\phantom{.0}   & $\sim$250\\
		105211\phantom{A} &  2  &                     &                      &          &         &     &          &  \015.4 $\pm$ \01.5 & \00.06  & S    &           &   \020 $\pm$ \03  &  \0\027\phantom{.0}   &\\
		109573A           &  3  &   195.3 $\pm$ 27.6  &   546.1 $\pm$ 219.0  &   0.92   &   3.43  & S   &          &   191.6 $\pm$ 12.9  & \01.92  & S    &           &    107 $\pm$ 19   &    2350\phantom{.0}   &\\
		115892\phantom{A} &  3  &                     &                      &          &         &     &          &  \033.3 $\pm$ \03.6 & \01.11  & S    &           &   \017 $\pm$ \03  &  \0\0\00.4            & *\\
		161868\phantom{A} & 12  & \050.5 $\pm$ \03.2  &   484.7 $\pm$ \081.7 &   0.81   &   0.83  & S   &          &  \050.9 $\pm$ \02.4 & \00.77  & S    &           &   \040 $\pm$ \07  &  \0\050\phantom{.0}   &\\
		181327\phantom{A} &  5  &  123.9 $\pm$ 35.7   &   717.4 $\pm$ 199.0  &   1.20   &   2.91  & R   & >95$\%$  &   106.4 $\pm$ 16.8  & \02.92  & R    &  >95$\%$  &   \073 $\pm$ 13   &   1465\phantom{.0}    &\\
		188228\phantom{A} &  6  & \063.7 $\pm$ \05.1  &   603.1 $\pm$ 116.3  &   1.01   &   2.03  & S   &          &  \064.1 $\pm$ \04.8 & \02.08  & S    &           &   \039 $\pm$ \07  & \0\0\02.5             &\\
		197481\phantom{A} &  3  &                     &                      &          &         &     &          &   248.7 $\pm$ 55.4  & \08.95  & R    &  >99$\%$  &  \0\08 $\pm$ \01  &   \0195\phantom{.0}   & \0289\\
		202628\phantom{A} &  8  &  192.4 $\pm$ 13.9   &   557.4 $\pm$ 103.7  &   0.94   &   2.05  & S   &          &   190.3 $\pm$ 11.2  & \01.90  & S    &           &   \027 $\pm$ \04  &  \0\070\phantom{.0}   & \phantom{-}$\sim$12\\
		216956\phantom{A} &  8  &                     &                      &          &         &     &          &  \019.4 $\pm$ \01.7 & \00.94  & S    &           &  \0\06 $\pm$ \01  &  \0\040\phantom{.0}   &\\
	\hline
	\end{tabular}
	\begin{flushleft}
	\textbf{Notes:} Shown are the best fit values to a Serkowski curve for each system. Fits, made to the $n_o$ observations of each system, are either 2-parameter: $p_{\rm max}$ and $\lambda_{\rm max}$, with $K$ calculated using equation \ref{eq:Whit}, or 1-parameter with $\lambda_{\rm max} =$ 470~nm and $K =$ 0.79 -- consistent with our expectations of the interstellar medium close to the Sun. 2-parameter fits are shown only for systems where a fitted $\lambda_{\rm max}$ was found between 200 and 1200~nm -- values consistent with interstellar polarisation. Using Pearson's $\chi^2$-test, we then test the null hypotheses, $H_0$, that the data are described by the given interstellar polarisation function, which we either sustain (S) or reject (R) at either 95$\%$ or 99$\%$ confidence level. Also tabulated, for easy comparison with $p_{\rm max}$, is the predicted interstellar polarisation from \citet{2017Cotton}, half the fractional reflected light signal from table \ref{tab:sample} and quadruple the magnetic activity indicator $|B_\ell|_{\rm max}$ (also from from table~\ref{tab:sample}) -- a prediction for the activity induced polarisation from \citet{2019Cotton}. * Activity likely \citep{2006Hubrig}, but not significantly measured, so difficult to estimate.\\
	\end{flushleft}	
\end{table*}

\begin{table}
	\centering
	\caption{PA Fits and Comparisons}
	\label{tab:PA}
	\tabcolsep 2.5 pt
	\begin{tabular}{rccrrlcc} 
		\hline
		        &                   &   \multicolumn{4}{c}{Observed}                                        &   \multicolumn{1}{c}{ISM Pred.}   &   \multicolumn{1}{c}{Disc}  \\
		HD\phantom{00}    & $n_b$   &   PA          & $\chi^2_r$    &   \multicolumn{2}{c}{$H_0$}           &   PA$_i$                          &   PA$_\perp$, PA$_\parallel$                              \\
		                  &         & ($\degr$)     &               &   \multicolumn{2}{c}{(S/R)}   &   ($\degr$)                       &   ($\degr$)                       \\
		\hline
		377\phantom{A}    &  5  &  163.3 $\pm$ 7.6  &    31.19      &   R       &   >99$\%$                 &  114   &  137,  \047 $\pm$ \04  \\ 
		39060\phantom{A}  &  6  &  114.7 $\pm$ 1.3  &     1.44      &   S       &                           & \029   &  120,  \030 $\pm$ \01  \\
		92945\phantom{A}  &  3  &  168.4 $\pm$ 7.9  &     0.12      &   S       &                           & \070   & \010,  100 $\pm$ \01  \\
		105211\phantom{A} &  2  & \081.1 $\pm$ 5.0  &     0.42      &   S       &                           & \056   &  122,  \032 $\pm$ \02  \\
		109573A           &  3  &  101.1 $\pm$ 1.4  &     0.71      &   S       &                           & \070   &  116,  \026 $\pm$ \01  \\
		115892\phantom{A} &  3  & \062.8 $\pm$ 4.2  &     3.89      &   R       &   >95$\%$                 & \067   &\\
		161868\phantom{A} &  5  & \027.1 $\pm$ 1.8  &     4.66      &   R       &   >99$\%$                 & \033   & \028,  118 $\pm$ \03  \\
		181327\phantom{A} &  5  &  146.4 $\pm$ 3.4  &     0.99      &   S       &                           &  154   & \017,  107 $\pm$ \02  \\
		188228\phantom{A} &  4  &  171.3 $\pm$ 2.1  &     3.93      &   R       &   >99$\%$                 &  147   &  101, \011 $\pm$ 15   \\
		197481\phantom{A} &  3  &  132.5 $\pm$ 4.7  &     3.08      &   R       &   >95$\%$                 &  110   & \039,  129 $\pm$ \01  \\
		202628\phantom{A} &  5  &  117.1 $\pm$ 2.5  &    11.14      &   R       &   >99$\%$                 &  110   & \044,  134 $\pm$ \02  \\
		216956\phantom{A} &  4  &  101.6 $\pm$ 2.8  &     0.15      &   S       &                           & \096   & \066,  156 $\pm$ \01  \\
	\hline
	\end{tabular}
	\begin{flushleft}
	\textbf{Notes:} The null hypotheses, $H_0$, is that the observed mean position angle, PA, is constant with wavelength: this is assessed by applying Pearson's $\chi^2$ test to the $n_b$ band-averaged observations. The null hypothesis is then either sustained (S), or rejected (R) at the $>95\%$ or $>99\%$ confidence level.  For the purposes of the test, we have added an additional 1$^\circ$ RMS uncertainty to each band measurement to account for PA calibration uncertainty not otherwise included in the nominal error. The observed PA can be compared with the prediction made for pure interstellar polarisation, as suggested by measurements of nearby stars, and the polarisation angle expected from a disc -- that perpendicular to the disc major axis. In \citet{2017Cotton} we found the error in our ISM PA prediction method to typically be 29$^\circ$, but that this could be improved with more control stars -- as we have in many cases here -- and might be as good as 10-15$^\circ$ with sufficient close controls (i.e. within 10$^\circ$). \\
	\end{flushleft}	
\end{table}

\subsection{A Test of Serkowski-Wilking Behaviour} \label{sec:Serk} We use a Pearson's $\chi^2$ test \citep{1900Pearson, 1983Plackett}, in which the null hypothesis, $H_0$, is that the wavelength dependence of the data is described by the empirically determined \textit{Serkowski-Wilking Law} for interstellar polarisation. 

Interstellar polarisation has a characteristic wavelength dependence given by the empirically determined by \citet{1971Serkowski, 1973Serkowski, 1975Serkowski} as: \begin{equation}\label{eq:Serk} \frac{p(\lambda)}{p_{\rm max}}=exp\left[-Kln^2\left (\frac{\lambda_{\rm max}}{\lambda} \right )  \right ] ,\end{equation} known as the \textit{Serkowski Law}, where $p(\lambda)$ is the polarisation at wavelength $\lambda$, $p_{\rm max}$ is the maximum polarisation occurring at wavelength $\lambda_{\rm max}$. The dimensionless constant $K$ describes the inverse width of the polarisation curve peaked around $\lambda_{\rm max}$; \citet{1975Serkowski} gave its value as 1.15. \citet{1980wilking} later described $K$ in terms of a linear function of $\lambda_{\rm max}$. Using this form, \citet{1992Whittet} found $K$ to be: \begin{equation}\label{eq:Whit} K=(0.01\pm0.05)+(1.66\pm0.09)\lambda_{\rm max},\end{equation} (where $\lambda_{\rm max}$ is given in $\mu$m) -- the form of equation \ref{eq:Serk} that uses the relation of equation \ref{eq:Whit} is referred to as the Serkowski-Wilking Law.

It should be remarked that polarisation, $p$, is a positive definite quantity, and as such not strictly Gaussian. For $S/N>4$ the difference is insignificant, but at lower $S/N$ a correction is required to debias the data. For this purpose we used the method of \citet{1974Wardle}, which is recommended for $S/N\geq0.7$ \citep{1985Simmons} \begin{equation}\label{eq:debias}\hat{p}=\begin{cases}\sqrt{p^2-\sigma_p^2} & \text{ for } p>\sigma_p \\  0 & \text{ for } p<\sigma_p\end{cases}.\end{equation} where $\sigma_p$ is the error in $p$. 

We carry out two forms of this test, in which each data set is fit by equations \ref{eq:Serk} and \ref{eq:Whit} with either one or two free parameters. In the two parameter form we fit for $p_{\rm max}$ and $\lambda_{\rm max}$, in the one parameter form we fix $\lambda_{\rm max} =$ 470~nm. This is the value deemed most likely by the first small study to look at the polarimetric colour of the ISM close to the Sun \citep{2016Marshall}; similar values have since been found for a small number of other nearby stars \citep{2019Cotton, 2020Bailey, 2020Marshall}. A redder value of $\lambda_{\rm max} =$ 550~nm has been found for stars in the wall of the Local Hot Bubble ($\approx$75--150~pc from the Sun, \citealp{2019Cotton}); this is a typical value for the Galaxy \citep{1975Serkowski}, but a broad range of values have been found for specific regions of space. The 2-parameter version of our test is intended to be agnostic about this value, but we limit ourselves to presenting curves with realistic values: 200~nm $\leq$ $\lambda_{\rm max}$ $\leq$ 1200~nm.

\begin{description}
    \item[\textbf{Test Outcomes}]
    \item[\textit{Reject:}] If in both cases $H_0$ is rejected, then we can conclude that the polarisation for the tested system is not purely interstellar. Here we fit each observation, and not just the band-average polarisation, so $H_0$ might be rejected on account of variability beyond that expected from the nominal errors if multiple observations were made in the same band -- this might result from stellar activity \citep{2017Cotton, 2019Cotton}.
    \item[\textit{Sustain:}] If $H_0$ is sustained, it could mean that interstellar polarisation is responsible for the polarisation wavelength dependence of the data, that the errors are too large to rule that out, or that whatever intrinsic polarisation is present mimics the Serkowski-Wilking Law sufficiently. Scattering from debris disc dust with the right properties could fall into this later category.
\end{description}

Another outcome of the Serkowski-Wilking curve fitting is that we get a value for $p_{\rm max}$ that can be more robustly compared with our expectations for the interstellar medium than might a single observation in the $g^\prime$ band.

The Serkowski-Wilking fits are presented in figure \ref{fig:serk}, the corresponding fit parameters along with the results of the Pearson's $\chi^2$ test of Serkowski-Wilking behaviour are shown in table \ref{tab:2p1p_Serk}.

\subsection{PA Constancy Test} \label{sec:PA} We use a Pearson's $\chi^2$ test, in which the null hypothesis, $H_0$, is that polarisation position angle, PA, is constant with wavelength.

To conduct this test, we first fit the $q$ and $u$ values with the 1-parameter Serkowski-Wilking determined parameters, with PA as a free parameter -- this gives the same answer but a better determination of the error in the mean position angle than does error-weighting each observation. To calculate $\chi^2$ we then compare this PA value to the band-average observed PAs. We have used band-averaged values for this purpose because the position angle distribution function is not Gaussian at low $S/N$ -- exhibiting kurtosis in the wings -- but approaches and essentially becomes Gaussian at $p/\sigma_p>6$ \citep{1993Naghizadeh-Khouei}, and band averaging brings most of our data into this regime. The PA errors reported in tables \ref{tab:observations1}--\ref{tab:controls} and presented in figure \ref{fig:PA} are derived from Fig.~2 of \citet{1993Naghizadeh-Khouei}, as is our usual practice. However, for the purposes of conducting the the Pearson's $\chi^2$ test, we cautiously add an additional 1$^\circ$ RMS error per band in recognition of both the small deviations from Gaussian behaviour and the uncertainties in PA calibration. The calibration uncertainties, usually neglected, come both from the uncertainties in the literature PA values of the standards each being about 1$^\circ$ and these varying slightly as a function of wavelength.

\begin{description}
    \item[\textbf{Test Outcomes}]
    \item[\textit{Sustain:}] If $H_0$ is sustained then either a single polarisation mechanism is dominant, or multiple mechanisms are acting along the same axis or acting with equal strengths at all wavelengths. The possibilities for a single mechanism include interstellar polarisation and polarisation due to scattering from the disc.
    \item[\textit{Reject:}] If $H_0$ is rejected, then most likely there are two competing polarisation mechanisms with different characteristic PAs that vary in strength with wavelength -- in our case this suggests interstellar and disc polarisation of comparable strengths. Two differently aligned dust clouds producing interstellar polarisation with different $\lambda_{\rm max}$ is also conceivable, but unlikely for nearby objects. A second possibility is that a single non-variable polarisation mechanism including a sign-flip is responsible -- if so, this should be evident in a plot of PA vs. $\lambda$. Scattering from a debris disc is one mechanism that might display this behaviour (see for e.g. the modelling in \citealp{2020Marshall}). Another mechanism that displays such behaviour is rapid rotation \citep{2017bCotton}. Finally, it is possible that a variable polarisation mechanism, like activity, sampled at different times with different wavelengths is responsible.  
\end{description}

If $H_0$ is sustained then we can compare the observed PA to our expectations for the ISM and the disc, to determine which is more likely to be dominant (along with considerations in Section \ref{sec:Serk}). For a symmetric, edge-on disc the polarisation PA will either be parallel or perpendicular to the disc major-axis, with the latter being more likely as scattering from the disc ansae dominates the polarisation signal. However, if the edge of the aperture cuts through the disc, this will reverse the sign of the polarisation. 

\subsection{System Assessments}
\label{sec:assessments}

In this section we primarily use the tests described in Sections \ref{sec:Serk} and \ref{sec:PA} along with comparisons to known disc and predicted ISM PAs, and to predicted ISM $p_{\rm max}$ to make an assessment of each system's polarisation as either disc dominated, interstellar dominated or having multiple (comparable) contributions from each. However, the assessments are often complicated by uncertainties, especially in the properties of the ISM, but also the potential for contamination from other intrinsic polarigenics. We mitigate these difficulties by also considering control star maps, along with what is naively implied by disc fractional luminosity, $L_{\rm dust}/L_{\star}$, and stellar activity indices (mainly $|B_\ell|_{\rm max}$). The assessments unavoidably have a degree of subjectivity, and so we give our detailed reasoning below. We develop further the cases where we are confident of a disc detection robustly separable from interstellar and other polarigenics later in Section \ref{sec:subtractions}.

\subsubsection{HD~377: disc dominated}

$H_0$ of both forms of the Serkowski-Wilking (S-W) Test are rejected at the 99\% level; it is clear from figure \ref{fig:serk} that the polarisation cannot be interstellar. The PA Constancy Test null hypothesis is also rejected at the 99\% level; figure \ref{fig:PA} shows a large change in PA with wavelength, at the extremes $\Delta$PA approaches 90$^\circ$, which leads us to suspect a significant disc contribution where the sign of the disc polarisation flips. This hypothesis is strengthened by the redder bands corresponding closely with perpendicular to the disc PA, and the bluer bands approaching a PA parallel with the disc, while the interstellar PA is intermediate. There is some magnetic activity in this system, but when taking account of the likely interstellar contribution, not enough to explain the PA behaviour seen. There is still likely to be a significant interstellar contribution to the polarisation, but we conclude that the polarisation from the disc dominates.

\subsubsection{HD~39060: disc dominated}

$H_0$ of both forms of the S-W Test are rejected at the 99\% level, and the polarisation is much greater than expected for the ISM, so we conclude that the system is intrinsically polarised. The disc has a large fractional infrared excess, making the disc the primary candidate for the intrinsic polarisation. $H_0$ for the PA Constancy test is sustained -- indicating one polarisation mechanism dominates -- with the observed PA a good match for perpendicular to the disc. We therefore conclude that the signal from the disc dominates the observed polarisation.  

\subsubsection{HD~92945: possible activity contamination}

$H_0$ is sustained for the PA Constancy Test, but rejected at the 99\% level for the S-W Test: this suggests disc dominated polarisation. This hypothesis is supported by the observed PA aligning perpendicular to the disc, and the observed polarisation being much higher than expected for the ISM. In the absence of any other potential sources of intrinsic polarisation, this would be enough for us to conclude the polarisation is disc dominated. 

Fig. \ref{fig:serk} shows polarisation that increases from red to blue, this might be reflective of the disc properties, but it is also what is expected from stellar activity induced polarisation \citep{1993Saar}. As discussed in Section \ref{sec:sample}, the star is known to be active, and we expect the associated polarisation to be $\simeq$100~ppm or more, making inferences about the disc polarisation speculative, at best. 

\subsubsection{HD~105211: interstellar dominated}

We only have two bands for HD~105211, but $H_0$ is sustained for both the PA Constancy Test and the S-W Test, which indicates interstellar dominated polarisation. the polarisation is small, and is in line with what is predicted for the ISM -- at 20~pc distance, the prediction is likely to be accurate. The observed PA is not a particularly good match for either the disc or the ISM but the ISM PA determination is not well constrained (figure \ref{fig:ism2}, top). Consequently, we conclude that the polarisation in this system is interstellar dominated.

\subsubsection{HD~109573A: interstellar dominated}

HD~109573 is the most distant star in our survey, and so might be expected to have the largest interstellar polarisation component. Our tests are consistent with a dominant interstellar polarisation: $H_0$ of the S-W Test and PA Constancy Test are both sustained. The 2-parameter Serkowski-Wilking fit gives $\lambda_{\rm max}$ consistent with being in the wall of the LHB (albeit with a large uncertainty) as is reasonable for a star at this distance. The interstellar polarisation prediction is not accurate at larger distances, and we have few distant control stars. It is notable that the most distant control star, HD~125473 (Star 14 in figure \ref{fig:ism2}, middle), has a PA that matches HD~109573A's, even though the nearer stars don't match. Despite its distance, with the very large $L_{\rm dust}/L_{\star}$ for this system, we might have expected to detect the disc, and there is a hint of wavelength dependence in figure \ref{fig:PA}, but it is not statistically significant. So, we conclude that the polarisation measured is interstellar dominated.

\subsubsection{HD~115892: possible activity contamination}

Although a trend is observed in figure \ref{fig:serk} whereby polarisation decreases with wavelength, this is not sufficiently significant to reject $H_0$ in the 1-parameter S-W Test. The obtained $p_{\rm max}$ value is about double the predicted value for the ISM (33 vs. 17 ppm), but is a very good match for the $\hat{p}/d$ values of the two nearest control stars shown in figure \ref{fig:ism2}. $H_0$ is rejected at the 95\% level in the PA Constancy Test. Yet, the observed PA matches the interstellar prediction well. Unfortunately we can not compare with the disc PA, because that information is not available. Nevertheless, the test results suggest similar interstellar and disc contributions, with the disc axis offset from the ISM. This is surprising because $L_{\rm dust}/L_{\star}$ is very small. However, unlike most of the systems we have observed, the HD~115892 disc is an asteroid belt analogue, so the measured fractional luminosity might not be a consistent diagnostic. The most probably explanation for the conflicting results though, is magnetic activity on the basis of the \citet{2006Hubrig} measurement.

\subsubsection{HD~161868: interstellar dominated}

$H_0$ is sustained in both S-W Tests, and the value of $\lambda_{\rm max}$ obtained from the 2-parameter version is a good match for our expectations of the colour of the ISM. The $p_{\rm max}$ value obtained is also a good match for the ISM prediction; on this basis dominant interstellar polarisation seems likely. However, the PA obtained is both perpendicular to the disc and a good match both for the ISM prediction, and the PA constancy test is rejected at the 99\% confidence level. These latter two facts are at odds, for there to be a position angle wavelength dependence, the two polarisation mechanisms cannot be aligned. On further inspection of figure \ref{fig:PA} we see that the PA Constancy test is being failed largely on account of a single measurement -- the $r^\prime$\,(R) band observation disagrees with both the $r^\prime$\,(B) and BRB observations that overlap it. We therefore conclude that observation is probably spurious and that this system is dominated by interstellar polarisation.

\subsubsection{HD~181327: interstellar dominated}

$H_0$ is rejected for both S-W Tests -- the polarisation is clearly stronger in the red, and the 2-parameter Serkowski-Wilking fit has the better $\chi^2_r$ but is still rejected at the 95\% confidence level. $H_0$ is, however, sustained in the PA Constancy Test. Together these two results suggest a dominating intrinsic polarisation. Yet, it does not seem likely this can be ascribed to the disc because the observed PA lies between the angles parallel and perpendicular to the disc. The PA is a good match for the ISM prediction but, as can be seen in figure \ref{fig:ism3}, the properties of the interstellar medium are very poorly constrained, especially so beyond 50~pc. A number of nearby stars have very high $\hat{p}/d$ values, but none are as close as 5$^\circ$. Multiple dust clouds with differing grain size characteristics, but similar PA could explain the conflicting results, however the polarisation with wavelength behaviour in figure \ref{fig:serk} is not well matched to this scenario. A more mundane possibility is that the data is noisier than accounted for by the nominal errors, and that since the 2-parameter S-W Test $H_0$ is only marginally rejected, the polarisation is interstellar dominated. We favour this latter conclusion, but have low confidence in it.

\subsubsection{HD~188228: multiple contributions}

$H_0$ is sustained for both S-W Tests. The polarisation is clearly stronger in the red, and the 2-parameter Serkowski-Wilking fit has the better $\chi^2_r$; this is unexpected ISM behaviour given the distance of the system at only 32~pc. The calculated $p_{\rm max}$ value is also higher than expected from the ISM, but there are more polarised control stars nearby. $H_0$ is rejected for the PA Constancy Test at the 99\% level. Figure\ref{fig:PA} shows this is entirely due to the 650LP band -- the two observations, which are in good agreement, were made on sequential nights during the 2017AUG run. This indicates multiple contributions to the observed polarisation. The observed best-fit PA is intermediate of the ISM prediction, and a PA parallel to the disc -- which is less likely for a disc wholly within the aperture. The 650LP PA is smaller, which would mean it had a lesser disc contribution, if it is the disc contributing to the system polarisation.

The very low disc fractional luminosity for this system makes a polarisation detection unlikely: the disc would have to be more reflective in optical light than it is in the infrared. 
Another possibility is multiple dust clouds along the line of sight with different $\lambda_{\rm max}$ and PA. The more polarised regions of the ISM further from the Sun have a redder $\lambda_{\rm max}$ than those closer \citep{2019Cotton}. The higher polarisation of some of the nearby control stars suggests this as a viable possibility. Ultimately, we can conclude there are multiple contributions, but cannot say whether these are most likely to be the disc and the ISM, or two regions of the ISM with different characteristics.

\subsubsection{HD~197481: possible activity contamination}

This system is clearly intrinsically polarised. $H_0$ is rejected at the 99\% level for the S-W Test and sustained for the PA Constancy Test. Further, the observed polarisation is larger in all three bands than expected from the ISM (which at such a close distance is well defined). The disc of HD~197481 does not lie wholly within the instrument aperture. As such, we expect a PA aligned parallel with the disc major axis, and this is the case within error. This indicates the polarisation is dominated by the disc contribution. The polarisation is larger than half $L_{\rm dust}/L_{\star}$, which calls for further examination.

The polarisation is largest at blue wavelengths. This is also what is expected from stellar activity \citep{1993Saar}. HD~197481 is a very active star. Naively using the result from \citet{2019Cotton} to predict the maximum linear polarisation from activity for the system using $|B_\ell|_{\rm max}$, as we do in table \ref{tab:2p1p_Serk}, suggests activity can account for the entire polarisation of the system. However, no detailed study of broadband linear polarisation has been carried out with an M-type star with a modern high precision polarimeter, and extrapolating from the earlier type stars with less line blanketing is speculative. Without monitoring and/or simultaneous spectropolarimetric observations we cannot gauge the level of contamination for stellar activity.

\subsubsection{HD~202628: multiple contributions}

$H_0$ is sustained for the S-W Test, but $H_0$ is rejected at the 99\% level for the PA Constancy Test. The observed PA is a good match for the ISM prediction, and inconsistent with the disc, which suggests mostly interstellar polarisation. However, the polarisation magnitude is very high for a star at 24~pc distance, suggesting multiple contributions. Despite the $H_0$ being sustained for the S-W Test, the data is not a good fit to either curve in figure \ref{fig:serk} ($H_0$ would be rejected at the 90\% level) -- which supports a multiple contributions assessment. $L_{\rm dust}/L_{\star}$ would seem to be on the low side for a large disc contribution, but HD~202628 has an eccentric disc \citep{2010Krist,2019Faramaz}, which can be expected to increase the observable polarisation through reduced cancellation.

\subsubsection{HD~216956: interstellar dominated}

$H_0$ is sustained for the S-W Test and for the PA Constancy Test. The measured PA is a fair match for the ISM prediction, and inconsistent with what would be expected from the disc. This is enough to conclude that the polarisation is interstellar dominated. There are hints of another trend with wavelength in both Figs. \ref{fig:serk} and \ref{fig:PA}, but neither is significant.

\subsection{Interstellar Subtractions}
\label{sec:subtractions}
Interstellar subtraction is possible where a clear detection of the disc has been made, and the PAs of the disc and ISM are different enough to make a determination of interstellar polarisation parameters. Our method for determining the interstellar parameters assumes that the interstellar PA does not vary with wavelength, that the disc polarisation is parallel or perpendicular to the disc major axis, and that there are no other intrinsic polarisation sources.

The core of our approach is to fit the interstellar PA and $p_{\rm max}$ (but fix $\lambda_{\rm max}=$ 470~nm) with a computer script that makes use of the {\sc Python} package {\sc Scipy}'s `curve\_fit' function \citep{2020Virtanen}\footnote{Because of later changes made to `curve\_fit', the program only works in {\sc Python 2} without modification}. The `fit function' first calculates the interstellar polarisation for the bandpass of the data point, using the Serkowski-Wilking Law, rotates it by $PA_i$ and subtracts it from the data; then the remainder is rotated by minus the disc PA as given in table \ref{tab:sample} to align the disc polarisation with $q^{\prime}$, and then returns the $u^{\prime}$ Stokes parameter, which the algorithm seeks to minimise.

However, without additional information, an infinite array of solutions are possible, as the algorithm can minimise $u^{\prime}$ for any PA$_i$. To tether the result to reality, we treat the ISM predictions for $p_i$ and PA$_i$ in Tables \ref{tab:2p1p_Serk} and \ref{tab:PA} respectively, as data points, which the algorithm simultaneously tries to match. For this purpose we take the error in PA$_i$ to be 29$^\circ$ -- which was the average deviation from the model to the actual measurement found in \citet{2017Cotton} -- although the fits turn out not to be sensitive to this value. The initial values are the predicted values from tables \ref{tab:2p1p_Serk} and \ref{fig:PA}.

\begin{figure*}
    \includegraphics[clip, trim={0.3cm, 0.2cm, 0.0cm, 0.0cm}, width=0.48\textwidth]{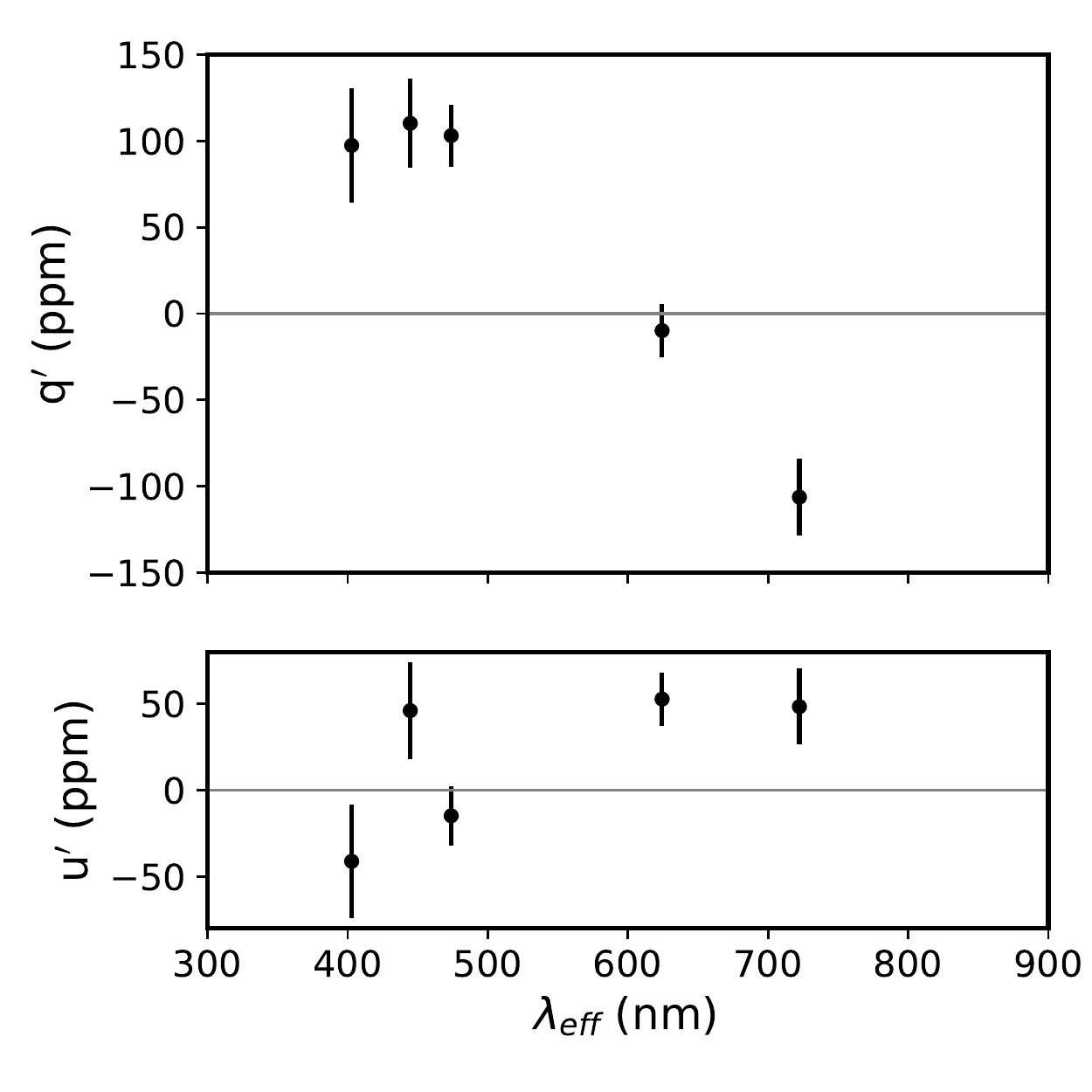}
    \includegraphics[clip, trim={0.3cm, 0.2cm, 0.0cm, 0.0cm}, width=0.48\textwidth]{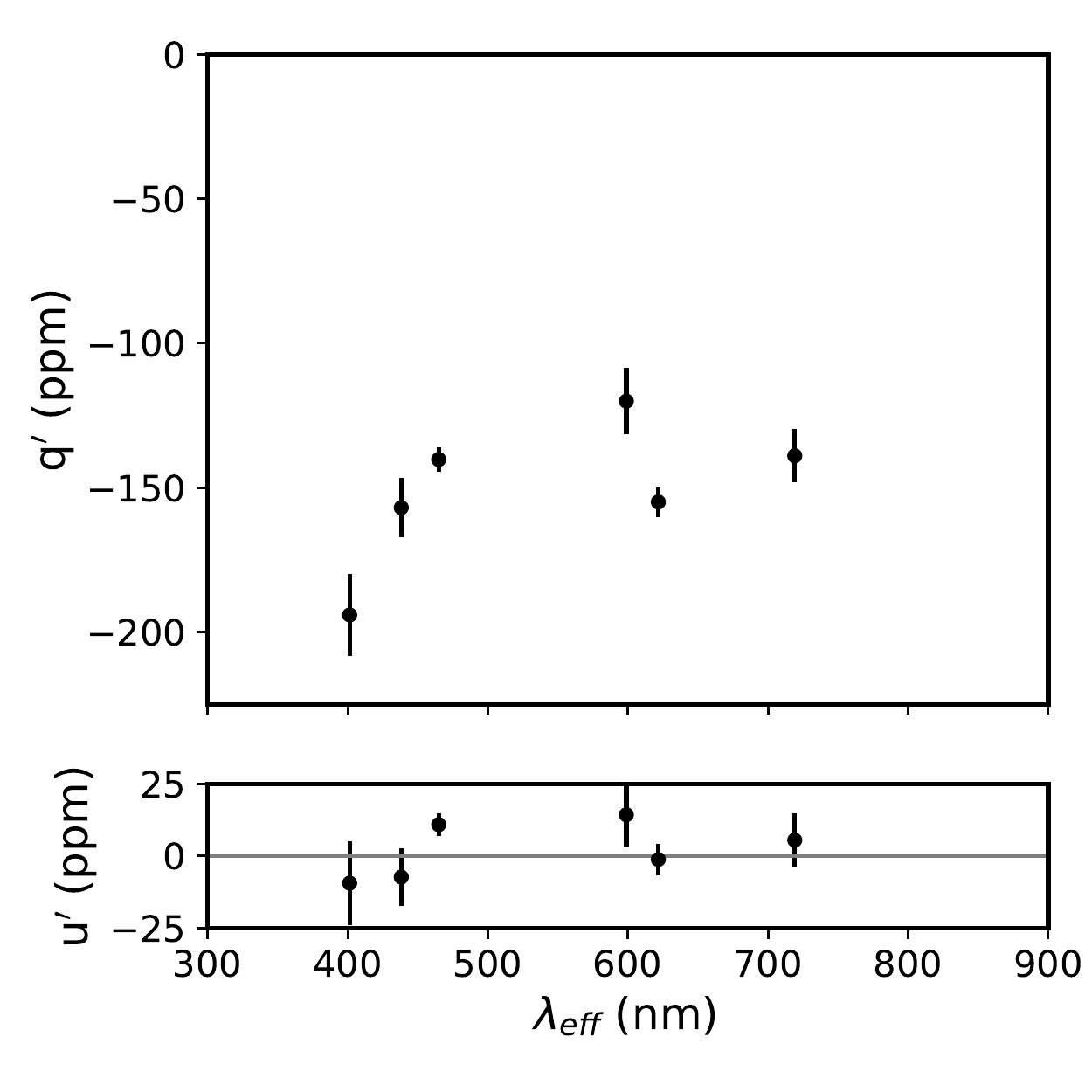}\\
    \includegraphics[clip, trim={0.3cm, 0.2cm, 0.0cm, 0.0cm}, width=0.48\textwidth]{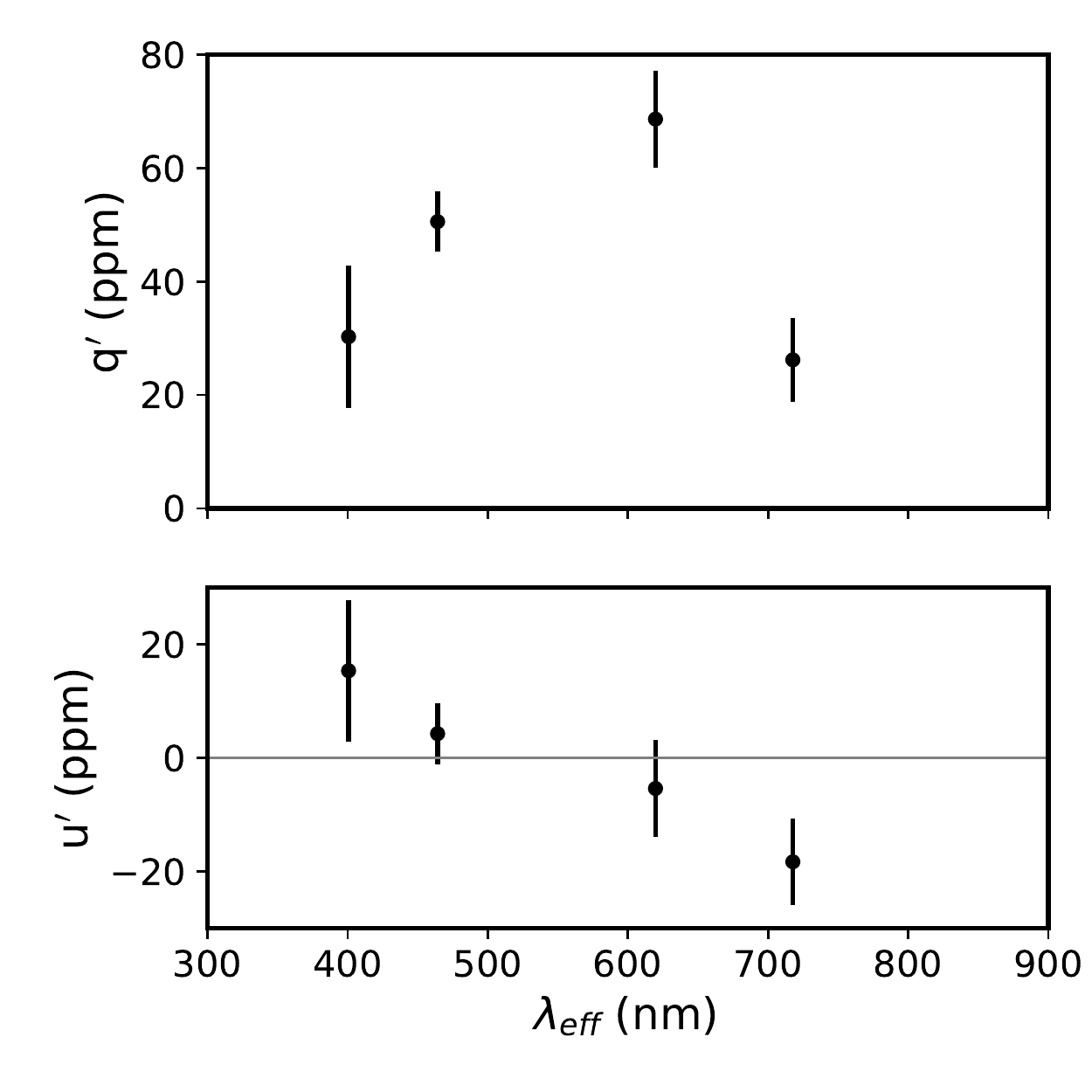}
    \includegraphics[clip, trim={0.3cm, 0.2cm, 0.0cm, 0.0cm}, width=0.48\textwidth]{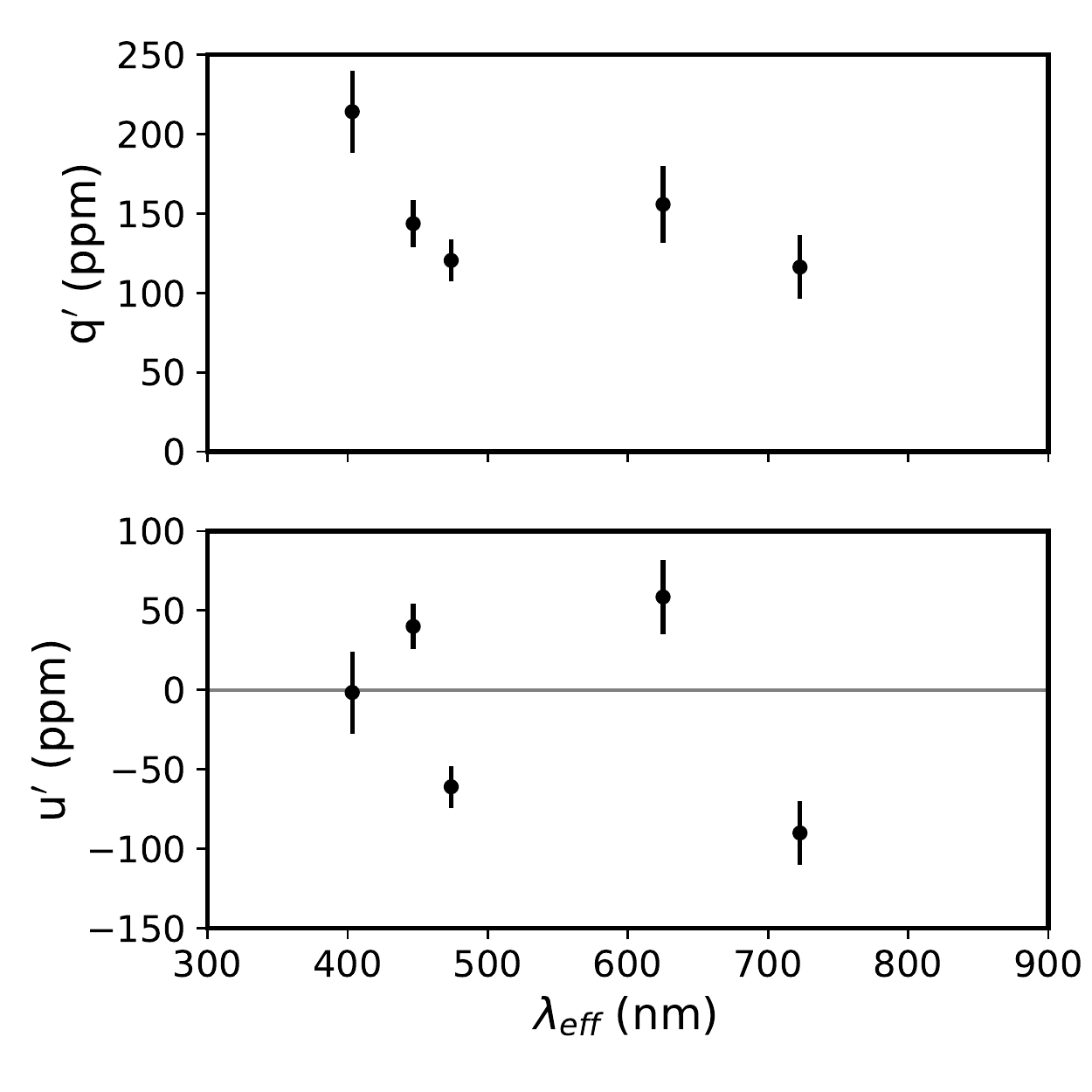}
    \caption{Band-averaged polarisation plots for HD~377 (top left), HD~39060 (top right), HD~188228 (bottom left), and HD~202628 (bottom right). The band-averaged polarisation has had an interstellar contribution subtracted, which was scaled to the stellar distance and position angle of nearby standards. The details of the interstellar contribution are given in the text.}
    \label{fig:ISM_sub}
\end{figure*}

\subsubsection{HD 377}

The fit result for HD~377 returns an interstellar polarisation at 470~nm with $p_{\rm max} =$ 67.8 $\pm$ 0.7 ppm, PA$_i$ $=$ 158.7$^\circ$ $\pm$ 0.4. Figure\ref{fig:ISM_sub} (top left) presents the data with the calculated interstellar polarisation removed. The interstellar contribution was rotated by $-47^{\circ}$ to align $q^{\prime}$ with the disc major axis -- thus the upper panel ($q^{\prime}$) is representative of the disc polarisation.

\subsubsection{HD 39060}

The fit result for HD~39060 returns an interstellar polarisation at 470~nm with $p_{\rm max} =$ 30.7 $\pm$ 0.9 ppm, PA$_i$ $=$ 46.6$^\circ$ $\pm$ 0.6. Figure\ref{fig:ISM_sub} (top right) presents the data with the calculated interstellar polarisation removed. The interstellar contribution was rotated by $-30^\circ$ to align $q^{\prime}$ to the disc major axis -- thus the upper panel ($q^{\prime}$) is proportional to the inversion of the disc polarisation -- given that not all of the disc falls within the aperture.

\subsubsection{HD 188228}

The fit result for HD~188228 returns an interstellar polarisation at 470~nm with $p_{\rm max} =$ 40.9 $\pm$ 0.4 ppm, PA$_i$ $=$ 146.9$^\circ$ $\pm$ 1.0. Figure\ref{fig:ISM_sub} (bottom left) presents the data with the calculated interstellar polarisation removed. The interstellar contribution was rotated by $-11^\circ$ to align $q^{\prime}$ to the disc major axis -- thus the upper panel ($q^{\prime}$) is representative of the disc polarisation.

\subsubsection{HD 202628}

The fit result for HD~202628 returns an interstellar polarisation at 470~nm with $p_{\rm max} =$ 84.2 $\pm$ 0.4 ppm, PA$_i$ $=$ 92.0$^\circ$ $\pm$ 1.0. Figure\ref{fig:ISM_sub} (bottom right) presents the data with the calculated interstellar polarisation removed. The interstellar contribution was rotated by $-134^\circ$ to align $q^{\prime}$ to the disc major axis -- thus the upper panel ($q^{\prime}$) is representative of the disc polarisation.

\subsection{Comparison with literature measurements}

\begin{figure*}
    \centering
    \includegraphics[width=0.48\textwidth]{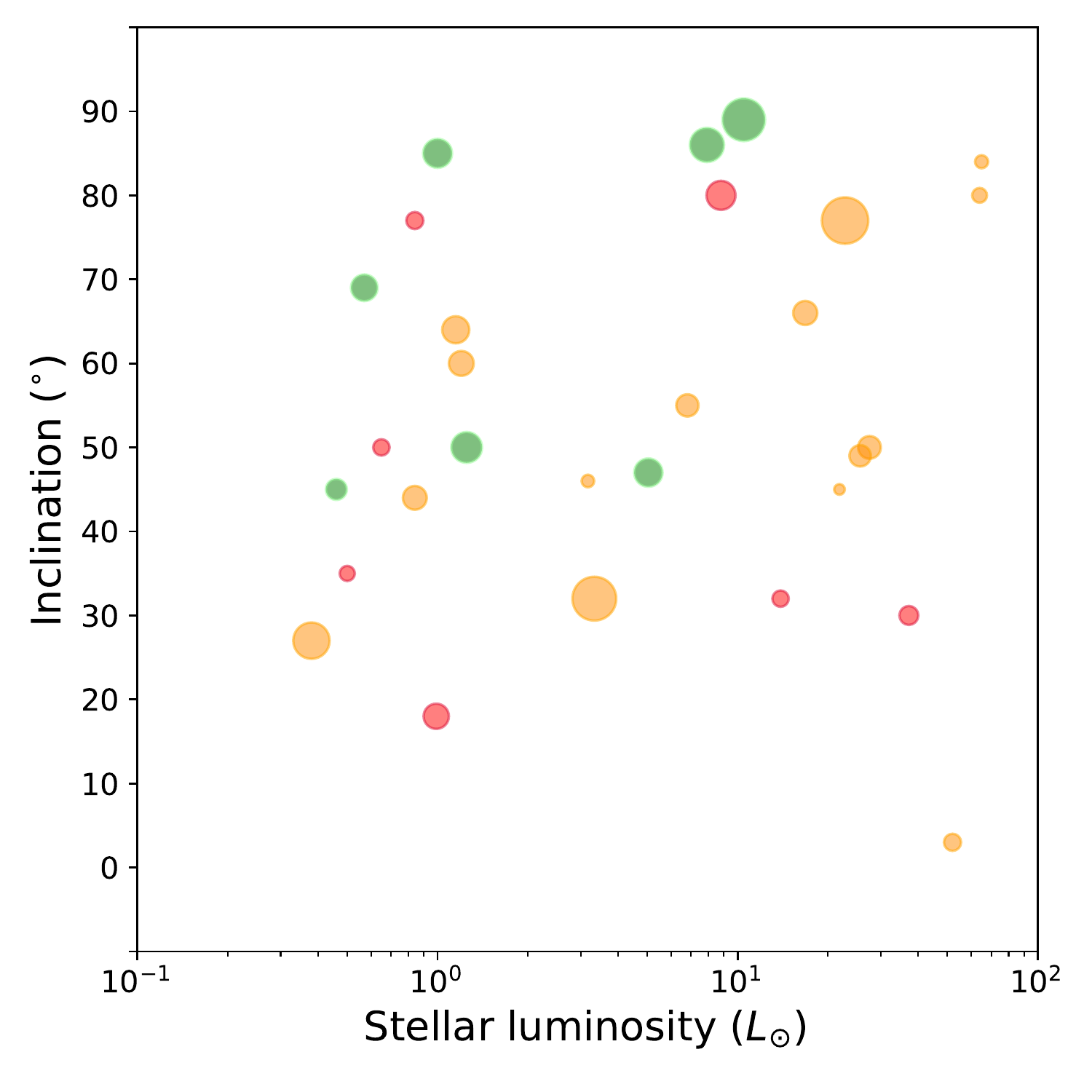}
    \includegraphics[width=0.48\textwidth]{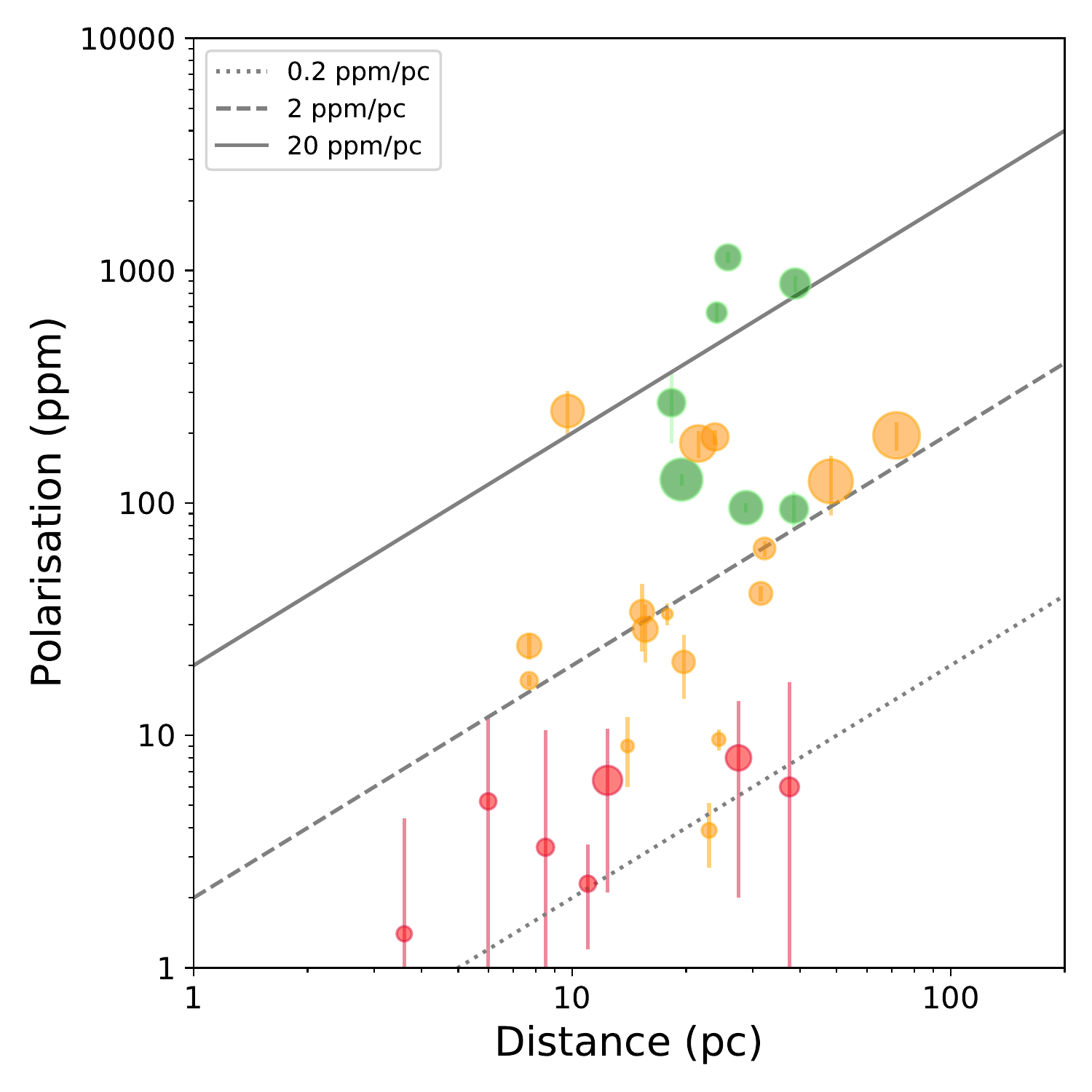}    
    \caption{Plots illustrating polarisation detections with aperture polarimetry in stellar luminosity vs. disc inclination (left) and distance vs. polarisation (right). We plot the debris disc detections using multi-wavelength measurements as green circles, orange circles denote debris discs identified from a single wavelength measurement, or targets with significant polarisation not attributable to a debris disc, and red circles denote non-detections ($P$/d$P \leq 3$). Data point size is scaled to the disc fractional luminosity. On the left we see that the detections are strongly biased toward higher inclination systems ($i > 45\degr$), likely due to the fact that these are also the discs most easily imaged in scattered light. On the right we see that detection generally requires a high disc polarisation ($> 100~$ppm), although multi-wavelength analysis can dig out discs at levels of 10s ppm if the ISM component is accurately quantified. The ISM component is denoted by the dotted, dashed, and solid grey lines at 0.2, 2, and 20 ppm/pc. These values are representative of N hemisphere targets within the LHB, S hemisphere targets within the LHB, and regions outside the LHB, respectively.}
    \label{fig:pol_correlations}
\end{figure*}

We combine these new results with the previously published HIPPI polarimetric detections from \citet{2016Cotton} (4), \citet{2017Cotton} (6), and \cite{2020Marshall} (1), and literature measurements of optical polarisation of debris discs from \citet{2010Bailey} (5), \cite{2015GarciaGomez} (4), and \citet{2020Piirola}\footnote{We identified several additional debris disc host stars in the samples of \cite{2015GarciaGomez} and \cite{2020Piirola}, but only spatially resolved discs were appropriate for inclusion here. We also note \cite{2020Piirola} identify HD~101805, with a substantial polarisation of 223~$\pm$~11~ppm (comparable to $\beta$ Pic), as having a circumstellar origin. However, no infrared excess has been detected for this star.} (4). Using the enlarged sample of 30 unique targets with 25 detections of significant polarisation, we search for trends in the behaviour of the polarimetric signal as a function of the disc fractional luminosity and the host star luminosity. The properties of the compiled sample are summarised in table \ref{tab:pol_props}, and plotted in figure \ref{fig:pol_correlations}. The lack of any strong correlation between total intensity scattered light brightness and continuum emission has been previously reported \citep{2014Schneider}. It is immediately clear that the detections are, in general, biased toward brighter systems at more edge-on inclinations \citep[identified in ][]{2020Esposito}. This is due to the obfuscating factors of the disc orientation, which affects the amount of light scattered in the direction of the observer, and the dust grain size and shape, which affect the dust scattering properties. 

Without multi-wavelength data disc polarisation is usually indistinguishable from the ISM. The right hand panel of figure \ref{fig:pol_correlations} shows that polarisation measured from many disc systems is similar to the interstellar polarisation expected. The utility of multi-band measurements and decent determinations of interstellar polarisation in surrounding stars can be seen here, in that some of the detections are actually less polarised than other systems at similar distances. But overall disc detections are for systems with higher polarisation than is probable for their distance -- which is reassuring.

The left hand panel of figure \ref{fig:pol_correlations}, reveals no trends in terms of stellar/disc luminosity or inclination. More systems with higher inclinations have detections, but this is only because more of those have been identified as disc systems by other means biasing the sample. Logically greater inclination is better for detection, but higher infrared excess should also be a positive factor. We infer that the ISM is obfuscating any impact that these factors would have on the detectability of debris discs in aperture polarimetry.

\begin{table}
\caption{Properties of additional debris disc stars detected in aperture polarisation. If a target was observed more than once, we quote the highest precision value here. If a target was observed at multiple wavelengths, we quote the polarisation measured in SDSS $g^{\prime}$ value or Johnson B band here. \label{tab:pol_props}}
\centering
\begin{tabular}{rrrcrrc}
    \hline\hline
    \multicolumn{1}{c}{HD} & \multicolumn{1}{c}{$L_{\star}$} & \multicolumn{1}{c}{$d_{\star}$} & \multicolumn{1}{c}{$L_{\rm dust}/L_{\star}$} & \multicolumn{1}{c}{$i$} &   \multicolumn{1}{c}{$p$}  & Ref. \\
         &($L_{\odot}$)& (pc) &   (ppm)  &($\degr$)&\multicolumn{1}{c}{(ppm)}& \\ 
    \hline
    105    & 1.25& 38.8 &   280& 50& 880$\pm$70 & 1 \\
    10700  & 0.50&  3.6 &   \phantom{0}16& 35&   1.4$\pm$ 3.0& 2 \\
    20794  & 0.65&  6.0 &   \phantom{0}20& 50&   5.2$\pm$ 6.7& 2 \\
    22484  & 3.17& 14.0 &   \phantom{0}10& 46& \phantom{00}9~$\pm$~3\phantom{.0}  & 3 \\
    31392  & 0.57& 25.8 &   130& 69& 1140$\pm$60 & 1 \\
    38858  & 0.84& 15.3 &   \phantom{0}77& 44& \phantom{0}34~$\pm$~11\phantom{.0} & 3 \\
    71155  & 37.2& 37.5 &   \phantom{0}32& 30& \phantom{00}6~$\pm$~11\phantom{.0}  & 3 \\
    95418  & 65.0& 24.4 &   \phantom{0}11& 84&   9.6$\pm$ 1.0& 4 \\
    102647 & 13.9& 11.0 &   \phantom{0}20& 32&   2.3$\pm$ 1.1& 4 \\
    105211 & 6.80& 19.7 &   \phantom{0}58& 55&  20.7$\pm$ 6.3& 2 \\
    107146 & 0.99& 27.5 &   100& 18& \phantom{0}8~$\pm$~6\phantom{.0}  & 3 \\
    109085 & 5.04& 18.3 &   177& 47& 270$\pm$90\phantom{.0} & 1 \\
    115617 & 0.84&  8.5 &   \phantom{0}23& 77&   3.3$\pm$ 7.2& 2 \\
    139006 & 64.0& 23.0 &   \phantom{0}15& 80&   3.9$\pm$ 1.2& 4 \\
    141891 & 8.80& 12.4 &   200& 80&   6.4$\pm$ 4.3& 5 \\
    161868 & 27.5& 31.5 &   \phantom{0}63& 50&  40.8$\pm$ 3.1& 4 \\
    172167 & 52.0&  7.7 &   \phantom{0}23&  3&  17.2$\pm$ 1.0& 4 \\
    172555 & 7.90& 28.8 &   \phantom{0}52& 86& 95.4$\pm$ 4.7& 6 \\
    207129 & 1.20& 15.6 &   \phantom{0}93& 60&  28.6$\pm$ 8.0& 2 \\
    216956 & 16.8&  7.7 &   \phantom{0}81& 66& 111.5$\pm$ 7.4& 5 \\
    \hline
\end{tabular}
\begin{flushleft}
\textbf{References:} 
1. \cite{2015GarciaGomez}, 
2. \cite{2017Cotton}, 
3. \cite{2020Piirola}, 
4. \cite{2010Bailey}, 
5. \cite{2016Cotton},
6. \cite{2020Marshall}. \\
\end{flushleft}
\end{table}

\section{Discussion}
\label{sec:discussion}

Here we discuss the measurements of individual sources where a detection was made, and summarize the findings for the non-detections in a separate sub-section.

\subsection{Detections}

We detected significant polarisation from seven of the~\ntgt targets, but can only unambiguously assign the origin of the measured polarisation to the circumstellar debris disc in two cases, namely HD~377 and HD~39060. Another two, HD~188228 and HD~202628, have evidence for multiple contributions (of similar weight) to the measured polarisation. We model the interstellar polarisation and subtract it for all four systems, presenting the polarisation as a function of wavelength. The three of these systems contained wholly within the aperture may now be suitable for radiative transfer modelling to establish grain properties, as was done for HD 172555 \citep{2020Marshall}.

HD~377 is the most complicated case of detection in the sample. A high signal-to-noise \textit{HST} image exists of the disc in a close to edge-on orientation, the disc is relatively bright in fractional luminosity, and is compact enough to lie fully within the HIPPI aperture \citep{2016Choquet}. The multi-wavelength HIPPI observations show some curious behaviour, namely an inversion of the polarisation sign at wavelengths beyond 600~nm which would be consistent with the polarisation predominantly arising from dust grains several microns in size.

HD~39060 represents the singular unambiguous detection of polarisation from circumstellar dust within the sample observed here. Given its proximity, disc brightness, and edge-on orientation, along with existing detections of the system \citep{1982Tinbergen,1991Gledhill,2000Krivova}, this result was expected. The magnitude of the polarisation measured here is consistent with these previous measurements at comparable wavelengths, and the multi-wavelength polarimetry presented here reveals a turn-over in the magnitude of $q^{\prime}$ polarisation around 600~nm.

HD~188228 exhibited measurable polarisation in our measurements, which is remarkable given the relative faintness of its circumstellar disc, and the intermediate inclination of its presentation toward us. The observed polarisation has a distinct fall-off in the longest wavelength, 650LP band, along with a rotation in position angle. It is upon these two points that the case for detection of polarisation intrinsic to the system rests. If this detection were attributable to circumstellar dust rather than the intervening ISM, the implication is that the dust grains have a higher albedo in the optical than at infrared wavelengths, which may not be implausible given the high albedoes inferred for some young debris discs \citep{2018Choquet}.

We also obtained a strong detection of polarisation from HD~202628, despite its relative faintness both in fractional luminosity and host star brightness. This system has been imaged in scattered light and continuum emission \citep{2010Krist,2019Faramaz}, revealing the debris disc to be a narrow, inclined and eccentric belt. As symmetry breaking increases polarimetric signals, this explains the detection where all other factors would disfavour such a result. There is no clear trend in the polarisation as a function of wavelength, and there is significant scatter in $u^{\prime}$ -- which may be a consequence of the disc eccentricity.

\subsection{Non-detections}

We did not detect substantial polarisation that could be ascribed to the disc from the remaining eight targets. Of the remaining systems not consistent with polarisation arising solely from the interstellar medium, the young, active stars HD~92945 and HD~197481 are likely contaminated by stellar activity, as is HD~115892. In these cases a non-detection is not meaningful. It is worth making some specifc remarks about the other systems, as well as others presented previously \citep{2016Cotton, 2017Cotton}.

Several of the stars which were not detected in these observations have relatively bright debris discs ($L_{\rm dust}/L_{\star} \geq 10^{-4}$), previously imaged in scattered light, oriented at moderate inclinations which should be favourable to detection. However, in most cases the non-detection can be explained as a result of the target's angular extent and orientation, or its intrinsic faintness and the obfuscating factor of the ISM foreground. 

If we consider the monochromatic observations of systems reported in \citet{2016Cotton, 2017Cotton} as well, then HD~102647 and HD~115892 are amongst the faintest discs targeted with aperture polarimetry, with $L_{\rm dust}/L_{\star} \leq 10^{-5}$. Their detection would have been remarkable given the precision of HIPPI and the expectation that the polarimetric signal would be some fraction of the disc brightness. 

HD~10700 is a regularly used as a low polarisation standard for HIPPI observations. It is similarly faint, despite its proximity, and therefore despite the absence of ISM foreground and with the compact architecture of the disc, it exhibits no significant polarisation (e.g. \citealp{2017Cotton}). The cause for non-detection of this system is made ambiguous as the inner edge of HD~10700's debris disc is ill-defined \citep[e.g.][]{2014Lawler,2016Macgregor} and the HIPPI aperture may have been too small to encompass the disc.

HD~105211's debris belt has an angular size ($\theta_{\rm R} \simeq 9\arcsec$), greater than the HIPPI aperture (diameter of 6\arcsec). At its inclination, only a small fraction of the disc would be enclosed in the aperture centred on the star, reducing the potential signal from the circumstellar dust. We can therefore rule out a substantial asteroid belt analogue contributing to the emission from this system, as has been previously postulated in studies of the system \citep{2017Hengst}.

HD~102647 and HD~161868 are large debris discs, both relatively faint, and both with A-type host stars. Similar to HD~10700, HD~102647 has also been previously used as a low polarisation standard to calibrate other HIPPI observations \citep[e.g.]{2017bCotton,2020Cotton}. The absence of any detectable polarisation from these two discs could be due to the combination of disc angular extent and orientation, as with HD~105211, or the larger minimum grain size around A-type stars reducing the polarimetric signal at optical wavelengths.

We also detect no polarimetric signal from HD~216956. In this case, the HIPPI aperture is small enough to fully exclude the main outer belt of the system (similar to HD~105211), such that it is the signal from a proposed asteroid (or cometary) belt deep within the system is probed with our observations, rather than the main belt.

\section{Conclusions}
\label{sec:conclusions}

We observed a sample of twelve debris discs at optical wavelengths in multiple filter bands with parts-per-million sensitivity using the aperture polarimetry instruments HIPPI and HIPPI-2. The targets were selected to provide a representative cross-section of debris discs in stellar spectral type, stellar age, and disc fractional luminosity. 

We clearly identified a polarimetric signature attributable to circumstellar dust from HD~377 and HD~39060; this is the first detection of polarisation from HD~377's disc. Additionally, HD~188288 and HD~202628 exhibited mixed contributions to the polarisation suggestive of significant debris disc induced signal within their measurements. For these four systems we present plots of polarisation wavelength dependence, based on our best efforts to remove the interstellar polarisation component. With this modelling of the grains may be attempted, as it was for HD~172555 \citep{2020Marshall}. For the remaining eight systems, the measured polarisation signals were variously dominated by components attributable to either the intervening interstellar medium or stellar activity.

The large number of non-detections within the selected sample, particularly cases that would be otherwise ideal systems for this approach e.g. HD~197481 (AU Mic) and HD~109573A (HR~4796A), demonstrates the fundamental problems associated with undertaking these kinds of observations. 

In this work we have attempted to provide a route map for the reliable measurement of debris disc polarisation, and present a systematic approach to robustly assess the available evidence, weighing the competing factors that may obfuscate extraction of that signal from the data. Disentangling intrinsic stellar polarisation from extrinsic circumstellar dust is presently too difficult, but removal of a potentially dominant and wavelength dependent ISM foreground is possible, yet requires delicate handling to be achieved.

\section*{Acknowledgements}

This research has made use of the SIMBAD database, operated at CDS, Strasbourg, France \citep{2000Wenger}. This research has made use of NASA's Astrophysics Data System.

The authors wish to thank Stephen Marsden for his collaborator recommendations.

JPM acknowledges research support by the Ministry of Science and Technology of Taiwan under grant MOST109-2112-M-001-036-MY3. DVC thanks the Friends of MIRA for their support.

\textit{Software:} {\sc Astropy} \citep{2013Astropy,2018Astropy}, {\sc Numpy} \citep{2020Numpy}, {\sc Matplotlib} \citep{2007Matplotlib}.

\textit{Facilities:} Anglo-Australian Telescope

\section*{Data availability}

The data underlying this article are available in the article and in its online supplementary material.

\bibliographystyle{mnras}
\bibliography{pol_discs}

\appendix

\section{Standard Observations}
\label{sec:std_obs}

\begin{table*}
\caption{Precision in position angle ($\phi$) by Observing Run.}
\centering
\tabcolsep 4 pt
\begin{tabular}{lc|cccccccccccc|r}
\hline
\hline
Run & & \multicolumn{11}{c|}{Position Angle Standard Observations} & S.D.\\
& \multicolumn{1}{r}{HD:} & 23512 & 80558 & 84810 & 111613 & 147084 & 154445 & 160529 & 161056 & 187929 & 210121 & 203532 & \multicolumn{1}{|c}{($^{\circ}$)} \\
\hline
2014MAYC && 0 & 0 & 0 & 0 & 0 & 1 & 0 & 0 & 0 & 0 & 0 & --   \\
2014AUG && 0 & 0 & 0 & 0 & 1 & 1 & 1 & 0 & \textit{1} & 0 & 0 & 0.24 \\
2015MAY && 0 & 0 & 0 & 0 & \textit{4} & 1 & 0 & 0 & 0 & 0 & 0 & 0.17 \\
2015JUN && 0 & 0 & 0 & 0 & 1 & 1 & 0 & 0 & 0 & 0 & 0 & 0.14 \\
2015OCT && 1 & 0 & 0 & 0 & 0 & \textit{2} & 0 & 0 & \textit{2} & 0 & 0 & 0.24 \\
2015NOV && 0 & 1 & 0 & 0 & 0 & 0 & 0 & 0 & 0 & 0 & 0 & --   \\
2016DEC && 0 & 1 & 2 & 0 & 0 & 0 & 0 & 0 & 0 & 0 & 0 & 0.49 \\
2017JUN && 0 & 0 & 0 & 0 & 2 & 1 & 1 & 0 & 0 & 0 & 0 & 1.11 \\
2017AUG && 0 & 0 & 0 & 0 & 1 & 1 & 0 & 0 & 1 & 0 & 0 & 0.53 \\
2018MAR && 0 & 1 & 0 & 1 & 1 & 0 & 0 & 0 & 1 & 0 & 0 & 0.26 \\
2018JUN && 0 & 0 & 0 & 0 & 1 & 1 & 0 & 1 & 0 & \textit{2} & 0 & 0.80 \\
2018JUL && 0 & 0 & 0 & 0 & 1 & 1 & 1 & 0 & 1 & 0 & 1 & 1.56 \\
2018AUG && 0 & 0 & 0 & 0 & 3 & 0 & 3 & 0 & \textit{5} & 0 & 0 & 0.86 \\
\hline
\end{tabular}
\begin{flushleft}
\textbf{Notes:} * All standards were observed in $g^\prime$ except where the number is italicised, in which case one instance was observed without a filter, except for HD~147084 in 2015MAY, where there are two instances. * One instance of HD~147084, HD~160529 and HD~187929 observed during 2018AUG used the R PMT, all other observations used the B PMT.\\
\end{flushleft}
\label{tab:pa_prec}
\end{table*}

\begin{table*}
\caption{Summary of telescope polarisation (TP) calibrations.}
\label{tab:tp}
\tabcolsep 2 pt
\begin{tabular}{lccrrr|ccccccc|rr}
\hline
\hline
Run(s) &   Fil  & PMT$^a$  & $\lambda_{\rm eff}$ & Eff.  & \multicolumn{1}{r}{} & \multicolumn{7}{c|}{Standard Observations} & \multicolumn{1}{c}{$q \pm \Delta q$} & \multicolumn{1}{c}{$u \pm \Delta u$}\\
&&&(nm)& (\%) & \multicolumn{1}{r}{HD:} & \phantom{00}2151 & \phantom{0}10700 & \phantom{0}48915 & 102647 & 102870 & 127762 & 140573 &\multicolumn{1}{c}{(ppm)} &\multicolumn{1}{c}{(ppm)} \\
\hline
2014MAYC \& 2014AUG &   500SP       &   B   & 445.3 & 83.4 && 3 & 0 & 0 & 0 & 0 & 0 & 1 & $-$42.1 $\pm$ \phantom{0}2.1  & $-$37.3 $\pm$ \phantom{0}2.1 \\   
                    &   $g^\prime$  &   B   & 472.1 & 90.5 && 3 & 0 & 2 & 2 & 1 & 0 & 4 & $-$39.9 $\pm$ \phantom{0}0.9  & $-$38.3 $\pm$ \phantom{0}0.9 \\
                    &   Clear       &   B   & 488.7 & 85.5 && 2 & 0 & 0 & 0 & 0 & 0 & 0 & $-$30.7 $\pm$ \phantom{0}2.0  & $-$42.9 $\pm$ \phantom{0}2.0 \\
                    &   $r^\prime$  &   B   & 601.1 & 84.8 && 0 & 0 & 0 & 2 & 0 & 0 & 3 & $-$17.7 $\pm$ \phantom{0}1.8  & $-$52.5 $\pm$ \phantom{0}1.8 \\
\hline
2015MAY \& 2015JUN  &   425SP       &   B   & 406.4 & 59.8 && 0 & 0 & 0 & 0 & 0 & 0 & 5 & $-$41.9 $\pm$ \phantom{0}6.1  & 18.5 $\pm$ \phantom{0}6.0 \\
                    &   $g^\prime$  &   B   & 471.8 & 90.1 && 0 & 0 & 2 & 1 & 0 & 0 & 4 & $-$35.8 $\pm$ \phantom{0}1.3  & 4.8 $\pm$ \phantom{0}1.3 \\
                    &   $r^\prime$  &   R   & 623.3 & 78.8 && 0 & 0 & 0 & 1 & 0 & 0 & 1 & $-$35.0 $\pm$ \phantom{0}3.5  & 6.2 $\pm$ \phantom{0}6.2 \\
\hline
2015OCT \& 2015NOV  &   $g^\prime$  &   B   & 468.4 & 90.2 && 3 & 0 & 3 & 0 & 0 & 0 & 0 & $-$50.4 $\pm$ \phantom{0}1.1  & $-$0.2 $\pm$ \phantom{0}1.1 \\
                    &   Clear       &   B   & 476.8 & 83.8 && 5 & 0 & 6 & 0 & 0 & 0 & 0 & $-$45.4 $\pm$ \phantom{0}0.8  & $-$0.5 $\pm$ \phantom{0}0.7 \\
                    &   $r^\prime$  &   R   & 621.2 & 79.3 && 1 & 0 & 2 & 0 & 0 & 0 & 0 & $-$34.4 $\pm$ \phantom{0}2.3  & $-$5.1 $\pm$ \phantom{0}2.5 \\
\hline
2016DEC             &   425SP       &   B   & 400.8 & 51.8 && 1 & 0 & 3 & 0 & 0 & 0 & 0 & $-$32.7 $\pm$ \phantom{0}3.5  & 2.1 $\pm$ \phantom{0}3.9 \\
                    &   $g^\prime$  &   B   & 466.9 & 87.5 && 1 & 0 & 2 & 0 & 0 & 0 & 0 & $-$27.3 $\pm$ \phantom{0}2.0  & 1.5 $\pm$ \phantom{0}1.9 \\
\hline
2017JUN \& 2017AUG  &   425SP       &   B   & 401.6 & 52.0 && 2 & 0 & 3 & 2 & 2 & 0 & 0 & $-$7.3 $\pm$ \phantom{0}3.6   & 8.5 $\pm$ \phantom{0}3.6 \\
                    &   500SP       &   B   & 440.1 & 77.0 && 2 & 0 & 3 & 2 & 2 & 0 & 0 & $-$10.0 $\pm$ \phantom{0}1.7  & $-$0.4 $\pm$ \phantom{0}1.6 \\
                    &   $g^\prime$  &   B   & 468.9 & 88.1 && 2 & 0 & 2 & 2 & 2 & 0 & 0 & $-$9.1 $\pm$ \phantom{0}1.5   & $-$2.6 $\pm$ \phantom{0}1.4 \\
                    &   $r^\prime$  &   R   & 622.1 & 82.2 && 3 & 0 & 3 & 2 & 2 & 0 & 0 & $-$10.4 $\pm$ \phantom{0}1.3  & $-$7.0 $\pm$ \phantom{0}1.3 \\
                    &   650LP       &   R   & 720.3 & 65.5 && 3 & 0 & 3 & 1 & 0 & 0 & 0 & $-$8.2 $\pm$ \phantom{0}2.3   & $-$5.1 $\pm$ \phantom{0}2.4 \\
\hline
2018MAR             &   $g^\prime$  &   B   & 465.8 & 81.9 && 0 & 0 & 3 & 3 & 3 & 0 & 0 & 130.1 $\pm$ \phantom{0}0.9    & 3.8 $\pm$ \phantom{0}0.9 \\
                    &   $r^\prime$  &   R   & 621.8 & 81.3 && 0 & 0 & 0 & 1 & 0 & 0 & 0 & 84.9 $\pm$ \phantom{0}3.8     & $-$9.5 $\pm$ \phantom{0}3.4 \\
\hline
2018JUN             &   $g^\prime$  &   B   & 475.7 & 92.8 && 0 & 0 & 0 & 2 & 0 & 3 & 4 & 1002.0 $\pm$ 14.3   & 1782.1 $\pm$ 14.3 \\
                    &   $r^\prime$  &   B   & 605.7 & 60.8 && 0 & 0 & 0 & 0 & 0 & 3 & 4 & 170.6 $\pm$ \phantom{0}4.8 & 369.8 $\pm$ \phantom{0}4.8\\
\hline
2018JUL \& 2018AUG  &   $g^\prime$  &   B   & 470.6 & 74.5 && 3 & 3 & 2 & 2 & 1 & 0 & 2 & $-$12.9 $\pm$ \phantom{0}1.1 & 4.1 $\pm$ \phantom{0}1.0\\
\hline
\end{tabular}
\begin{flushleft}
{$^a$} B, R indicate blue- and red-sensitive H10720-210 and H10720-20 photomultiplier-tube detectors, respectively.
\end{flushleft}
\label{tab:mod}
\end{table*}

The instrumental position angle, $\phi$, is calibrated with reference to high polarisation standard stars observed in $g^\prime$ or Clear. Table \ref{tab:pa_prec} lists the standards observed for each run. The errors associated with the literature values are around a degree. The column labelled `S.D.' gives the standard deviation of $\Delta \phi=\phi_{\rm obs} - \phi_{\rm lit}$, where $\phi_{\rm obs}$ is the $\theta$ for the observation after calibration, and $\phi_{\rm lit}$ the literature value as given in \citet{2020Bailey}. 

Tables \ref{tab:tp} gives a summary of observations used to calibrate the TP for each run in each band. The polarisation of each of these stars is assumed to be zero.

\section{Full list of interstellar controls}
\label{apx:controls}
The following is a complete list of interstellar controls as shown in figures \ref{fig:ism1}, \ref{fig:ism2}, \ref{fig:ism3} and \ref{fig:ism4}. 
\newline

\noindent\textbf{HD 377:} 1: HD 224617, 2: HD 225003, 3: HD 224156, 4: HD 222368, 5: HD 4628, 6: HD 101, 7: HD 7047, 8: HD 4915, 9: HD 1832, 10: HD 218687, 11: HD 225261, 12: HD 219877, 13: HD 216385, 14: HD 8648, 15: HD 4813, 16: HD 1388, 17: HD 5294, 18: HD 215648, 19: HD 6715, 20: HD 361, 21: HD 4307, 22: HD 217924, 23: HD 8262, 24: HD 7439, 25: HD 222345, 26: HD 693, 27: HD 9562, 28: HD 4128, 29: HD 222422, 30: HD 9472, 31: HD 5035, 32: HD 12235 33: HD 211476, 34: HD 203, 35: HD 4247, 36: HD 11171, 37: HD 3795, 38: HD 8350, 39: HD 1562, 40: HD 11964A, 41: HD 8129, 42: HD 14214, 43: HD 10700, 44: HD 12846, 45: HD 4208, 46: HD 212697, 47: HD 11007.\medbreak

\noindent\textbf{HD 39060:} 1: HD  40105, 2: HD  32743, 3: HD  33262, 4: HD  45289, 5: HD  31746, 6: HD  52298, 7: HD  50223, 8: HD  32820, 9: HD  28454, 10: HD  39014, 11: HD  59468, 12: HD  62848, 13: HD  29992, 14: HD  63008, 15: HD  64185, 16: HD  62644, 17: HD  65907, 18: HD  44447, 19: HD  49095, 20: HD  37495, 21: HD  69655, 22: HD  43834, 23: HD  50806, 24: HD  20794, 25: HD  48938, 26: HD  45588, 27: HD  74956, 28: HD  77370, 29: HD  38393, 30: HD  43745, 31: HD  71243, 32: HD  51733, 33: HD  80007, 34: HD  25945, 35: HD  70060, 36: HD  38382, 37: HD  73524, 38: HD  12311, 39: HD  33095, 40: HD  34721.\medbreak

\noindent\textbf{HD 92945:} 1: HD  88742, 2: HD 100407, 3: HD 86629, 4: HD 84117, 5: HD 101614, 6: HD 91889, 7: HD 102365, 8: HD 104982, 9: HD 88215, 10: HD 105452, 11: HD 99610, 12: HD 93497, 13: HD 101198, 14: HD 104731, 15: HD 82241, 16: HD 91324, 17: HD 104304, 18: HD 76932, 19: HD 92588, 20: HD 108767, 21: HD 73524, 22: HD 109085, 23: HD 106516, 24: HD 78612, 25: HD 81809, 26: HD 108799, 27: HD 109141, 28: HD 110304, 29: HD 70060, 30: HD 81997, 31: HD 108510, 32: HD 114613, 33: HD 96937, 34: HD 71196, 35: HD 88072, 36: HD 113415, 37: HD 88725, 38: HD 114432, 39: HD 74956, 40: HD 106116, 41: HD 113720, 42: HD 100563, 43: HD 102870, 44: HD 77370, 45: HD 76151.\medbreak

\noindent\textbf{HD 105211:} 1: HD 93372, 2: HD 101805, 3: HD 90589, 4: HD 91324, 5: HD 110304, 6: HD 80007, 7: HD 128620J, 8: HD 93497, 9: HD 71243, 10: HD 104731, 11: HD 77370, 12: HD 101614, 13: HD 102365, 14: HD 147584, 15: HD 74956, 16: HD 65907, 17: HD 114613, 18: HD 64185, 19: HD 43834, 20: HD 82241, 21: HD 44447, 22: HD 136351, 23: HD 69655, 24: HD 100407, 25: HD 162521, 26: HD 86629, 27: HD 62848.\medbreak

\noindent\textbf{HD 109573A} 1: HD 104731, 2: HD 114613, 3: HD 110304, 4: HD 102365, 5: HD 101614, 6: HD 104982, 7: HD 100407, 8: HD 119756, 9: HD 105452, 10: HD 109379, 11: HD 114432, 12: HD 115659, 13: HD 113415, 14: HD 125473, 15: HD  93497, 16: HD 113720, 17: HD 108767, 18: HD 115617, 19: HD 109085, 20: HD 91324, 21: HD 109141, 22: HD 108799, 23: HD 128620J, 24: HD 88742, 25: HD 101198, 26: HD 135235, 27: HD 131342, 28: HD 106516, 29: HD 104304, 30: HD 136351, 31: HD 86629, 32: HD 108510, 33: HD 99610, 34: HD 117860, 35: HD 82241, 36: HD 126679.\medbreak

\noindent\textbf{HD 115892:} 1: HD 114613, 2: HD 119756, 3: HD 114432, 4: HD 115659, 5: HD 110304, 6: HD 104731, 7: HD 113415, 8: HD 104982, 9: HD 115617, 10: HD 102365, 11: HD 105452, 12: HD 101614, 13: HD 100407, 14: HD 108767, 15: HD 109085, 16: HD 136351, 17: HD 109141, 18: HD 108799, 19: HD 128620J, 20: HD 117860, 21: HD 106516, 22: HD 93497, 23: HD 108510, 24: HD 116568, 25: HD 104304, 26: HD 101198, 27: HD 91324, 28: HD 143114, 29: HD 127352.\medbreak

\noindent\textbf{HD 161868:} 1: HD 161096, 2: HD 162917, 3: HD 164651, 4: HD 164259, 5: HD 157347, 6: HD 165777, 7: HD 159561, 8: HD 171802, 9: HD 153210, 10: HD 159332, 11: HD 175638, 12: HD 150433, 13: HD 153631, 14: HD 155125, 15: HD 173880, 16: HD 147449, 17: HD 147512, 18: HD 173667, 19: HD 157172, 20: HD 156164, 21: HD 182640, 22: HD 160915, 23: HD 181391, 24: HD 161797, 25: HD 151192, 26: HD 145229, 27: HD 180409, 28: HD 168874, 29: HD 151504, 30: HD 163993, 31: HD 147547, 32: HD 164595, 33: HD 185124, 34: HD 169916, 35: HD 153808, 36: HD 141004, 37: HD 155060, 38: HD 140573, 39: HD 142093, 40: HD 187691, 41: HD 145518, 42: HD 152598, 43: HD 140667, 44: HD 176377, 45: HD 144766, 46: HD 150680, 47: HD 149890, 48: HD 165135, 49: HD 190412, 50: HD 169586, 51: HD 138573, 52: HD 137898.\medbreak

\noindent\textbf{HD 181327:} 1: HD 173168, 2: HD 165499, 3: HD 167425, 4: HD 197157, 5: HD 190248, 6: HD 188887, 7: HD 188114, 8: HD 177389, 9: HD 162521, 10: HD 183414, 11: HD 186219, 12: HD 199288, 13: HD 160928, 14: HD 153580, 15: HD 209100, 16: HD 147584, 17: HD 176687, 18: HD 212330, 19: HD 169586, 20: HD 194640, 21: HD 165135, 22: HD 156384, 23: HD 138538, 24: HD 212132, 25: HD 169916, 26: HD 217364, 27: HD 174309, 28: HD 197692, 29: HD 151680, 30: HD 2151, 31: HD 131342.\medbreak

\noindent\textbf{HD 188228:} 1: HD 186219, 2: HD 183414, 3: HD 177389, 4: HD 190248, 5: HD 167425, 6: HD 162521, 7: HD 165499, 8: HD 147584, 9: HD 2151, 10: HD 209100, 11: HD 212330, 12: HD 197157, 13: HD 153580, 14: HD 217364, 15: HD 101805, 16: HD 199288, 17: HD 71243, 18: HD 131342, 19: HD 128620J, 20: HD 43834, 21: HD 90589, 22: HD 212132, 23: HD 93372, 24: HD 12311, 25: HD 44447.\medbreak

\noindent\textbf{HD 197481:} 1: HD 194640, 2: HD 197692, 3: HD 199288, 4: HD 205289, 5: HD 197157, 6: HD 207098, 7: HD 176687, 8: HD 207958, 9: HD 197210, 10: HD 213845, 11: HD 212697, 12: HD 209100, 13: HD 180409, 14: HD 169916, 15: HD 185124, 16: HD 212330, 17: HD 190412, 18: HD 165135.\medbreak

\noindent\textbf{HD 202628:} 1: HD 199288, 2: HD 197157, 3: HD 212132, 4: HD 209100, 5: HD 194640, 6: HD 212330, 7: HD 197692, 8: HD 217364, 9: HD 205289, 10: HD 190248, 11: HD 213845, 12: HD 207098, 13: HD 183414, 14: HD 176687, 15: HD 177389, 16: HD 212697, 17: HD 207958, 18: HD 186219, 19: HD 198075, 20: HD 167425, 21: HD 165499, 22: HD 2261, 23: HD 739.\medbreak

\noindent\textbf{HD 216956:} 1: HD 213845, 2: HD 222422, 3: HD 212697, 4: HD 739, 5: HD 205289, 6: HD 207098, 7: HD 2261, 8: HD 207958, 9: HD 361, 10: HD 693, 11: HD 3795, 12: HD 1388, 13: HD 4247, 14: HD 4128, 15: HD 199288, 16: HD 212330, 17: HD 209100, 18: HD 197692, 19: HD 4307, 20: HD 194640, 21: HD 4813, 22: HD 197157, 23: HD 8129.

\bsp	
\label{lastpage}
\end{document}